\UseRawInputEncoding

\documentclass[twocolumn, astrosymb]{aastex631}
\usepackage[T1]{fontenc}
\usepackage{tipa}
\usepackage{accents}
\newcommand{\ubar}[1]{\underaccent{\bar}{#1}}
\usepackage{tablefootnote}

\newcommand{\kms}{km\,s$^{-1}$}

\newcommand{\teff}{T$_{\rm eff}$ }
\newcommand{\logg}{log\,$g$ }
\newcommand{\msun}{M$_{\odot}$}

\submitjournal{ApJ}

\shorttitle{Sgr2Aqu2}
\shortauthors{Zaremba et al.}

\begin{document}

\title{GHOST commissioning science results - IV: Chemodynamical analyses of Milky Way satellites Sagittarius II and Aquarius II}

\correspondingauthor{Dasha Zaremba}
\email{dariazaremba@uvic.ca}

\author{Daria Zaremba}
\affiliation{Department of Physics and Astronomy, University of Victoria, PO Box 3055, STN CSC, Victoria BC V8W 3P6, Canada}

\author{Kim Venn}
\affiliation{Department of Physics and Astronomy, University of Victoria, PO Box 3055, STN CSC, Victoria BC V8W 3P6, Canada}

\author{Christian R. Hayes}
\affiliation{NRC Herzberg Astronomy \& Astrophysics, 5071 West Saanich Road, Victoria, BC V9E 2E7, Canada}
\affiliation{Space Telescope Science Institute, 3700 San Martin Drive, Baltimore, MD 21218, USA}

\author{Rapha\"{e}l Errani}
\affiliation{McWilliams Center for Cosmology, Department of Physics, Carnegie Mellon University, Pittsburgh, PA 15213, USA}

\author{Triana Cornejo}
\affiliation{Department of Physics and Astronomy, University of Victoria, PO Box 3055, STN CSC, Victoria BC V8W 3P6, Canada}

\author{Jennifer Glover}
\affiliation{Department of Physics and Astronomy, University of Victoria, PO Box 3055, STN CSC, Victoria BC V8W 3P6, Canada}
\affiliation{Department of Physics, McGill University, 3600 Rue University, Montreal, QC, Canada}

\author{Jaclyn Jensen}
\affiliation{Department of Physics and Astronomy, University of Victoria, PO Box 3055, STN CSC, Victoria BC V8W 3P6, Canada}

\author{Alan W. McConnachie}
\affiliation{NRC Herzberg Astronomy \& Astrophysics, 5071 West Saanich Road, Victoria, BC V9E 2E7, Canada}
\affiliation{Department of Physics and Astronomy, University of Victoria, PO Box 3055, STN CSC, Victoria BC V8W 3P6, Canada}

\author{Julio F. Navarro}
\affiliation{Department of Physics and Astronomy, University of Victoria, PO Box 3055, STN CSC, Victoria BC V8W 3P6, Canada}

\author{John Pazder}
\affiliation{NRC Herzberg Astronomy \& Astrophysics, 5071 West Saanich Road, Victoria, BC V9E 2E7, Canada}
\affiliation{Department of Physics and Astronomy, University of Victoria, PO Box 3055, STN CSC, Victoria BC V8W 3P6, Canada}

\author{Federico Sestito}
\affiliation{Department of Physics and Astronomy, University of Victoria, PO Box 3055, STN CSC, Victoria BC V8W 3P6, Canada}
\affiliation{Department of Physics, Astronomy and Mathematics, University of Hertfordshire, Hatfield, AL10 9AB, UK}

\author{Andr\'{e} Anthony}
\affiliation{NRC Herzberg Astronomy \& Astrophysics, 5071 West Saanich Road, Victoria, BC V9E 2E7, Canada}

\author{Dave Andersen}
\affiliation{TMT International Observatory, 100 W. Walnut Street, Suite 300, Pasadena, CA 91124, USA}
\affiliation{Department of Physics and Astronomy, University of Victoria, PO Box 3055, STN CSC, Victoria BC V8W 3P6, Canada}

\author{Gabriella Baker}
\affiliation{Australian Astronomical Optics, Macquarie University, 105 Delhi Road, North Ryde, NSW 2113, Australia}








\author{Timothy Chin}
\affiliation{Australian Astronomical Optics, Macquarie University, 105 Delhi Road, North Ryde, NSW 2113, Australia}


\author{Vladimir Churilov}
\affiliation{Australian Astronomical Optics, Macquarie University, 105 Delhi Road, North Ryde, NSW 2113, Australia}


\author{Ruben Diaz}
\affiliation{Gemini Observatory/NSF's NOIRLab, Casilla 603, La Serena, Chile}



\author{Tony Farrell}
\affiliation{Australian Astronomical Optics, Macquarie University, 105 Delhi Road, North Ryde, NSW 2113, Australia}

\author{Veronica Firpo}
\affiliation{Gemini Observatory/NSF's NOIRLab, Av. J. Cisternas 1500 N, 1720236, La Serena, Chile}





\author{Manuel Gomez-Jimenez}
\affiliation{Gemini Observatory/NSF's NOIRLab, Casilla 603, La Serena, Chile}


\author{David Henderson}
\affiliation{Gemini Observatory/NSF's NOIRLab, 670 North A\textquotesingle{}oh\=ok\=u Place, Hilo, HI 96720, USA}




\author{Venu M. Kalari}
\affiliation{Gemini Observatory/NSF's NOIRLab, Casilla 603, La Serena, Chile}







\author{Jon Lawrence}
\affiliation{Australian Astronomical Optics, Macquarie University, 105 Delhi Road, North Ryde, NSW 2113, Australia}




\author{Steve Margheim}
\affiliation{Rubin Observatory/NSF's NOIRLab, Casilla 603, La Serena, Chile}



\author{Bryan Miller}
\affiliation{Gemini Observatory/NSF's NOIRLab, Casilla 603, La Serena, Chile}












\author{J. Gordon Robertson}
\affiliation{Australian Astronomical Optics, Macquarie University, 105 Delhi Road, North Ryde, NSW 2113, Australia}
\affiliation{Sydney Institute for Astronomy, School of Physics, University of Sydney, NSW 2006, Australia}

\author{Roque Ruiz-Carmona}
\affiliation{Gemini Observatory/NSF's NOIRLab, Casilla 603, La Serena, Chile}






\author{Katherine Silversides}
\affiliation{NRC Herzberg Astronomy \& Astrophysics, 5071 West Saanich Road, Victoria, BC V9E 2E7, Canada}

\author{Karleyne Silva}
\affiliation{Gemini Observatory/NSF's NOIRLab, Casilla 603, La Serena, Chile}

\author{Peter J. Young}
\affiliation{Research School of Astronomy and Astrophysics, College of Science, Australian National University, Canberra 2611, Australia}

\author{Ross Zhelem}
\affiliation{Australian Astronomical Optics, Macquarie University, 105 Delhi Road, North Ryde, NSW 2113, Australia}


\begin{abstract}

We present Gemini/GHOST high-resolution spectra of five stars observed in two low surface brightness Milky Way satellites, Sagittarius II (Sgr2) and Aquarius II (Aqu2). 
For Aqu2, the velocities and metallicities of the two stars are consistent with membership in a dark matter-dominated ultra faint dwarf galaxy (UFD). The chemical abundance ratios suggest inefficient star formation from only one or a few supernovae (e.g., low Na, Sr, Ba), and enriched potassium (K) from super-AGB stars.
For Sgr2, the velocity and metallicity dispersions of its members are not clearly resolved and
our detailed chemical abundances show typical ratios for metal-poor stars, with low dispersions. There is only one exception - we report the discovery of an r-process enhanced star (Sgr2584, [Eu/Fe]$ = +0.7 \pm 0.2$; thus, an r-I star). As r-I stars are found in both UFDs (Tuc~III, Tuc~IV, Grus~II) and globular clusters (M15 and M92), then this does not help to further classify the nature of Sgr2. 
Our exploration of Sgr2 demonstrates the difficulty in classifying some of the faintest (ambiguous) satellites.
We advocate for additional diagnostics in analysing the ambiguous systems, such as exploring radial segregation (by mass and/or chemistry), N-body simulations, and the need for dark matter to survive Galactic tidal effects. 
The spectra analysed in this paper were taken as part of the GHOST commissioning observations, testing faint observation limits (G$<18.8$) and the single and double IFU observing modes.

\end{abstract}

\keywords{Milky Way Galaxy -- dwarf galaxies -- star clusters -- stellar abundances -- astronomical instrumentation: GHOST}

\section{Introduction} 
\label{sec:intro}

With the advent of extensive and deep photometric surveys\footnote{Photometric surveys
such as the Sloan Digital Sky Survey  \citep[SDSS:][]{York2000, Abazajian2009}, the Dark Energy Survey \citep[DES:][]{Abbott18}, DELVE \citep{Drlica-Wagner21}, Panoramic Survey Telescope and Rapid Response System  \citep[PS1:][]{chambers2016PS}, UNIONS \citep[see][]{Jensen21, Smith2023}, and Euclid \citep[e.g.,][]{Hunt2024Euclid}},
it has become possible to discover extremely faint satellites of the Milky Way with $M_V > -5$. 
Identifying the physical nature of these new systems has proven to be a challenge,
(e.g., UNIONS1/UMaIII, \citealt{Smith2024}; Eridanus~III and DELVE~1, \citealt{Simon24}) 

Some of the new systems are likely  ultra faint dwarf galaxies (UFDs), with  $M_\star < 10^5 M_\odot$ \citep{Simon19}.  
UFDs provide valuable insights into the faint end of the galaxy luminosity function, galaxy formation models, ancient star formation histories, and chemical evolution in the early universe \citep[][]{Koposov09, Starkenburg13, Frebel_2015}. The abundance and distribution of dark matter in the halos of these dim, dense systems are also vital for constraining the nature and properties of dark matter and cosmological models \citep[][]{Springel2008, Brooks2014, Bullock_2017, Sales22}.
Alternatively, some of the new systems are likely to be  (potentially disrupted) star clusters \citep[e.g.,][]{Malhan2018, JiA2020S5, LiT2022S5, MartinVenn2022}, 
Faint star clusters are also valuable probes but for different scientific purposes, e.g., as
tracers of their host galaxy properties, ranging from the host's gravitational potential and assembly history, to its star formation and chemical enrichment history \citep[][]{Kravtsov2005, meszaros2015exploring, Helmi_2018, Li_2019}

\begin{figure}
    \centering
    \includegraphics[width=1\linewidth]
    {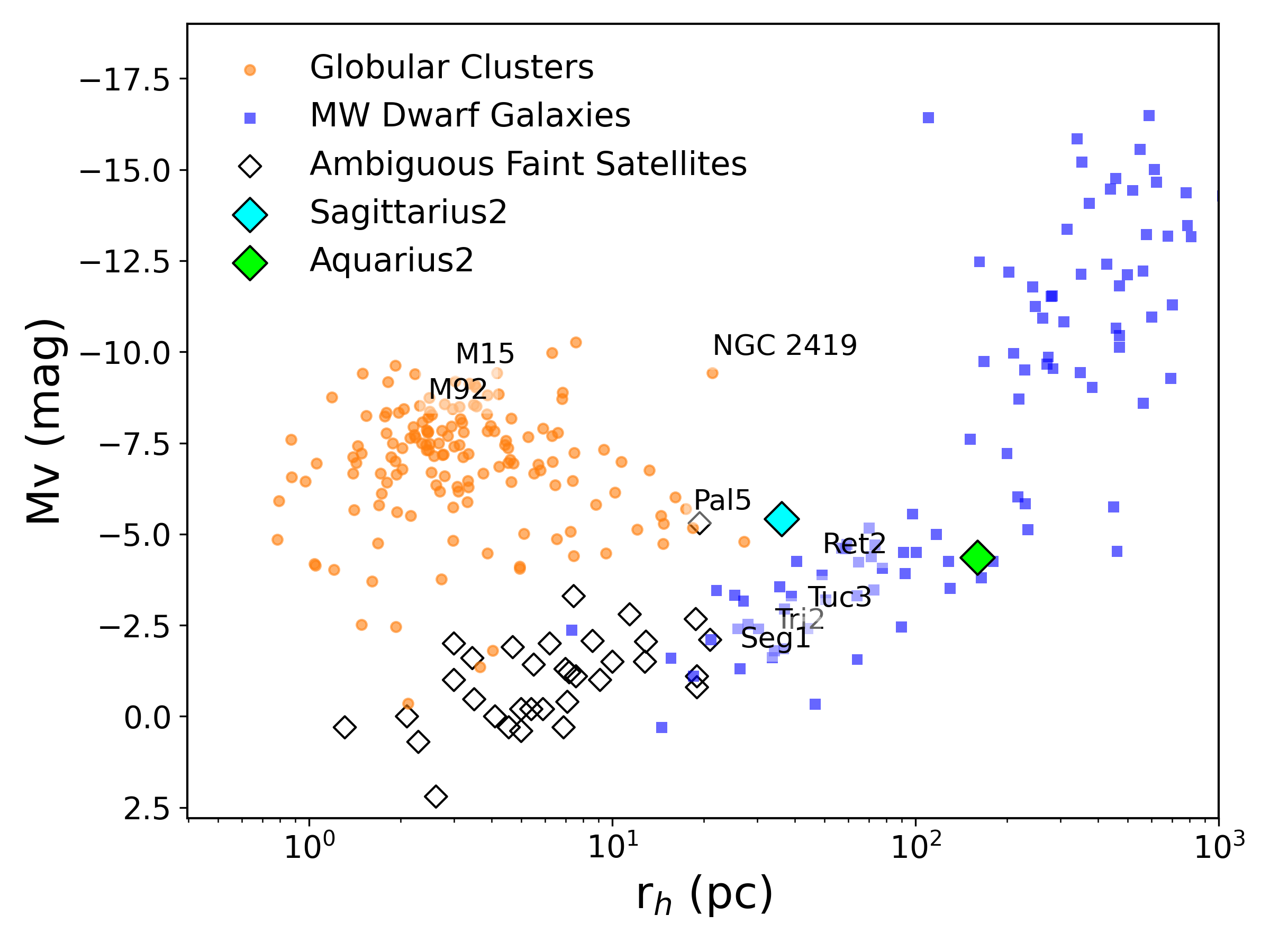}
    \caption{$M_V$ vs. $r_h$ diagram for Milky Way satellites. Globular clusters from \cite{Harris10} are plotted as orange circles, dwarf galaxies from \cite{Mcconnachie12} are shown as blue squares. Diamonds represent satellites with ambiguous classifications (see \citealt{McVenn2020} and references therein, as well as \citealt{Mau_2020, Cerny_2021, Gatto_2021, Cerny_2023_6sats, Cerny_2023_Delve6}). Aqu2 and Sgr2 are depicted in lime and cyan, respectively.}
    \label{fig:Mv_rh}
\end{figure}

As more faint systems are discovered, the distinction between these two classes (UFD versus star cluster) has become more puzzling. Their photometric properties place them precisely at the boundary between globular clusters and dwarf galaxies in the size-luminosity plane (see Fig.~\ref{fig:Mv_rh}), leading to the \textit{"valley of ambiguity"} \citep{Gilmore2007} or \textit{"trough of uncertainty"} \citep{Conn2018}.
One way to distinguish dark-matter-dominated UFDs from self-gravitating faint star clusters is to measure their dispersions in radial velocity, a characteristic directly correlated with dark matter content \citep{William12, Walker2023}. 
Another is to measure their dispersions in metallicity, associated with ongoing star formation and chemical evolution \citep[e.g.,][]{Leaman2012, Hasselquist2021, Walker2023}, where 
deeper gravitational potentials in dwarf galaxies allow them to retain the products of stellar feedback, preserving the signatures of self-enrichment in the form of significant metallicity dispersions. 
However, in the new exceptionally faint systems, the robustness of these measurements is challenging. This is primarily due to: (i) the small number of confirmed members, (ii) only a handful of stars bright enough for spectroscopic follow-up, (iii) limited precision in individual [Fe/H] and radial velocities ($v_r$), and (iv) potential velocity dispersion inflation caused by unidentified binaries \citep[e.g.,][]{McConnachie2010}.
Additionally, \textit{"microgalaxies"}, i.e., heavily stripped remnants of early accreted satellites, which can reach arbitrarily low luminosities \citep[see][]{Errani20}, could have such small velocity dispersions that they are indistinguishable from kinematically cold globular clusters -- unless a precision of $<$100 m~s$^{-1}$ can be obtained \citep[see Fig. 7 in][]{Errani2024microgal}.

Given the critical role of the smallest galaxies in addressing key questions in cosmology, the faintest galaxies and galaxy candidates require a more complex approach to answering the question 
\textit{"Is there a dark matter halo?".}
One alternative is to indirectly infer the presence of a dark matter halo by studying the stability of the stellar system within the Milky Way tidal field. 
This method is particularly valuable when observational limits do not constrain the internal velocity dispersion sufficiently, or, where the internal velocity dispersion may be inflated by the presence of binary stars. \citep[e.g.,][]{Errani2024microgal, Errani24unionsI}. 
Another approach is to examine stellar mass segregation,
as expected in globular clusters due to energy equipartition, which redistributes stars based on their mass \citep{Baumgardt22}. 
In old globular clusters with relatively short relaxation times, massive stars sink toward the center, while less massive stars are pushed outward. In contrast, dark-matter-dominated UFDs often have relaxation times exceeding a Hubble time, making significant mass segregation unlikely.

Detailed chemical analyses of the brightest stars in an UFD galaxy or faint star cluster can also be invaluable in exploring the nature of an ambiguous system. 
High-resolution spectroscopy enables robust metallicity measurements through numerous iron lines, 
and detailed chemistry can be used to search for distinguishing features. 
In globular clusters, these may include specific star-to-star variations in light elements (C, N, O, Na, Mg, Al, and some s-process elements) due to multiple populations \citep{Gratton12, Bastian2018}.
In UFDs, typical chemical signatures found to-date include low ratios of $\alpha$-capture elements (e.g., O, Mg, Si, Ca), low ratios of some iron-group elements (e.g., Zn, Mn), low neutron-capture element abundances and/or ratios (e.g., Sr, Ba, and/or [Sr/Ba]), and carbon-rich stars \citep[e.g.,][]{Venn2004, Venn2012, Berg15, Frebel_2015, Salvadori15, Ji2019, Monty2020, Monty2024, Sitnova21, Tarumi21, delosReyes2022, Rossi_2023, Lucchesi24}. 

In this paper, we focus on two low surface brightness Milky Way satellites: Sagittarius II (Sgr2) and Aquarius II (Aqu2).  The position of these two systems are shown on the size-luminosity plane for MW satellites; see Fig.~\ref{fig:Mv_rh}. 
Sgr2 presents an intriguing scientific case as it is positioned precisely between star clusters and dwarf galaxies ($M_V = -5.7$, $r_h = 36$ pc). On the other hand, Aqr2 stands out for its unusually large half-light radius relative to its faintness ($M_V = -4.4$, $r_h = 159$ pc).

Sgr2 was discovered by \cite{Leavens15} in PS1, where it was identified as an old (12.5 Gyr), metal-poor ([Fe/H] = -2.20) dwarf galaxy candidate. A deeper photometric study with Magellan/Megacam \citep{Mutlu-Pakdil_2018} revealed structural parameters more consistent with a globular cluster classification.
Using DEIMOS spectroscopy and the metallicity-sensitive,
narrow-band photometry provided by the Pristine survey, \citet[][hereafter L20]{Longeard_2020}  measured a velocity dispersion $\sigma_{v_r}^{\text{L20}} = 2.7^{+1.3}_{-1.0}$ \kms\ suggesting the presence of a low-mass dark matter halo and therefore the UFD scenario; however, they also found a very low metallicity dispersion $\sigma_{[Fe/H]}^{\text{L20}} = 0.10^{+0.06}_{-0.04}$ dex. Subsequently, \citet[][hereafter L21]{Longeard_2021} supplemented the L20 dataset with 19 new members identified with \textit{VLT}/FLAMES spectroscopy. This revealed a lower velocity dispersion of $\sigma_{v_r}^{\text{L21}} = 1.7^{+0.5}_{-0.5}$ \kms, consistent with that of MW globular clusters of similar luminosity. The metallicity dispersion of this latter dataset was unresolved at $\sigma_{[Fe/H]}^{\text{L21}} <0.20$ dex at the 95\% confidence level.  To date, Sgr2 is classified as an exceptionally large stellar cluster, and an analysis by \citealt{Errani24unionsI} suggests that its high density contrasts with the Milky Way halo at pericentre (where the density of Sgr2 is $\bar \rho_\mathrm{h} \approx  3.7 \times 10^7 \,\frac{\mathrm{M}_\odot}{\mathrm{kpc^3}}$ vs. the density of the MW halo at pericentre, $\bar \rho_\mathrm{peri} = 1.0 \times 10^6 \,\frac{\mathrm{M}_\odot}{\mathrm{kpc^3}}$) such that Sgr2 would not be significantly affected by tidal perturbations, even without a dark matter halo.

Aqu2 is less ambiguous in its classification, yet still presents challenges. Discovered by \citet[][hereafter T16]{Torrealba16} using \textit{SDSS} and VST ATLAS photometry from \cite{Shanks2015} with \textit{Keck}/DEIMOS spectroscopy, the system was initially classified as an UFD simply due to its large $r_h$. Using 9 spectroscopically confirmed member stars (including 5 BHB), T16 derived a systemic velocity $v_{r}^{T16} = -71.1 \pm 2.5$ \kms with a large velocity dispersion $\sigma_{v_r}^{\text{T16}} =5.4^{+3.4}_{-0.9}$ \kms, and [Fe/H]$_{T16} = -2.3 \pm 0.5$ dex with no resolved metallicity dispersion.  
\citet[][hereafter B23]{Bruce23} revisited these measurements using \textit{Magellan}/IMACS spectroscopy for 8 RGB stars. They found a lower systemic velocity ($v_{r}^{B23} = -65.3 \pm 1.8 $ km s$^{-1}$)  and smaller velocity dispersion $\sigma_{v_r}^{\text{B23}} =4.7^{+1.8}_{-1.2}$ \kms, as well as slightly lower metallicity and metallicity dispersion ([Fe/H]$_{B23} = -2.57$ dex, 
with $\sigma_{[Fe/H]}^{\text{B23}} =0.36^{+0.2}_{-1.4}$ dex). Given these measurements, both groups classified the system as an very metal-poor and very dark-matter dominated UFD. 
B23 noted that
two stars (Gaia DR3 2609109756631321472 and Gaia DR3 2609061687357323776, hereafter Aqu2472 and Aqu2776 respectively) dominated their results. Removing Aqu2472 lowered the velocity dispersion by nearly half to  $\sigma_{v_r}^{\text{B23*}} =2.7^{+1.6}_{-1.2}$ \kms, while excluding Aqu2776 reduced the metallicity dispersion by nearly 10x, to $\sigma_{[Fe/H]}^{\text{B23*}} =0.04^{+0.08}_{-0.02}$ dex. The tidal resilience analysis by \citet{Errani24unionsI} reveals that without dark matter, Aqu2's mean density ($\bar \rho_\mathrm{h} \approx 1.2 \times 10^5 \frac{\mathrm{M}_\odot}{\mathrm{kpc^3}}$) closely matches the Milky Way's at pericentre ($\bar \rho_\mathrm{peri} = 1.4 \times 10^5 \frac{\mathrm{M}_\odot}{\mathrm{kpc^3}}$), which suggests that Aqu2 may show signs of tidal interaction.

\begin{table*}
\caption{Members of Aqu2 and Sgr2 with GHOST spectra. Heliocentric distances are: Sgr2 at $73.1^{+1.1}_{-0.7}$ kpc \citep{Longeard20}, Aqu2 at $107.9^{+3.3}_{-3.3}$ kpc \citep{Torrealba16}.}
\label{tab:targets}
\centering
\resizebox{1.0\textwidth}{!}{
\hspace{-3.0cm}
\begin{tabular}{llcccccccccc}
\toprule
Target  & Gaia DR3 sourceID            &  RA          &   DEC      & G     & BPRP & A$_G$\footnote{$\rm A_V=A_G/0.85926$ \citep{Marigo08,Evans18}.} & g$_0$\footnote{g$_0$, r$_0$ PS1 for Sgr 2 and g$_0$, i$_0$ SDSS for Aqu 2} & r$_0$/i$_0^{\textcolor{blue}{\rm b}}$ & pmra & pmdec \\
\hline

Aqu2776 & 2609061687357323776 & 338.5352 & -9.3278 & 18.78 & 1.4 & ... & 19.32 &  18.54 & -0.446 & -0.359 \\ 
Aqu2472 & 2609109756631321472 & 338.4696 & -9.2859 & 18.79 & 1.2 & 0.00 & 19.25 & 18.54 &  -0.552 & -0.541
 \\
Sgr2584 & 6864047652495955584 & 298.1624 & -22.0775 & 16.96 & 1.48 & 0.85 & 17.50 & 16.37 & -0.704 & -0.939\\
Sgr2656 & 6864047583776582656 & 298.1815 & -22.0773 & 18.41 & 1.19 & 0.04 & 18.72 & 17.92 & -0.837 & -0.911\\ 
Sgr2936 & 6864423788550679936 & 298.1534 & -22.0496 & 17.31 & 1.39 &  0.55 & 17.76 & 16.64 & -0.852 & -0.918 \\
\hline
\end{tabular}}
\end{table*}

Using high-resolution spectra from the newly commissioning \textit{Gemini}/GHOST spectrograph \citep{McConnachie2024, Kalari2024}, we revisit the analyses of Sgr2 and Aqu2.   Our targets include two previously confirmed members of Sgr2, one new member of Sgr2, and two of the brightest stars in Aqu2.
Our objectives include: (i) refining constraints on the velocity and metallicity dispersions of these systems, (ii) performing a detailed chemical analysis for the first time in these two faint systems, and (iii) increasing the observational epochs per star to assess any  binary characteristics. These endeavors are to contribute to the discussion on the classification of these ambiguous systems as dark matter dominated UFDs or faint stellar clusters.

\section{GHOST Observations} 
\label{sec:obs}

The high-resolution spectra for the five targets were obtained during the GHOST commissioning run in June 2022. The observations were conducted using the standard resolution mode with 2x4 binning. Targets Sgr2656 and Sgr2936, as well as Aqu2776 and Aqu472, were observed simultaneously using the two integrated field units (IFUs). The observation of Sgr2584 was performed separately in single IFU mode. For more details on the observations per exposure, see the Appendix.

\begin{figure*}
\centering
    \includegraphics[width=1\textwidth]{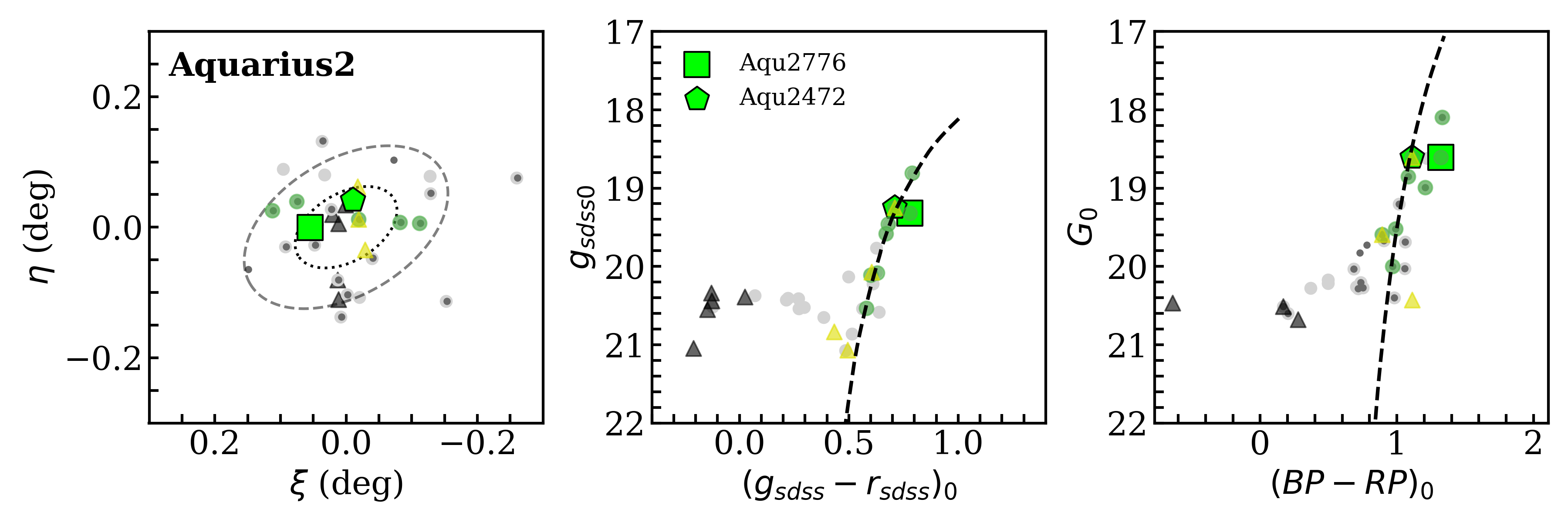}
    \caption{Spatial distribution (left; dotted and dashed ellipses indicate $2r_h$ and $3r_h$, respectively) and color-magnitude diagrams from SDSS  (middle) and Gaia DR3 (right) photometry for Aqu2 member stars.  Overlaid on the CMDs are Dartmouth isochrones of 12.5 Gyr and [Fe/H] $= -2.5$. Yellow triangles represent members identified by \cite{Torrealba16}, with BHB stars in black. Green circles are members from \cite{Bruce23}. Dark grey dots are stars with membership probability $P > 0.1$, selected using the algorithm described by \cite{Jensen23}, while pale grey are stars from \cite{Pace22}, also with $P > 0.1$, selected using DECaLS photometry. Targets analyzed in this paper are shown in larger lime markers.}
    \label{fig:Aqu_cmd}
\end{figure*}

\begin{figure*}
\centering
    \includegraphics[width=1\textwidth]{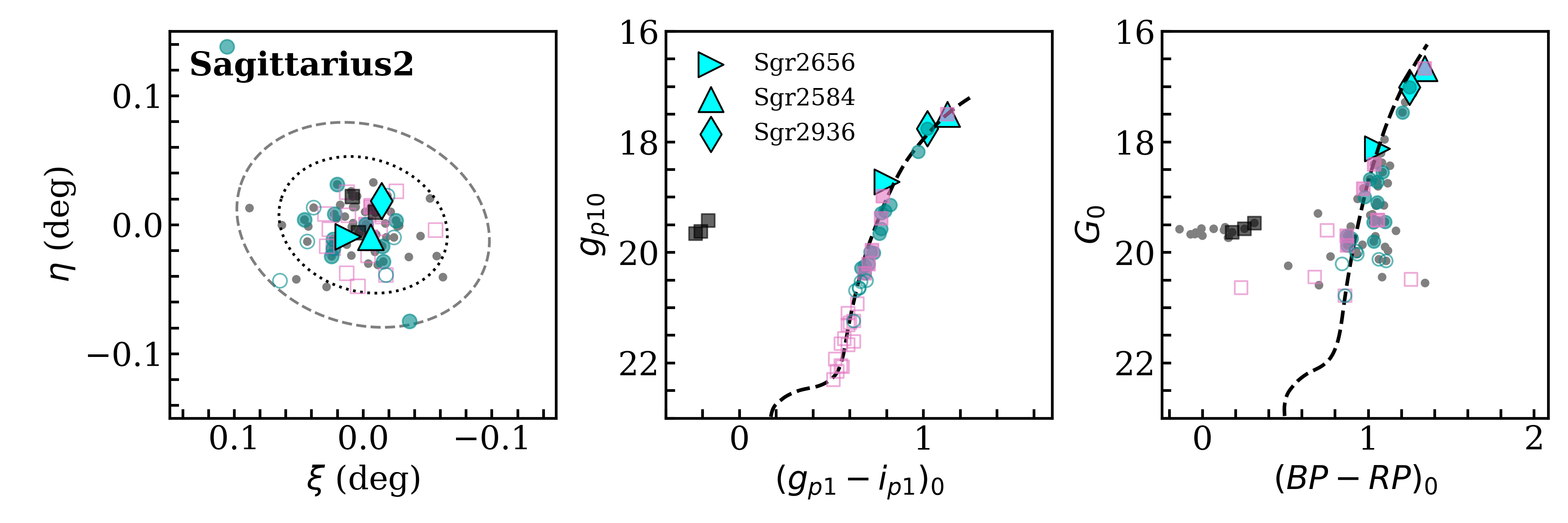}
    \caption{Spatial distribution (left; dotted and dashed ellipses indicate $2r_h$ and $3r_h$, respectively) and color-magnitude diagrams from PS1  (middle) and Gaia DR3 (right) photometry for Sgr2 member stars. Overlaid on the CMDs are Dartmouth isochrones of 12 Gyr and [Fe/H] $= -2.35$. Pink squares are from \cite{Longeard_2020}, black for BHB stars; cyan circles from \cite{Longeard_2021}. Open markers show stars with CaHK photometric metallicity only, solid markers indicate stars with spectroscopic metallicity from Ca triplet lines. Grey dots are high probability ($P>0.5$) members identified with the algorithm described by \cite{Jensen23}. Targets analyzed in this paper are highlighted with larger cyan markers.}
    \label{fig:Sgr_cmd}
\end{figure*}

\subsection {Target Selection}
\label{sec:targets}

The targets were selected using a Bayesian inference method to identify highly probable members in UFDs, as described in \cite{Jensen23}. The membership probabilities were estimated based on Gaia DR3 photometry and astrometry, considering projected spatial positions, systemic proper motion, and positions in the color-magnitude diagram (CMD) of the likely members. The Gaia DR3 source ID, RA, DEC, G, A$_G$, and BP-RP values for each target are presented in Table~\ref{tab:targets}. The spatial positions of the targets, along with their locations on the CMDs 
are shown in Fig~\ref{fig:Aqu_cmd} and Fig~\ref{fig:Sgr_cmd}. Similar way of selecting targets has been highly successful in selecting members without medium resolution spectroscopy \citep{McVenn2020, Sestito2023_Scl, Waller23, Hayes23, Sestito23Umi}. In addition, these targets tested the faint limits of GHOST acquisitions and science exposures as well as the single and dual IFU target modes.

\subsection{GHOST Data Reductions} 

The GHOST spectra were reduced using an early and developing version of the Gemini DRAGONS pipeline \citep[see comments in][]{Hayes23}. DRAGONS is a Python package that performs standard data reduction such as flat-fielding, bias subtraction, and corrections for heliocentric motion \citep{Labrie23}. A full list of the GHOST data files used for the reduction of each target in Aqu2 and Sgr2 is provided in Table~\ref{tab:exp}.

\begin{figure*}
    \centering
    \includegraphics[width=1\textwidth]{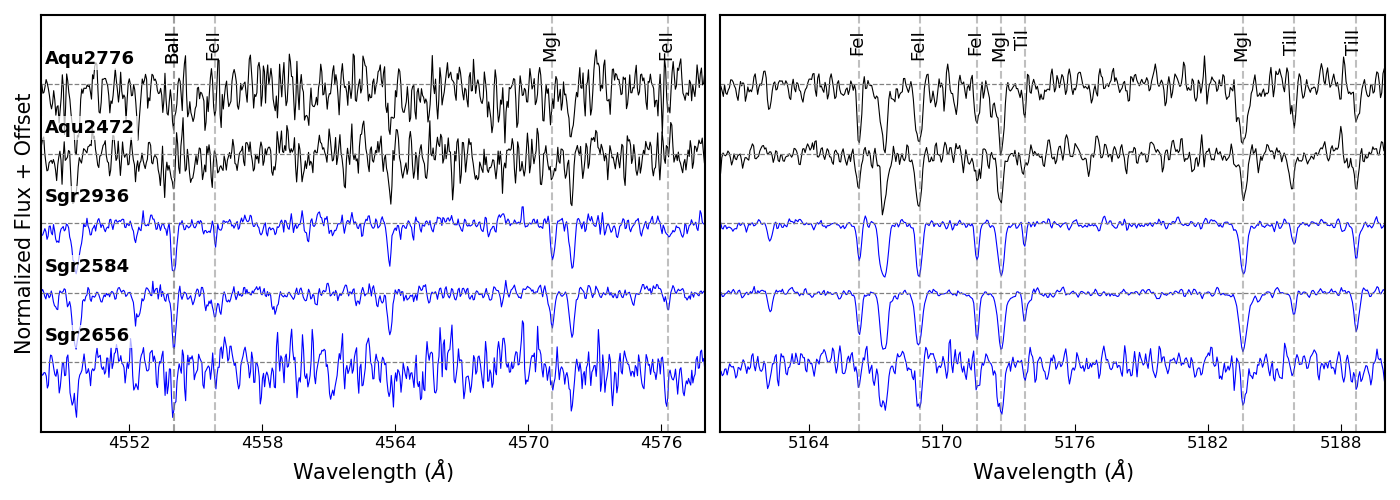}
    \caption{Samples of the GHOST spectra for our targets in Sgr2 and Aqu2, highlighting specific spectral lines (BaII $\lambda$4554, MgI $\lambda$4571, and the Mgb triplet).}
    \label{fig:spectra}
\end{figure*}

For each camera, the DRAGONS pipeline produced 1D spectra for each exposure of each object. For each object, the exposures were then co-added by taking their median, per camera. This resulted in two (blue and red) 1D spectra for each object, with improved signal-to-noise ratios.
The co-added spectra were continuum normalized by a two-step process. The first step estimates the continuum using median filtering and divide the spectrum by the estimated continuum. The second step would adjust the position of the continuum per wavelength via asymmetric k-sigma clipping, as the median filtering step underestimates the continuum in the presence of strong lines. Samples of the final spectra for all five targets are shown in Fig~\ref{fig:spectra}.

\subsection{Radial Velocities ($v_r$)}

Radial velocities were determined using cross-correlation (IRAF/{\sl fxcor}) with the GHOST spectrum of the standard star HD\,122563, commonly used as a benchmark for metal-poor stars.
The spectral region from 3800 to 6700 \AA\ was used, as the SNR worsens at shorter wavelengths and the sky/telluric lines are poorly correlated at longer wavelengths.
Our radial velocities per target are given in Table~\ref{tab:params}. 
The $v_r$ for our GHOST spectrum of HD122563 is $-26.3$ \kms\ \citep[see][]{Hayes23}, in excellent agreement with Gaia DR3 and results in the literature.  All target $v_r$ are measured from template fitting using this standard star as a reference. We did not apply any zero-point corrections to the velocity measurements, as our derived values show excellent agreement with those obtained from other spectrographs for the same stars. GHOST's velocity precision -- reaching meters per second \citep{McConnachie2024, Kalari2024} -- also surpasses that of comparable instruments.

\section{Method}
\label{sec:method}

The spectral analysis was conducted using a new set of jupyter notebooks, \texttt{Py\_Looper}\footnote{\url{https://github.com/dashazaremba/PyLooper}}, which offer a semi-automated routine for high-resolution spectral analysis based on equivalent widths (EW). This process starts by determining the metallicity ([Fe/H]) from iron lines, using an initial set of stellar parameters based on photometric calibrations (see Section~\ref{sec:params}). Spectral parameters ($T_{\text{eff}}, \, \log g, \, v_{\text{mic}}, \, [\text{M/H}]$) can be refined adopting both 1DLTE and 1DNLTE analyses. Subsequently, the routine measures other spectral lines, applying necessary corrections, including NLTE corrections from precalculated grids (see Section~\ref{sec:NLTE}),
and hyperfine structure corrections\footnote{HFS corrections were calculated for odd-Z elements only. For elements where isotopic ratios depend on r/s-process dominance (e.g., Ba II, Eu II), HFS corrections were calculated directly from spectrum syntheses.} (HFS) for odd-Z elements. Overall, this method includes error propagation from uncertainties in metallicity and the derived stellar parameters throughout the analysis. For the line measurement errors, we use the line-to-line scatter in Fe~I for the other elements as this is a good indicator of impact of the SNR. To validate the \texttt{Py\_Looper} method, we conducted a line-by-line comparison of EWs and abundances for Fe lines in the HD222925 standard star (see the Appendix)  against the literature values reported by \citet{roederer2018}. 

For measuring EWs, \texttt{PyLooper} employs a modified version of \texttt{pyEW}\footnote{\url{https://github.com/madamow/pyEW}}, which applies derivative spectroscopy to enhance weak signals and resolve blended lines. Higher-order derivatives narrow peak widths, making otherwise obscured features more distinct in derivative spectra\footnote{Each derivative is calculated by dividing the difference between the original spectrum $f(\lambda)$ and a shifted version $f(\lambda + \Delta\lambda)$ by $\Delta\lambda$, yielding $\frac{df}{d\lambda}(\lambda + \frac{1}{2} \Delta\lambda)$. Higher derivatives are obtained by iterating this process the required number of times.} \citep[and references therein]{Yu_2024}. After identifying lines, each fit is performed in small, user-defined spectral ranges around $\lambda$ $\pm$ offset using either single-Gaussian, multi-Gaussian, or Voigt profiles.

\texttt{PyLooper} provides local continuum re-normalization through either an automated or manual process. In the automatic routine, a polynomial of a specified order is fit to the spectrum using the random sample consensus (RANSAC) method, which iteratively excludes outliers based on a threshold calculated from local noise. This process continues until the set of retained points stabilizes or a maximum number of iterations is reached. The spectrum is then normalized by dividing the flux by the fitted continuum. This method proved to be more effective for higher SNR spectra compared to our targets.
For manual re-normalization, small local scaling adjustments are applied, raising or lowering the continuum by 2-5\% at a time to achieve a more accurate best fit.
In this analysis, Voigt profile fitting was disabled, and multi-Gaussian fits were applied only to blended lines. Most lines were fitted with a single Gaussian after manual continuum re-normalization (rescaling) within a wavelength range of $\pm$3-5 \AA. Abundances were then measured using the Python wrapper \texttt{q2} \citep{Ramirez2014} for MOOG (version 2019).

To reduce systematic errors, we  compare results for our target stars to two standard stars with well-established stellar parameters, HD122563 and HD222925, which differ in metallicity and T$_{\rm eff}$. 
The initial parameters were sourced from the literature \citep{roederer2018, Giribaldi2023} and subsequently analyzed using the same methodology applied to our targets for consistency. These parameters result in an excellent fit for our analysis, i.e., no correlations in the FeI line abundances with excitation potential (slope $-0.04\pm0.01$) or line equivalent widths (slope 
$0.10\pm0.05$), and with good ionization equilibrium between A(FeI) and A(FeII), as described in Section~\ref{sec:feh}; thus, we retain these parameters, but adopt our own error analysis. The stellar parameters with their associated uncertainties for the standards are presented in Table~\ref{tab:params}. The same Fe line list was used for both the standards and the targets. For other elements, the lines were first measured for the two standards, with only moderately strong and strong lines retained, as weaker lines could be too contaminated by noise in our spectra. This refined line list was then consistently applied to the target stars.

\subsection{Model Atmospheres Analysis}
\label{sec:atmos}

Chemical abundances are determined in this paper from a classical model atmospheres analysis of the spectral features in each star.
Model atmospheres from the MARCS website \citep{Gustafsson08} were adopted, particularly the OSMARCS spherical models given that all the targets are giants, with log$g< 3.5$.

The 1DLTE radiative transfer code 
MOOG\footnote{MOOG (2019) is available at \url{http://www.as.utexas.edu/~chris/moog.html}} 
\citep[][2019 version]{Sneden73, Sobeck11} 
was used to convert EWs into chemical abundances and to perform spectrum syntheses.

\subsection{Stellar Parameters}
\label{sec:params}

Surface temperatures (T$_{\rm eff}$) were found  using the colour-temperature calibrations for Gaia photometry from \cite{Mucciarelli2020}. The input parameters include the {\it Gaia} DR3 de-reddened (BP$-$RP) colour and a metallicity estimate. The 2D reddening map\footnote{2D reddening map at \url{https://irsa.ipac.caltech.edu/applications/DUST/}} from \citet{Schlafly11} was used to correct the photometry for extinction\footnote{To convert from the  E(B-V) map to  {\it Gaia} extinction coefficients,  the  $\rm A_V/E(B-V)= 3.1$ \citep{Schultz75} and the $\rm A_G/A_V = 0.85926$, $\rm A_{BP} /A_V = 1.06794$, $\rm A_{RP} /A_V = 0.65199$ relations \citep{Marigo08,Evans18} are used.}. As input metallicities, the mean [Fe/H] 
$= -2.6$ and $-2.3$ 
for Aqu2 and Sgr2 from \cite{Torrealba16} and \cite{Longeard_2020} were adopted, respectively.

Surface gravities were found using the Stefan-Boltzmann equation
\citep[e.g., see][]{Sestito2023_Scl}.
This step required T$_{\rm eff}$, the {\it Gaia} DR3 de-reddened G magnitude, bolometric corrections on the flux \citep[from][]{Andrae18}, and a heliocentric distance\footnote{Heliocentric distances are Sgr2 at $73.1^{+1.1}_{-0.7}$ kpc \citep{Longeard20}, Aqu2 at $107.9^{+3.3}_{-3.3}$ kpc \citep{Torrealba16}.}.
A Monte Carlo algorithm was employed to propagate uncertainties in the input parameters and estimate  the total and correlated uncertainties in the derived stellar parameters. The input uncertainties are as follows: 0.05 for the BP\_RP color index, 0.5 dex for metallicity, 100 K for T$_{\text{eff}}$, and 1.0 kpc (Sgr2) and 3.3 kpc (Aqu2) for the distance.
The input quantities were then randomised within $1\sigma$ each using a  Gaussian distribution, except for the stellar mass.  The latter is treated with a flat prior from 0.5 to 0.8 \msun, which is consistent with the mass of  long-lived very metal-poor stars.

Initial microturbulence values were estimated using the calibrations for red giants in MW satellites by \cite{Mashonkina17}.

\begin{figure}
    \includegraphics[width=0.5\textwidth]{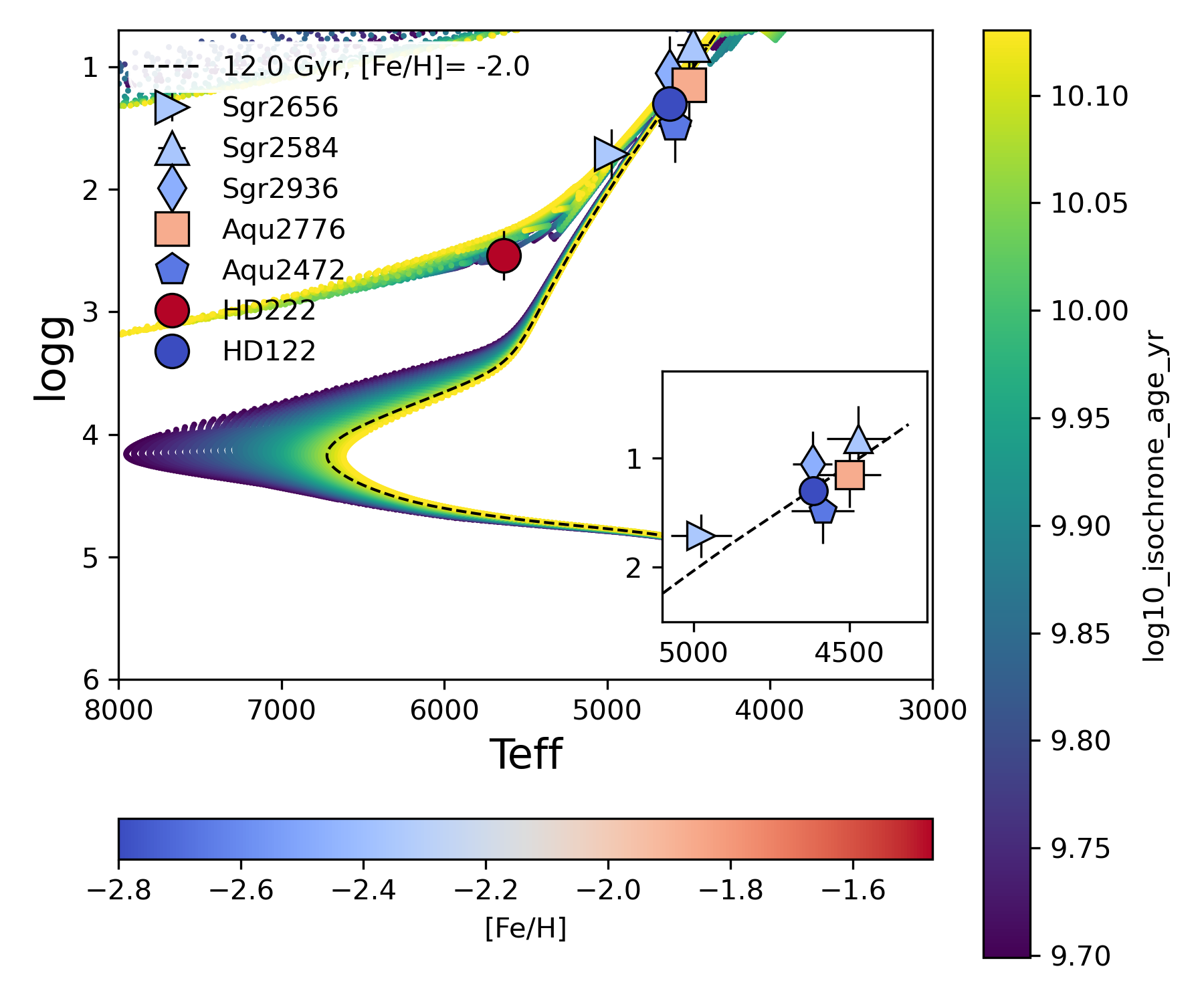}
    \caption{Aqu2 and Sgr2 targets with measured stellar parameters, along with two standards HD222925 and HD122563 used for differential analysis. Markers are color-coded by metallicity derived from LTE analysis of Fe lines. MIST isochrones span ages up to 13.5 Gyr at a fixed metallicity of [Fe/H] = -2. The dashed line is the 12 Gyr isochrone, the estimated age for Sgr2 \citep[$12 \pm 0.5$ Gyr]{Longeard_2020}.}
    \label{fig:pars_iso}
\end{figure}

The stellar parameters from these calculations are listed in Table \ref{tab:params}, and all targets with derived parameters are plotted on the Kiel diagram in Fig~\ref{fig:pars_iso}.

\begin{table*}
\caption{Stellar parameters for the targets in Aqu2 and Sgr2, and two standard stars}
\label{tab:params}
\centering
\hspace{-2.2 cm}
\begin{tabular}{lcccccccc}

\hline
Target & \teff$_{phot}$ & \logg$_{phot}$ & \teff$_{spec}$ & \logg$_{spec}$    & $\xi$ & [Fe/H]$_{LTE}$  & [Fe/H]$_{NLTE}$ & RV \\ 
 & (K) & (cgs) & (K) & (cgs) & (\kms) & (dex) & (dex) & (\kms) \\
\hline

Aqu2776 & $4499\pm78$ & $1.23\pm0.07$ & $4499\pm98$ & $1.15\pm0.25$ & $1.6\pm0.2$ & $-1.87\pm0.06$ & $-1.80\pm0.06$ & $-64.34\pm0.03$ \\

Aqu2472 & $4858 \pm94$ & $1.46\pm0.06$ & $4858\pm98$ & $1.48\pm0.30$  & $2.4\pm0.2$  & $-2.66\pm0.09$ & $-2.52\pm0.09$ & $-55.98\pm0.11$ \\

Sgr2656 & $4975\pm100$ & $1.72\pm0.05$ & $4975\pm102$ & $1.71\pm0.20$ & $2.2\pm0.1$ & $-2.35\pm0.08$ &$-2.23\pm0.08$ & $-177.33\pm0.10$ \\

Sgr2584 & $4472\pm77$ & $0.87\pm0.06$ & $4472\pm99$ & $0.82\pm0.30$ & $2.2\pm0.2$ & $-2.36\pm0.06$ & $-2.24\pm0.04$ & $-176.18\pm0.10$ \\

Sgr2936 & $4618\pm83$ & $1.12\pm0.06$ & $4618\pm63$ & $1.05\pm0.28$ & $2.4\pm0.2$ & $-2.47\pm0.04$ & $-2.36\pm0.04$ & $-175.70\pm0.10$ \\

\hline

HD122563 & ... & ... & $4615 \pm28$ & $1.30\pm0.12$ & $2.0\pm0.1$ & $-2.84\pm0.01$\footnote{\textbf{Stellar parameters are from \cite{Roederer14}}} & $-2.72\pm0.01$ & \, $-26.3\pm0.1$\footnote{\cite{Hayes23}} \\
HD222925 & ... & ... & $5636\pm99$ & $2.54\pm0.12$ & $2.2\pm0.1$ & $-1.47\pm0.01$\footnote{\textbf{Stellar parameters are from \cite{Giribaldi2023}}} & $-1.30\pm0.01$ & \ $-38.5 \pm 0.1$\\

\hline
\end{tabular}
\end{table*}

\subsection{Spectral Lines Analysis}
\label{sec:lines}

Our spectral analysis is based on the line list provided in the Appendix.
This line list was compiled from our analyses of two standard stars, HD122563 \citep{Venn2025} and HD222925 \citep{Hayes23}.
Spectral lines from a variety of sources \citep{Yong13, Norris17, Kielty2021, Lucchesi22, roederer2022, Sestito24} were examined in these standard stars and their equivalent widths (EWs) compared to those in the literature.
All atomic data was adopted from the \texttt{linemake} compilation\footnote{Linemake available at \url{https://github.com/vmplacco/linemake}} \citep{Placco21}.

Our analysis is primarily an EW analysis, with spectrum synthesis included in two ways: (1) as a check on the line profile fit in each EW measurement, and (2) to calculate some blended line abundances (e.g., Eu) or carbon (e.g., CH from the G band).
We note that strong lines with EW$>150$ m\AA\ were excluded from the 1DLTE abundances, except for certain species (e.g., NaI, KI, BaII, EuII) for which no weaker lines were available.

\subsection{Metallicity [Fe/H]}
\label{sec:feh}

FeI and FeII abundances were measured using both resonance and subordinate lines, with only those lines having equivalent widths in the range \( 20 < \text{EW} < 150 \, \text{m\AA} \) retained. Initial stellar parameters for the MARCS model atmospheres were estimated from photometry, as detailed in Sec~\ref{sec:params}. 
Subsequently, the stellar parameters were refined based on iron lines spectroscopy. Specifically, \( v_{\text{mic}} \) was adjusted to flatten the slope of the absolute iron abundances, A(Fe), vs the reduced equivalent width (\( \log \left( \frac{\text{EW}}{\lambda} \right) \)), using linear interpolation. Additionally, we examined the effect of adjusting \( \log g \) to achieve ionization equilibrium between FeI and FeII lines in NTE analysis. 
For most targets, the \( \log g \) values were kept nearly unchanged compared to the photometric estimates, as the Fe~I$_{\text{NLTE}}$ abundances were found to agree with the FeII abundances -- which are not affected by NLTE -- within \(\pm 1 \sigma\). However, the scatter was substantial, typically around 0.2--0.3 dex, due to the low SNR. The only exception was Aqu2776, where reducing \( \log g \) by 0.07 improved the ionization equilibrium by 0.04 dex, and also reduced the scatter of some derived elemental abundances. Thus, we adopted this slightly lower \( \log g \) value for Aqu2776. 
For Aqu2472, a 0.29 dex discrepancy was observed between the mean FeI$_{\text{NLTE}}$ and FeII abundances. Increasing \( \log g \) by 0.3 only reduced this discrepancy by 0.06 dex. Given the large intrinsic scatter (0.3 dex) among the Fe lines and that we only have 2 FeII lines for this star, we decided to retain the photometric \( \log g \) value.
\( T_{\text{eff}} \) was not adjusted in the spectroscopic analysis, as the slope of the linear interpolation of A(Fe) vs excitation potential (\( \chi \)) was found to be flat (within the range 0.00 - 0.05) for all targets.

Spectroscopic stellar parameter uncertainties were derived by adjusting each parameter with a $\Delta$ step (e.g., \( T_{\text{eff}} \pm \Delta T_{\text{eff}} \), \( \log g \pm \Delta \log g \), \( v_{\text{mic}} \pm \Delta v_{\text{mic}} \)), then recalculating iron lines statistics (absolute abundances A(FeI), A(FeII),  the slopes \texttt{ep\_slope}, \texttt{rew\_slope}, ionization equilibrium of A(FeI)$_{\text{NLTE}}$ and A(FeII) ) at each step. Changes in these statistics were averaged for each parameter adjustment and propagated to determine parameter errors. Final uncertainties in metallicity were calculated by averaging the differences from these parameter adjustments and adding the statistical error from the iron lines in quadrature.

The final derived stellar parameters with uncertainties, as well as the weighted average metallicities 
from FeI and FeII lines (in both LTE and NLTE), are presented in Table~\ref{tab:params} as: 

\smallskip
\(\text{[Fe/H]} = \frac{\text{[FeI/H]} \cdot N_1 + \text{[FeII/H]} \cdot N_2}{N_1 + N_2}\)

\smallskip
\(\text{err}_{\text{[FeI/H]}} = \frac{\sigma_{\text{feh1}}}{\sqrt{N_1}}\), 

\smallskip
\(\text{err}_{\text{[FeII/H]}} = \frac{\sigma_{\text{feh2}}}{\sqrt{N_2}}\),

\smallskip
\(\text{err}_{\text{[Fe/H]}} = \sqrt{\frac{\text{err}_{\text{[FeI/H]}}^2 \cdot N_1 + \text{err}_{\text{[FeII/H]}}^2 \cdot N_2}{N_1 + N_2}}\).

\subsection{Other Elements}

Spectral lines of C to Eu are available for chemical abundance measurements in this sample.  Some elements requires spectrum syntheses or additional corrections; e.g., due to isotopic splitting, hyperfine structure corrections, and/or NLTE corrections.  We do not consider 3D effects in this paper\footnote{Fully consistent 3D NLTE model atmospheres with line-by-line radiative transfer are not yet available. Furthermore, mean 3D (<3D>) NLTE has yet to demonstrate clear advantages over 1D NLTE in individual cases \citep[e.g.,][]{Mallinson2024}.}

\subsubsection{Isotopic \& Hyperfine corrections}
\label{sec:HFS}

Isotopic and hyperfine structure corrections for odd-Z elements (ScI, MnI) were applied automatically within \texttt{PyLooper} routine (see Section~\ref{sec:method}). For BaII and EuII, HFS corrections were determined through spectrum synthesis, adopting r-process isotopic ratios from \cite{Sneden2008}.

\subsubsection{NLTE corrections}
\label{sec:NLTE}

NLTE corrections for Ca, Mg, Ti, Mn, Si, and Fe are obtained from the MPIA webtool database\footnote{\url{http://nlte.mpia.de}}. Corrections for Na are from the \textsc{INSPECT} database\footnote{\url{http://inspect-stars.com}} \citep{Lind2011}. A python wrapper to extract these corrections for our stars is available\footnote{\url{https://github.com/anyadovgal/NLTE-correction}}.
For K, we apply the NLTE correction grid from \cite{Reggiani2019A&A...627A.177R}. Ba NLTE corrections are from \cite{Mashonkina19}.
All NLTE line corrections are presented in Table~\ref{tab:lbl_iron_abund} (Fe lines) and Table~\ref{tab:lbl_ab} (other elements) in the Appendix.

The average NLTE abundances for each species are presented in Table~\ref{tab:NLTE_ab}. While not all elements have NLTE corrections, they are included nonetheless, as the Fe NLTE corrections contribute to their [X/Fe] ratios.

\begin{figure}
    \centering
    \includegraphics[width=0.50\textwidth]{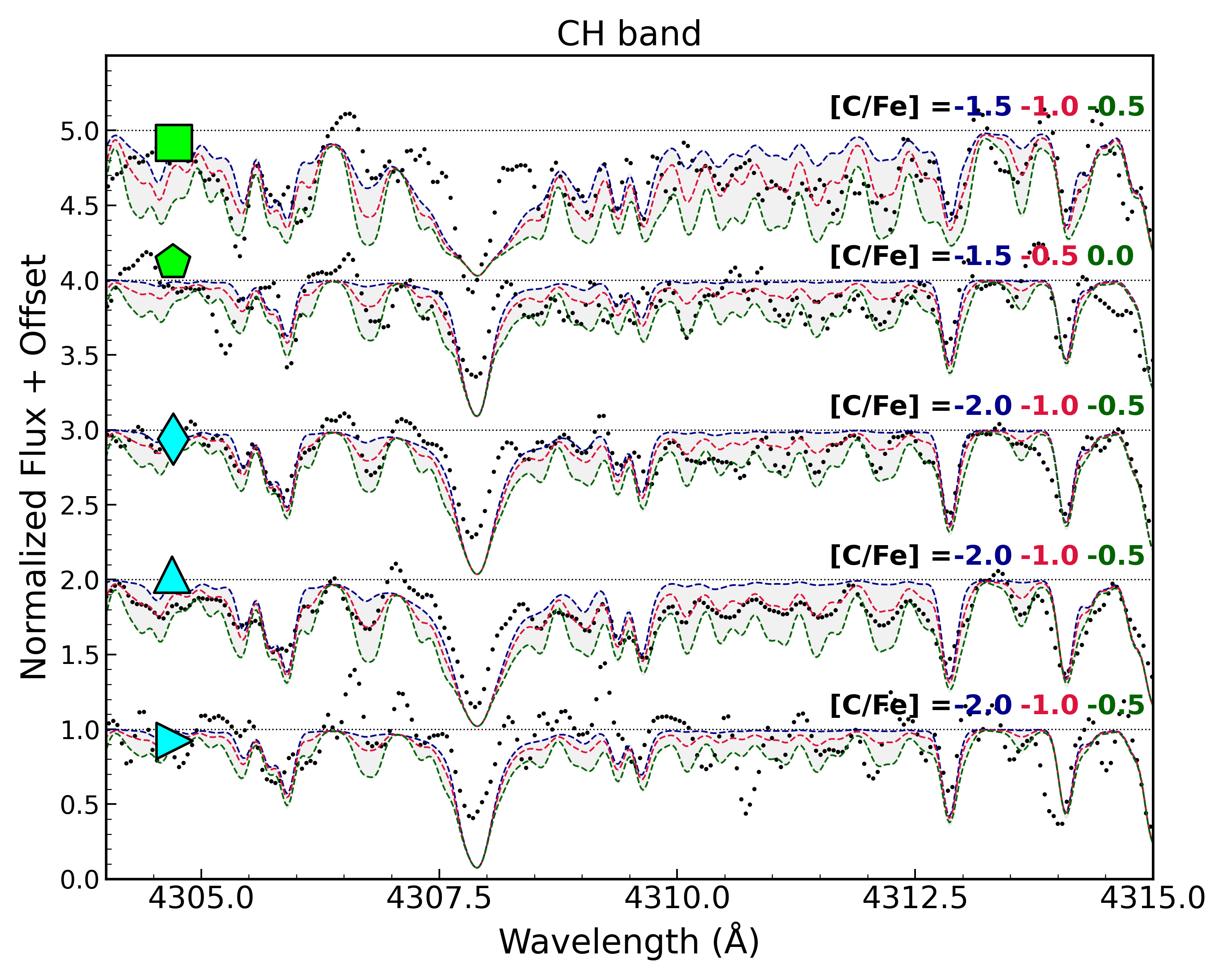}
    \caption{GHOST spectra of targets in Aqu2 and Sgr2, centered on the CH molecular band region. The colors represent synthetic spectra with varying [C/Fe] abundances, as indicated in the plots. The abundances shown are uncorrected for evolutionary effects }
    \label{fig:ch-synth}
\end{figure}

\section{Chemical Abundances}
\label{sec:chem}

Chemical abundances below are compared to the Sun using standard notation [X/Y] = log n(X)/n(Y)$_*$ $-$ log n(X)/n(Y)$_\odot$, where n(X) and n(Y) are column densities (in cm$^{-2}$).  We adopt the solar abundances from \citet{Asplund09}.

For comparison with Milky Way halo stars, we use abundance data from \cite{ Roederer13, Roederer14, Aoki13, Yong13, Yong2021, Li22}.

Measured LTE abundances for targets in both systems are presented in Table~\ref{tab:LTE_abund}.

\begin{table*}
\caption{1DLTE Abundances}
\centering
\resizebox{1\textwidth}{!}{
\hspace{-0.6cm}
\begin{tabular}{lccccc}
\toprule
\multicolumn{6}{c}{\text{[X/Fe] $\pm$ $\sigma_{err} (N_{lines})$ }}\\
species & Aqu2472 & Aqu2776 & Sgr2936 & Sgr2584 & Sgr2656 \\
\hline
CH\footnote{C abundances from CH after applying an evolutionary correction from \cite{Placco14}} & -0.16 $\pm$ 0.30 & -0.32 $\pm$ 0.30 & -0.22 $\pm$ 0.30 & -0.21 $\pm$ 0.10 & 0.00 $\pm$ 0.50 \\
NaI & -1.08 $\pm$ 0.35 (2) & -1.32 $\pm$ 0.26 (2) & -0.07 $\pm$ 0.20 (2) & 0.08 $\pm$ 0.21 (2) & 0.20 $\pm$ 0.41 (2) \\
MgI & 0.39 $\pm$ 0.37 (3) & 0.61 $\pm$ 0.39 (3) & 0.69 $\pm$ 0.20 (3) & 0.50 $\pm$ 0.17 (4) & 0.20 $\pm$ 0.08 (2) \\
KI & 0.79 $\pm$ 0.43 (2) & 1.12 $\pm$ 0.21 (1) & 0.51 $\pm$ 0.10 (1) & 0.73 $\pm$ 0.16 (1) & $<$ 0.91 \\
CaI & 0.59 $\pm$ 0.30 (5) & 0.21 $\pm$ 0.19 (10) & 0.43 $\pm$ 0.10 (10) & 0.48 $\pm$ 0.12 (14) & 0.69 $\pm$ 0.21 (8) \\
ScII & $<$ 0.11 & 0.28 $\pm$ 0.43 (2) & 0.26 $\pm$ 0.14 (4) & 0.27 $\pm$ 0.24 (3) & 0.36 $\pm$ 0.19 (3) \\
TiI & 1.04 $\pm$ 0.48 (6) & -0.29 $\pm$ 0.25 (5) & 0.33 $\pm$ 0.17 (9) & 0.26 $\pm$ 0.23 (11) & 0.60 $\pm$ 0.26 (4) \\
TiII & 0.67 $\pm$ 0.14 (4) & 0.81 $\pm$ 0.30 (4) & 0.56 $\pm$ 0.11 (11) & 0.71 $\pm$ 0.10 (14) & 0.25 $\pm$ 0.24 (3) \\
VI & ... & ... & -0.23 $\pm$ 0.13 (1) & 0.61 $\pm$ 0.24 (1) & ... \\
CrI & -0.15 $\pm$ 0.49 (4) & -0.28 $\pm$ 0.29 (9) & -0.16 $\pm$ 0.15 (7) & 0.00 $\pm$ 0.21 (11) & $<$ 0.06 \\
MnI & ... & -0.67 $\pm$ 0.17 (1) & -0.28 $\pm$ 0.10 (1) & -0.22 $\pm$ 0.18 (2) & ... \\
FeI & -2.66 $\pm$ 0.07 (19) & -1.88 $\pm$ 0.05 (31) & -2.48 $\pm$ 0.03 (59) & -2.36 $\pm$ 0.03 (42) & -2.35 $\pm$ 0.05 (23) \\
FeII & -2.89 $\pm$ 0.17 (2) & -1.63 $\pm$ 0.11 (5) & -2.48 $\pm$ 0.09 (4) & -2.23 $\pm$ 0.16 (5) & -2.10 $\pm$ 0.18 (3) \\
NiI & 0.16 $\pm$ 0.53 (3) & -0.28 $\pm$ 0.15 (8) & 0.25 $\pm$ 0.13 (7) & 0.20 $\pm$ 0.14 (10) & 0.19 $\pm$ 0.13 (2) \\
NdII & $<$ 1.60 & $<$ 0.76 & $<$ 0.52 & 0.31 $\pm$ 0.08 (2) & $<$ 1.83 \\
ZnI & ... & 0.29 $\pm$ 0.10 (1) & 0.44 $\pm$ 0.04 (1) & 0.46 $\pm$ 0.04 (2) & ... \\
SrII & $<$ -1.58 & $<$ -2.10 & 0.08 $\pm$ 0.12 (1) & 0.09 $\pm$ 0.06 (1) & 0.37 $\pm$ 0.10 (1) \\
YII & $<$ 0.07 & ... & 0.03 $\pm$ 0.10 (1) & -0.18 $\pm$ 0.05 (1) & $<$ 0.41 \\
BaII & $<$ -0.90 & $<$ -1.33 & -0.13 $\pm$ 0.16 (3) & -0.14 $\pm$ 0.11 (4) & -0.26 $\pm$ 0.14 (2) \\
LaII & ... & ... & $<$0.92 $\pm$ 0.11 (1) & $<$ 0.44 $\pm$ 0.07 (1) & $<$ 1.39 \\
EuII & ... & ... & $<$ 0.30 & 0.65 $\pm$ 0.20 (2) & $<$ 0.60 \\
\hline
\label{tab:LTE_abund}
\end{tabular}}
\end{table*}

\begin{table*}
\caption{1DNLTE Abundances}
\centering
\resizebox{1\textwidth}{!}{
\hspace{-0.6cm}
\begin{tabular}{lccccc}
\toprule
\multicolumn{6}{c}{\text{[X/Fe]$_{\text{NLTE}}$ $\pm$ $\sigma_{err} (N_{lines})$ }}\\
species & Aqu2472 & Aqu2776 & Sgr2936 & Sgr2584 & Sgr2656 \\
\hline
CH & -0.30 $\pm$ 0.30 & -0.39 $\pm$ 0.30 & -0.33 $\pm$ 0.30 & -0.33 $\pm$ 0.10 & 0.12 $\pm$ 0.50 \\
NaI & -1.32 $\pm$ 0.35 (2) & -1.52 $\pm$ 0.26 (2) & -0.53 $\pm$ 0.20 (2) & -0.53 $\pm$ 0.34 (2) & -0.45 $\pm$ 0.41 (2) \\
MgI & 0.32 $\pm$ 0.37 (3) & 0.55 $\pm$ 0.39 (3) & 0.54 $\pm$ 0.20 (3) & 0.49 $\pm$ 0.17 (4) & 0.21 $\pm$ 0.09 (2) \\
KI & 0.48 $\pm$ 0.43 (2) & 0.90 $\pm$ 0.21 (1) & 0.16 $\pm$ 0.10 (1) & 0.43 $\pm$ 0.16 (1) & $<$ 0.54 \\
CaI & 0.45 $\pm$ 0.30 (5) & 0.20 $\pm$ 0.19 (10) & 0.41 $\pm$ 0.10 (10) & 0.48 $\pm$ 0.12 (14) & 0.73 $\pm$ 0.21 (8) \\
ScII & $<$ -0.10 & 0.27 $\pm$ 0.43 (2) & 0.12 $\pm$ 0.14 (4) & 0.13 $\pm$ 0.24 (3) & 0.23 $\pm$ 0.19 (3) \\
TiI & 1.44 $\pm$ 0.48 (6) & 0.28 $\pm$ 0.25 (5) & 0.75 $\pm$ 0.17 (9) & 0.77 $\pm$ 0.23 (11) & 1.16 $\pm$ 0.27 (4) \\
TiII & 0.55 $\pm$ 0.14 (4) & 0.71 $\pm$ 0.30 (4) & 0.48 $\pm$ 0.11 (11) & 0.63 $\pm$ 0.10 (14) & 0.14 $\pm$ 0.24 (3) \\
VI & ... & ... & -0.37 $\pm$ 0.13 (1) & 0.48 $\pm$ 0.24 (1) & ... \\
CrI & 0.01 $\pm$ 0.49 (4) & -0.03 $\pm$ 0.29 (9) & 0.15 $\pm$ 0.15 (7) & 0.35 $\pm$ 0.21 (11) & $<$ 0.28 \\
MnI & ... & -0.68 $\pm$ 0.17 (1) & -0.42 $\pm$ 0.10 (1) & -0.13 $\pm$ 0.18 (2) & ... \\
FeI & -2.51 $\pm$ 0.07 (19) & -1.79 $\pm$ 0.05 (31) & -2.35 $\pm$ 0.03 (59) & -2.22 $\pm$ 0.03 (42) & -2.21 $\pm$ 0.05 (23) \\
FeII & -2.86 $\pm$ 0.17 (2) & -1.63 $\pm$ 0.11 (5) & -2.48 $\pm$ 0.09 (4) & -2.22 $\pm$ 0.16 (5) & -2.10 $\pm$ 0.18 (3) \\
NiI & -0.05 $\pm$ 0.53 (3) & -0.29 $\pm$ 0.15 (8) & 0.11 $\pm$ 0.13 (7) & 0.07 $\pm$ 0.14 (10) & 0.06 $\pm$ 0.13 (2) \\
NdII & $<$ 1.38 & $<$ 0.74 & $<$ 0.38 & 0.18 $\pm$ 0.08 (2) & $<$ 1.71 \\
ZnI & ... & 0.28 $\pm$ 0.10 (1) & 0.30 $\pm$ 0.04 (1) & 0.33 $\pm$ 0.04 (2) & ... \\
SrII & $<$ -1.79 & $<$ -2.12 & -0.06 $\pm$ 0.12 (1) & -0.04 $\pm$ 0.06 (1) & 0.25 $\pm$ 0.10 (1) \\
YII & $<$ -0.14 & ... & -0.12 $\pm$ 0.10 (1) & -0.32 $\pm$ 0.05 (1) & $<$ 0.29 \\
BaII & $<$ -1.11 & $<$ -1.34 & -0.24 $\pm$ 0.16 (3) & -0.14 $\pm$ 0.15 (5) & -0.39 $\pm$ 0.20 (2) \\
LaII & ... & ... & $<$0.78 $\pm$ 0.11 (1) & $<$0.33 $\pm$ 0.07 (1) & $<$ 1.27 \\
EuII & ... & ... & $<$ 0.19 & 0.53 $\pm$ 0.20 (2) & $<$ 0.48 \\
\hline
\label{tab:NLTE_ab}
\end{tabular}}
\end{table*}

\subsection{Carbon}
\label{sec:carbon}

Carbon abundances are determined by fitting the CH G-band (4290--4315 \AA). A $^{12}C/^{13}C$ ratio between 6 and 11 was included in our synthetic spectra, consistent with values expected for stars at the tip of RGB \citep[e.g.,][]{Szigeti18}. To account for the evolutionary depletion of carbon, the [C/Fe] corrections from \cite{Placco14} were applied to derive the natal carbon abundances. The uncertainties were estimated by varying the carbon abundances within $\pm ~0.5$ dex and examining synthetic fits; see Fig~\ref{fig:ch-synth}. 

\subsection{Alpha Elements (Mg, Ca, Ti)}

The production of $\alpha$-elements primarily occurs through core-collapse supernovae, with a smaller contribution to some from Type Ia supernovae  \citep[e.g., approximately 39\% of Ca;][]{Kobayashi20}.

The MgI, CaI, TiI, and TiII abundances in these stars are determined from a combination of strong resonance and weak subordinate lines.  NLTE and HFS corrections help to reduce line-to-line scatter per element, however we still find an offset between TiI and TiII.  This has been discussed as overly simple NLTE corrections and 3D stellar model effects by \cite{Mallinson2022}. As these effects are expected to be smaller in the TiII lines, then we prioritize the use of TiII throughout the rest of this analysis.

\subsection{Odd-Z Elements (Na, K, Sc)}

Odd-Z elements are excellent tracers of metal-poor core-collapse supernovae due to the odd-even effect in the predicted yields \citep[e.g.,][]{HegerWoosley2010, Nomoto13, Kobayashi20, Ebinger20}. We discuss the formation sites for Na and K in more detail in Section~\ref{sec:disc}.

Sodium abundances are derived from the two strong Na I D resonance lines near 5890 and 5895 \AA, via EW and spectrum synthesis analyses.  These lines typically exhibit strong NLTE departures: for Sgr2936/Sgr2656, $\Delta_{NLTE}$ for Na is $-0.3$/ $-0.5$, respectively. For Sgr2584, the value is smaller ($-0.2$), however we had to calculate the NLTE corrections with a higher logg value (by 0.2) due to unavailability in the NLTE parameters grid. 

The Na I abundances in Aqu2 are extremely low when compared to other red giants in the MW halo and nearby dwarf galaxies; see Fig.~\ref{fig:Abund_plot}.
The more metal-poor star, Aqu 2472, has a very small NLTE correction (only $\Delta_{NLTE} = -0.03$), while the more metal-rich star, Aqu 2776, was slightly larger ($-0.2$). This latter value may be an overcorrection however as we had to increase EWs to be withihn the NLTE grid parameter space (i.e., 30 and 70 m\AA\ more) of INSPECT. Regardless of uncertainties in these NLTE Na corrections, the two targets in Aqu2 have remarkably low Na abundances.  This is discussed further (below) in Section~\ref{sec:chem_aqu_kna}.

On the other hand, potassium in both Aqu2 and Sgr2 appears to be larger than in the MW halo red giants.
The KI resonance line at 7699 \AA\ has been corrected for NLTE effects. For Aqu2776, we also use the KI line at 7664 \AA, however this line is blended with telluric in the other targets.

The ScII lines at 4324.996, 4415.557, 5031.01, 5526.77 \AA\  have been corrected for isotopic and HFS corrections.

\subsection{Iron-Peak Elements (Cr, Mn, Ni, Zn)}

Iron-peak elements are synthesized during the thermonuclear explosions of Type Ia supernovae, as well as during incomplete or complete Si-burning in core-collapse supernovae \citep{Kobayashi2006}.   Abundances have Mn and Ni have recently been reviewed as potential ways to identify MW halo stars that formed in dwarf galaxies \citep[e.g.,][]{delosReyes2022}.

MnI is determined from both resonance and subordinate lines, with both isotopic and HFS corrections.

Our CrI, MnI, and NiI look similar to the MW halo, to within errors.  Possibly the higher metallicity star in Aqu2 shows slightly lower abundances.

\begin{figure*}
    \includegraphics[width=1\textwidth]{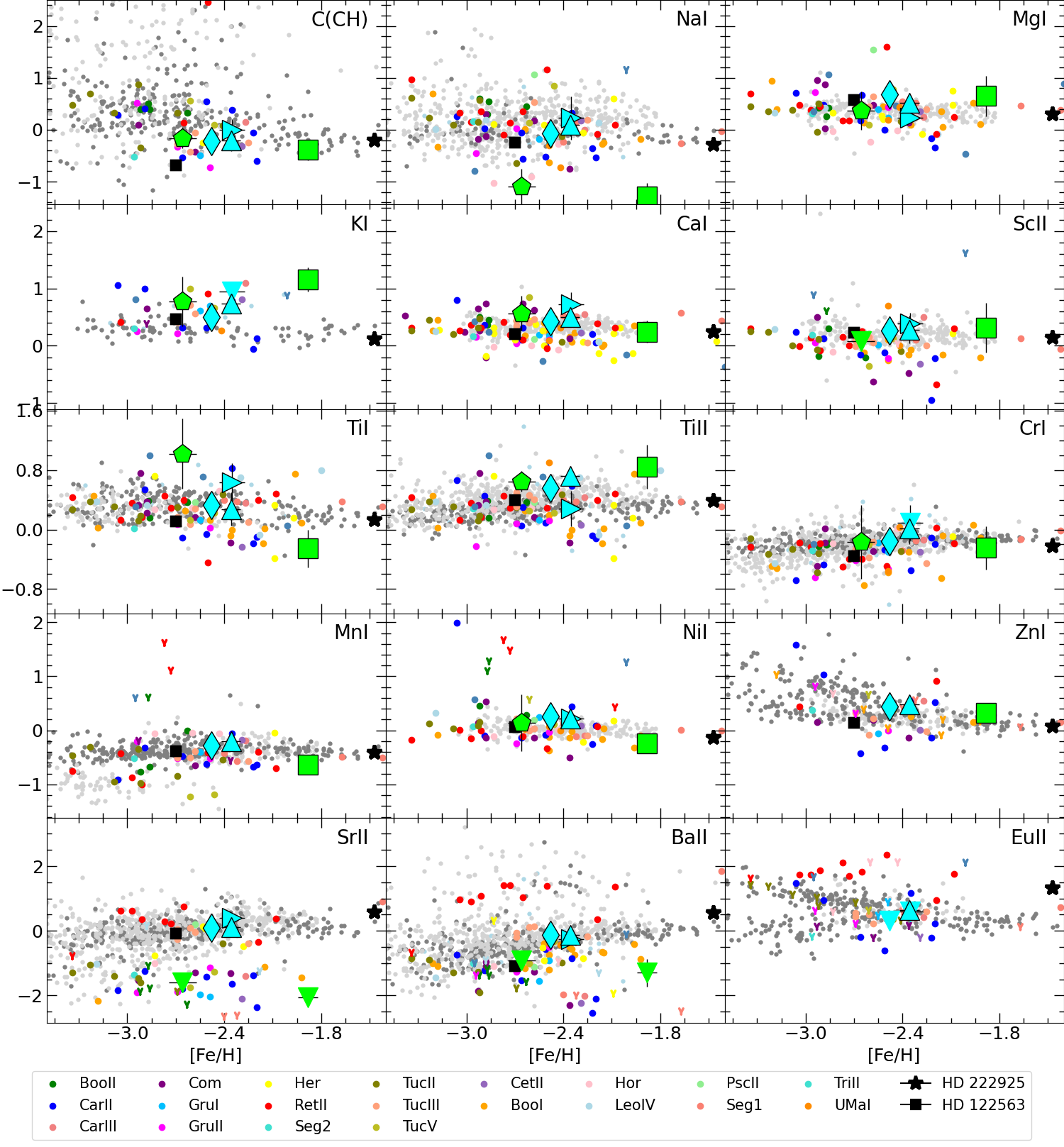}
    \caption{Derived 1DLTE abundances for Aqu2 (lime markers) and Sgr2 (cyan markers) compared to stellar abundances from the MW halo from \cite{Li22, Roederer13, Aoki13, Yong13, Yong2021} (light gray) and \cite{Roederer14} (dark gray), and other UFD galaxies (colored dots according to legend, see text for references). Upper limits are marked with downward pointing triangles. 
    }
    
    \label{fig:Abund_plot}
\end{figure*}

We add ZnI to this discussion, determined from two lines at 4722 and 4810 \AA. No corrections are applied, and it seems well correlated with other metal-poor stars in the MW and dGs (see Fig.~\ref{fig:Abund_plot}).

\subsection{Neutron-capture Elements (Sr, Ba, Eu)}

Neutron-capture elements from through both slow and rapid neutron capture events (merging neutron star binaries and a range of Type II supernovae).  Eu has been shown to be 98\% r-process \citep{Simmerer2004ApJ...617.1091S}, such that [Sr, Ba/Eu] can be used to examine the rise of the s-process in metal-poor galactic stars.  

Europium has two naturally occurring isotopes ${}^{151}\text{Eu}$ and ${}^{153}\text{Eu}$. Using the r-process isotopic fractions from \cite{Sneden2008}, we synthesize Eu II lines at 4129.725 \AA\ and 4205.04 \AA. 
The strong Eu II lines at 4129 and 4205 \AA\ are prominent in Sgr 2584, while only an upper limits are  available from these lines in Sgr 2936.
For Sgr 2656, which has the lowest SNR in the sample, the 4129 \AA\ line is heavily contaminated by noise, so that the upper limit is  estimated from only Eu II 4205 \AA.
The Eu II 4435 \AA\ line is blended with Ca I 4435 \AA\ throughout, so we exclude it from our analysis.
The Eu~II 6645~\AA\ was too weak in our spectra to be detected.

The Ba II 4554 \AA\ line is quite strong, and has significant isotopic shifts and HFS corrections that must be included. For Sgr2584, the HFS correction for this line is $-0.2$ dex, which brings it into better agreement with other Ba II lines at 5853, 6141, and 6496 \AA .  The Ba II 4934 line is blended with an iron line, so we excluded it from the analysis. The largest NLTE correction for Sgr2 members is observed for the line at 6496 \AA, with a value of $\sim -0.2$ dex for all members, whereas for other lines, the correction is $< -0.1$ dex.

Sr II is determined from the two resonance lines at 4077 and 4215 \AA, with no corrections applied.

The heavy element abundances in Aqu2 and Sgr2 (including sample spectrum syntheses) are discussed further in Sections~\ref{sec:aqu2_chem} and \ref{sect:sgr2-rI}.

\subsection{Comparison Stars in UFDs}
\label{sec:disc}

We also compare to velocities, metallicities, and chemistries of red giant stars in other dwarf galaxies; shown in Fig.~\ref{fig:Abund_plot}.
The UFD galaxies with literature abundance measurements are: Bootes I \citep{ Feltzing09,  Norris10, Gilmore13, Ishigaki14, Waller23}, 
Bootes II \citep{Ji16},
Carina II \citep{Ji20}, Carina III \citep{Ji20}, Cetus II \citep{Webber2023ApJ...959..141W}, Coma Berenices \citep{Frebel10, Vargas13, Waller23}, Grus I \citep{Ji2019}, Grus II \citep{Hansen2020}, Hercules \citep{Koch08, Aden11, Vargas13,  Francois16}, Horologium I \citep{Nagasawa_2018}, Leo IV \citep{Simon2010ApJ...716..446S, Francois16, Vargas13}, Pisces II \citep{Spite2018A&A...617A..56S}, Reticulum II \citep{Ji16, Hayes23}, Segue 1 \citep{Norris10, Frebel14}, Segue 2 \citep{RoedererKirbby2014}, Triangulum II \citep{Venn2017, Kirby_2017, Ji2019}, Tucana II \citep{Ji16, Chiti18, Chiti_2023}, Tucana III \citep{Hansen2017, Marshall_2019}, Tucana V \citep{Hansen2024}, Ursa Major I \citep{Waller23}, Ursa Major II \citep{Frebel10}.

\section{Results for Aqu2}

\begin{figure}
    \centering
    \includegraphics[width=1\linewidth]{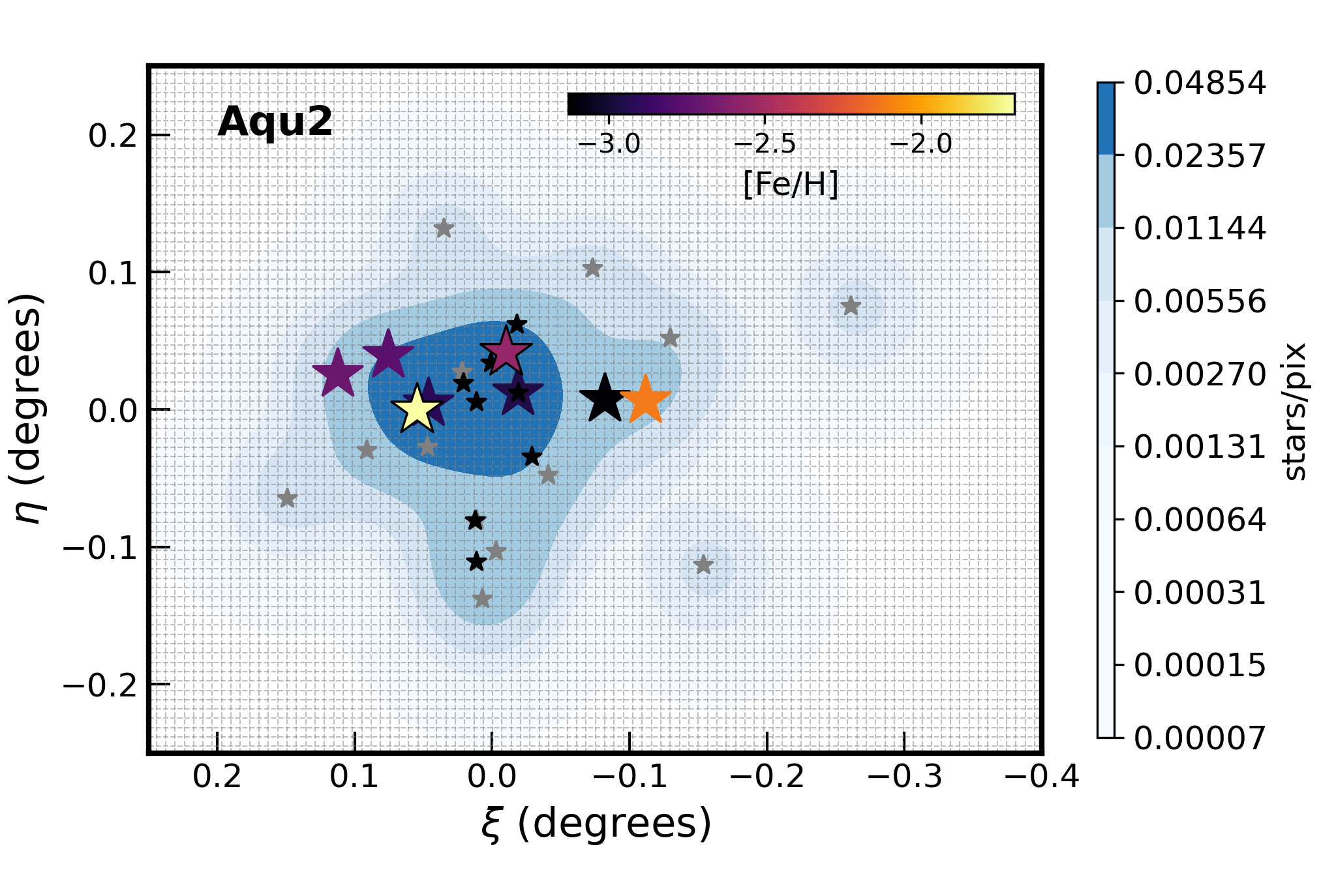}
    \caption{Isophote contour map of Aqu2, generated from the projected coordinates of member stars. Member stars are from \cite{Torrealba16} (black), \cite{Bruce23} (larger stars color-coded by metallicity), and member candidates from \cite{Jensen23} with membership probability $>$ 0.1 (grey). Targets from this study are also color-coded by metallicity, with black edges for distinction. 
    The density of stars per pixel is represented by contour levels in shades of blue, with levels calculated logarithmically from approximately 0.0015 to the maximum density. The central surface brightness, based on the total magnitude, $m_V$, of 15.8, is approximately 26.0 mag arcsec$^{-2}$, while the outermost contour corresponds to a surface brightness of approximately 33.1 mag arcsec$^{-2}$. Each pixel in the map measures 0.5 arcmin $\times$ 0.5 arcmin, inferred from binning the projected coordinates $(\xi, \eta)$ into a grid of 50 x 50 bins, shown with gray lines. 
    }
    \label{fig:aqu2_isophots}
\end{figure}

Our results for two stars in Aqu2 are compared to measurements from lower resolution spectra in the literature - 
specifically, \citealt{Bruce23} (B23) and \citealt{Torrealba16} (T16). Targets are shown in an isophotal contour map of Aqu2 in Fig.~\ref{fig:aqu2_isophots}.

\subsection{[Fe/H] and $v_r$ dispersions}
\label{sec:aqu2_feh_rv}

\begin{figure}
\includegraphics[width=1\linewidth]{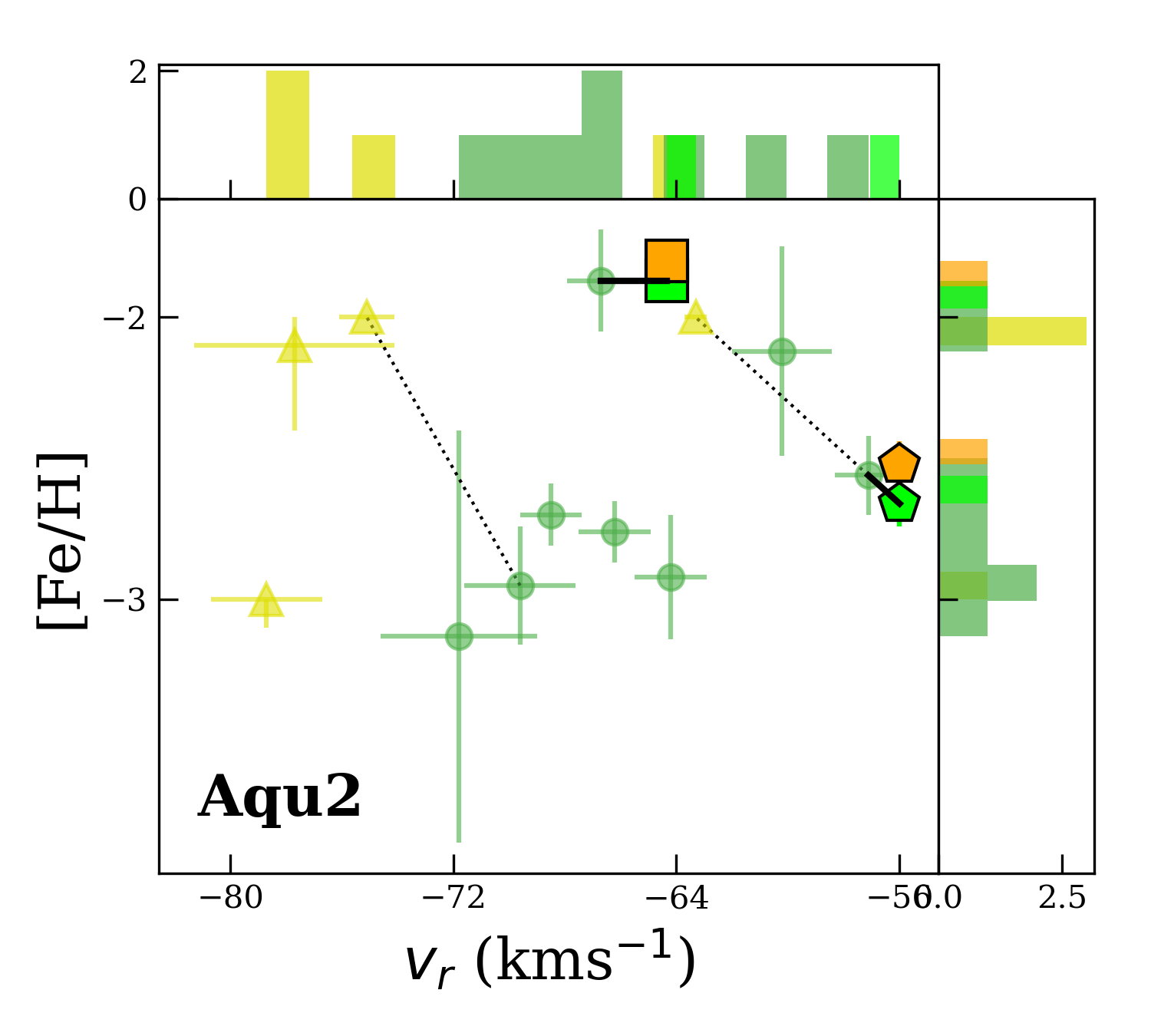}
    \caption{Radial velocity ($v_r$) and metallicity ([Fe/H]) distribution of Aqu2 members (Aqu2776/square,  Aqu2472/pentagon). Yellow triangles represent data from T16 (excluding BHB stars), while green symbols indicate members from B23. Dotted lines connect overlapping stars between T16 and B23, and solid black lines highlight stars overlapping between our sample and B23. Lime/orange symbols are for our 1DLTE/NLTE metallicities.}
    \label{fig:Aqu2_rv_feh}
\end{figure}

The two targets analyzed in this work were previously observed by B23, with one of them (Aqu 2472) also observed by T16. 
Our [Fe/H] and v$_{\text{r}}$  measurements (see Table~\ref{tab:params}) for both stars show excellent agreement with those of B23, as illustrated in Fig~\ref{fig:Aqu2_rv_feh}. This agreement allows us to combine B23's data with our measurements for overlapping stars to derive systemic dispersions in [Fe/H] and v$_{\text{r}}$ for Aqu2. 
Interestingly, both targets emerge as outliers, with Aqu2776 in metallicity and Aqu2472 in velocity.

\begin{figure*}
\includegraphics[width=1\textwidth]{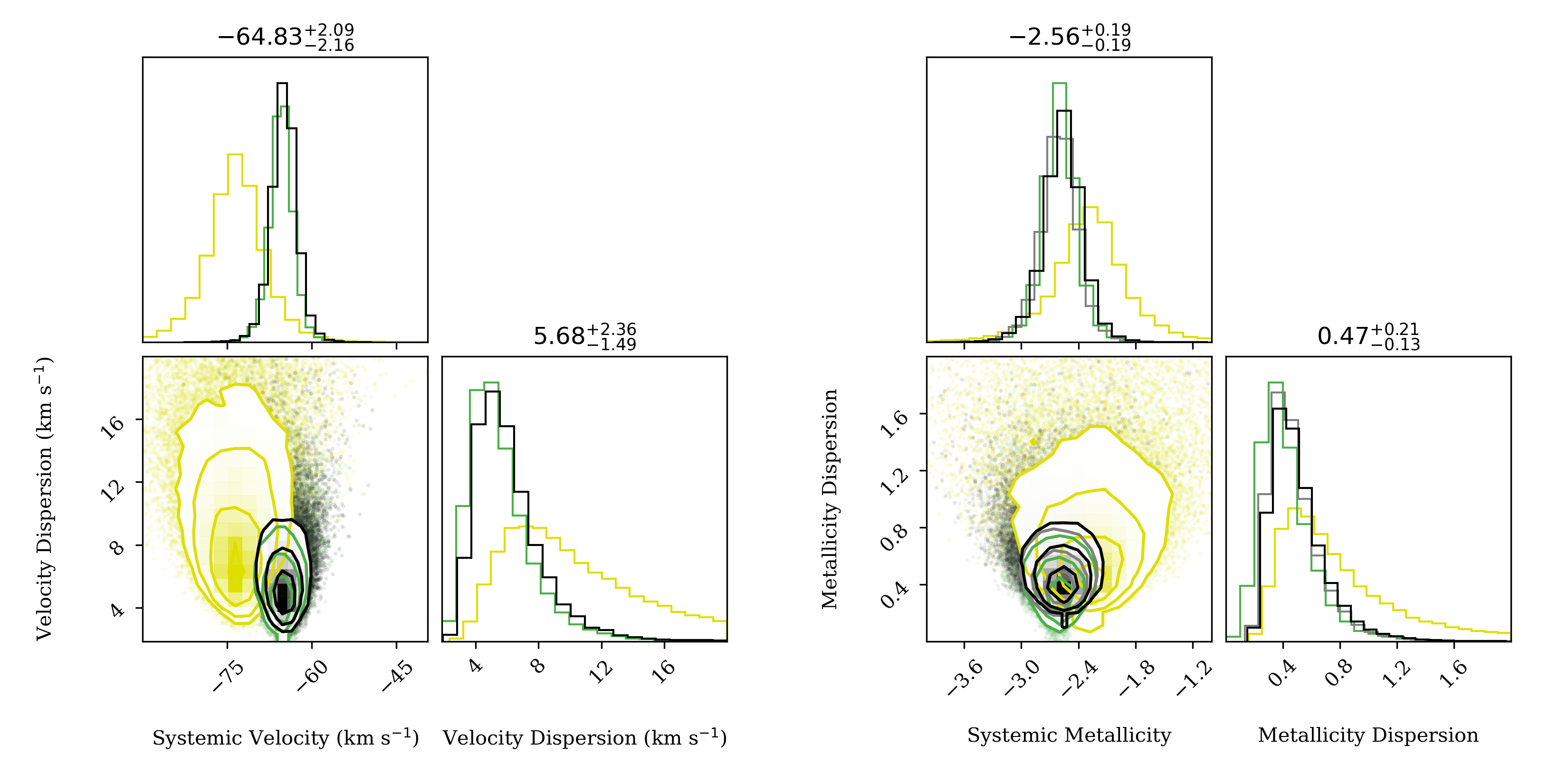}
    \caption{Two-dimensional joint PDFs of systemic velocity and metallicity with their dispersions for Aqu2, derived from running an MCMC sampler and likelihood function described in Section~\ref{sec:aqu2_feh_rv}.  Data from T16 (excluding BHB stars) is shown in yellow, and B23 data in green. Black represents the combined dataset from B23 and this work, with our measurements used for overlapping stars. In the metallicity panel, grey contours represent LTE, while black contours represent NLTE metallicities. The printed values correspond to the black PDFs}
    \label{fig:Aqu2_corn_plots}
\end{figure*}

For Aqu2472, the radial velocity is offset by +9~\kms\ from the systemic velocity,
consistent with the findings in B23.
For Aqu2776, the radial velocity is also consistent with B23 and 
in good agreement with the other members; however, it is offset in metallicity by +0.6~dex above the systemic value.
When compared to B23, we find no evidence for binarity (which may have affected these parameters), validating B23's results which were based on only a single exposure. 
B23 previously noted the substantial impact of Aqu2776 on the systemic metallicity dispersion of Aqu2, which decreases by a factor of 9 when this star is excluded.
The elevated metallicity of Aqu2776 also accounts for the star's offset from the best-fit isochrone on both CMDs in Fig.~\ref{fig:Aqu_cmd}. Contrary to B23's assumption, this deviation is not due to excess carbon, as we show in Section~\ref{sec:carbon}. Alternatively, we suggest this elevated [Fe/H] is due to inhomogeneous mixing in an unevolved UFD galaxy (see Section~\ref{sec:chem_aqu_kna}).

We recalculate the systemic velocity, \( v_{\text{sys}} \), and metallicity, \(\text{[Fe/H]}_{\text{sys}}\), along with their dispersions \( \sigma_v \) and \(\sigma_{\text{[Fe/H]}}\), employing a Bayesian approach using Markov Chain Monte Carlo (\citealt{Hastings1970}, MCMC) to sample from a posterior distribution of these parameters.
The log-likelihood function is given by
\[
\text{log $\mathcal{L}$} = -0.5 \times \sum \left( \frac{(\text{value} - \text{value}_{\text{sys}})^2}{\sigma^2} + \ln(\sigma^2) \right)
\]
where "value" corresponds to either \( v_{\text{sys}} \) or \(\text{[Fe/H]}_{\text{sys}}\), and
\(\sigma^2 = \text{value}_{error}^2 + \sigma_{\text{value}}^2\).
Uniform priors are applied to these parameters, constraining the systemic velocity between \(-90\) $<$ \( v_{\text{sys}} \) $<$ \(-30\), systemic velocity dispersion between 0 $<$ \( \sigma_v \) $<$ 20 \kms, 
metallicity between \(-4 < \text{[Fe/H]}_{\text{sys}} < -1\), and metallicity dispersion  from \(0 < \sigma_{\text{[Fe/H]}} < 2.0\). 
Any values outside these ranges are assigned \(-\infty\) in the prior.
The resultant posterior distribution functions (PDFs) for different data samples are shown in Fig.~\ref{fig:Aqu2_corn_plots}.

When combining this work with other targets from B23, our systemic velocity and velocity dispersion for Aqu2 are \(v_{\text{sys}} \) $=-64.83^{+2.09}_{-2.16}$ \kms\ and \(\sigma_v \) $=5.68^{+2.36}_{-1.49}$ \kms, in good agreement with B23. 
Similarly, our systemic metallicity and metallicity dispersion are
[Fe/H]$_{\text{sys}} = -2.56 \pm 0.19$, and
$\sigma_{\text{[Fe/H]}} = 0.47^{+0.21}_{-0.13}$, which are also in excellent agreement with the values reported by B23. 
Finally, we note no significant difference in the systemic metallicity nor its dispersion when comparing measurements with and without NLTE corrections applied to our data.

\subsection{Chemistry in Aqu2} 
\label{sec:aqu2_chem}

The derived 1DLTE abundances for Aqu2 targets are shown in Fig.~\ref{fig:Abund_plot} in comparison to those of red giants in the MW halo and UFDs.
Aqu2 has a clear metallicity range, one of our targets has [Fe/H] = -2.66, while the other is [Fe/H] = -1.87 dex.  Yet the majority of the element ratios [X/Fe] resemble stars in the MWG. The most clear exceptions are low Na, Sr, Ba, and high K.
These chemical features are 
similar to other unevolved UFDs
with a ``one-shot" enrichment event, i.e., where the Pop II stars retain the chemical yields of the initial Pop III supernovae, without subsequent pollution from Type Ia SNe or AGB stars \citep[e.g.,][]{Frebel12}.  The intense feedback from core-collapse supernovae halts further star formation, leaving Pop II stars with a unique chemical signature preserved in their atmospheres.

As an unevolved UFD, then the observed high [Fe/H] variation between our two stars could  result from incomplete metal mixing, due to the rapid formation of Pop II stars at the center of the proto-galaxy. Indeed, both stars are located near the system's core, as shown by the isophotes in Fig~\ref{fig:aqu2_isophots}. 
The disruption in the isophotes for this small system are also consistent with evacuation from supernova feedback in a shallow potential.
Thus, the central positioning and brief Pop II star formation timescales support that the metallicity variations reflect gas inhomogeneities rather than late-stage chemical evolution.

\begin{figure}
    \centering    \includegraphics[width=0.50\textwidth]{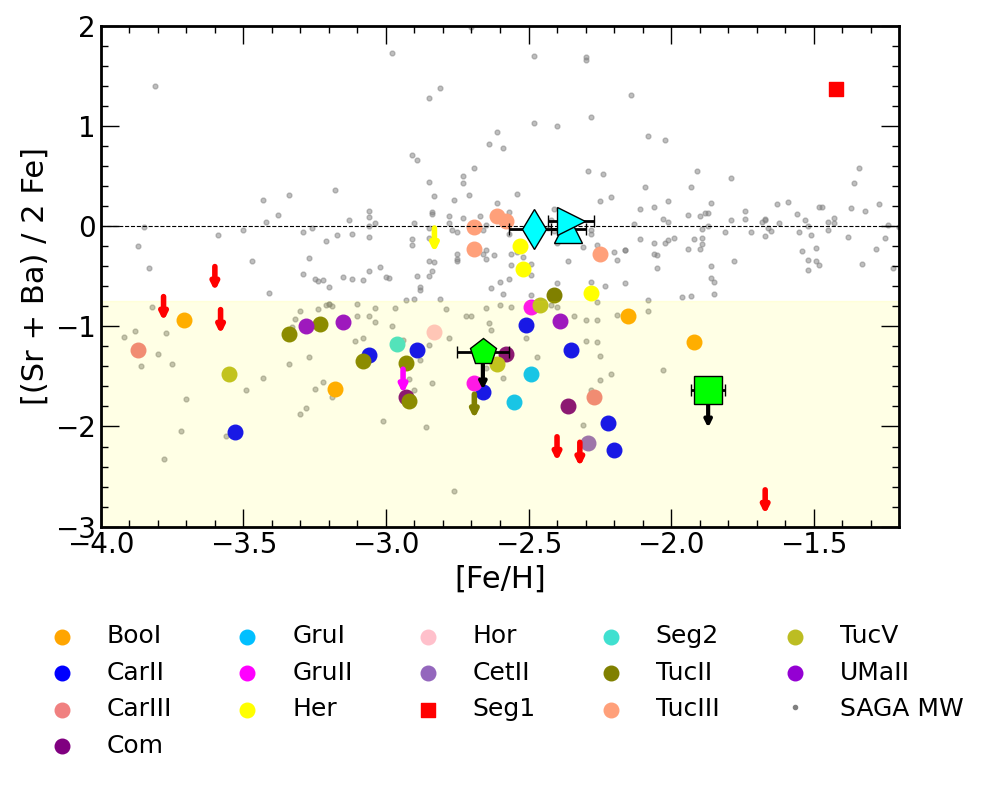}
    \caption{Average n-capture abundances, expressed as [(Sr + Ba)/2 Fe], plotted against [Fe/H] for selected UFDs and MW halo stars (references in Fig.~\ref{fig:Abund_plot}). Upper limits for Aqu2 are indicated in lime, Sgr2 stars are in cyan, and Segue~1 is highlighted in red. The pale yellow region indicates the level predicted for unevolved systems \citep{Frebel12}, with [Sr/Fe] $<$ -0.5 and [Ba/Fe] $<$ -1.
    }
    \label{fig:n-capt_Aqu2}
\end{figure}

Other examples of unevolved systems are  Segue~1 \citep{Frebel14}, Coma Berenices \citep{Frebel18, Waller23}, Hercules \citep{Koch08}. In Fig.~\ref{fig:n-capt_Aqu2}, we present the average abundances of Sr and Ba for selected UFDs, along with the predicted levels for unevolved systems. We highlight Segue~1 (in red) as an example of an unevolved system with a relatively large sample of spectroscopic members with detailed abundance measurements across a wide range of metallicities. This aligns with the suggestion by \cite{Ji2019} that low n-capture abundances can serve as a defining characteristic of the faintest dwarfs. Systems in Fig.\ref{fig:n-capt_Aqu2} that do not exhibit such deficiencies (Sgr2, Tuc~III) will be discussed in the following sections.

\begin{figure}     
\hspace{-0.4cm}\includegraphics[width=0.54\textwidth]{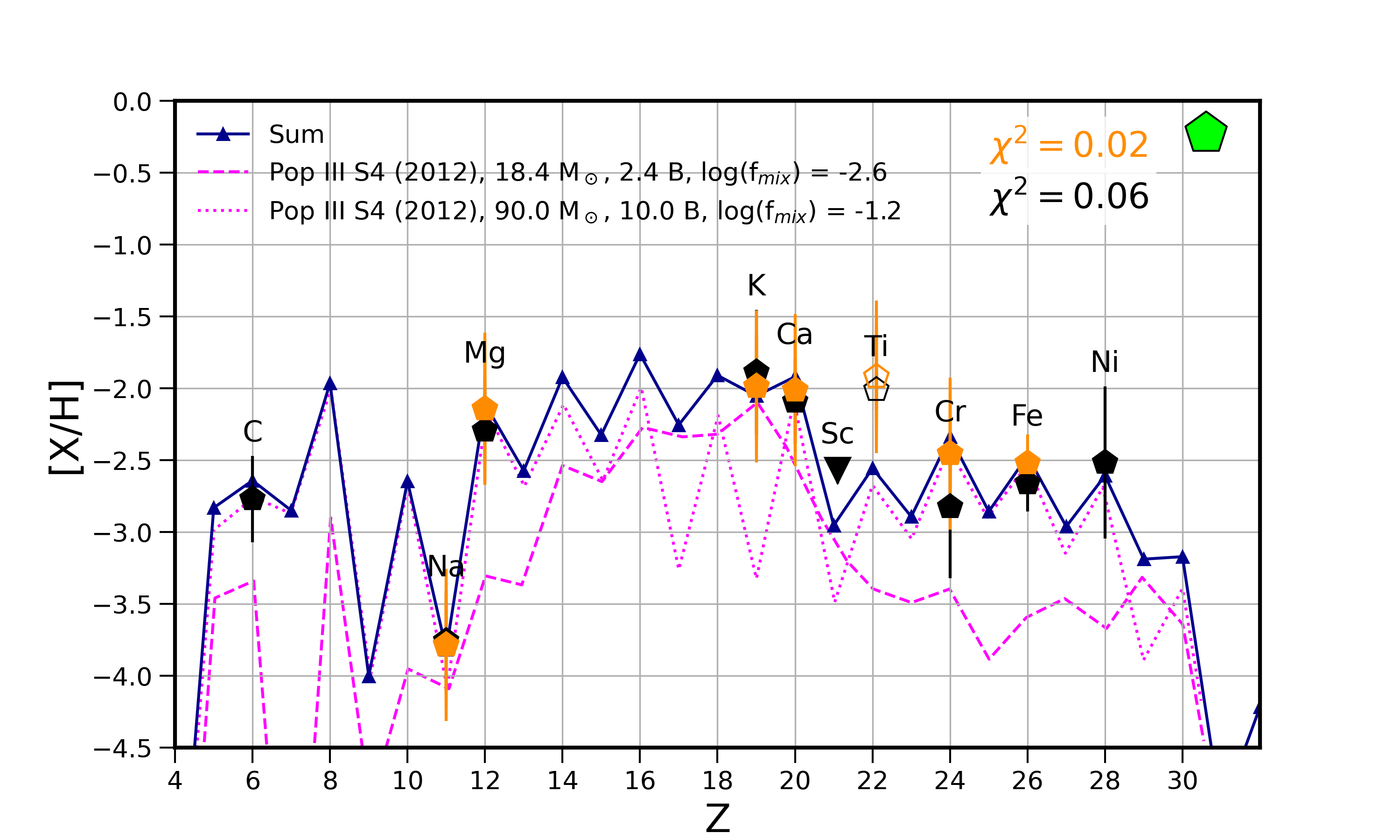}
    \caption{Chemical abundances in Aqu2472 are compared to Pop III models from \citet[][S4 with 2012 updates]{HegerWoosley2010} using \textsc{StarFit}.  The Aqu2472 1DLTE and NLTE abundances from Tables~\ref{tab:LTE_abund} and \ref{tab:NLTE_ab} are shown in black and orange, respectively, including the goodness of the fit ($\chi^2$). Ti was not included in the fit (see text).
    }
    \label{fig:Aqu2StarFit}
\end{figure}

We compare our chemical abundances for Aqu2472 to Population III model yields from \citet[][2012 update]{HegerWoosley2010}. In Fig.~\ref{fig:Aqu2StarFit}, an excellent fit was found when combining the yields from two Pop III SN using StarFit\footnote{StarFit: \url{https://starfit.org/}}. These two stars included an 18.4 M$_\odot$ SN with typical energy (2.4 x 10$^{51}$ erg), combined with a 90 M$_\odot$ SN with excess energy (10 x 10$^{51}$ erg), both with mild mixing fractions.  
Ti was excluded from the fit in Fig.~\ref{fig:Aqu2StarFit}, as its NLTE corrections may be insufficient \cite[e.g., see][]{Mallinson2022, Mallinson2024}; 
including Ti increases the $\chi^2$ value to 0.82. 
We note that our Ti abundance can be reproduced by including a third source, particularly with a significant r-process yield; however, the required additional model would also lead to an overproduction of Sr and Ba relative to our upper limit estimates, making such a scenario unlikely.
For Aqu2776, none of the tested combinations of 2-3 Pop III SN yields provided a satisfactory fit (with $\chi^2 \ge0.8$). Given its proximity to the system's center, as shown in Fig.~\ref{fig:aqu2_isophots}, and its overall higher metallicity, we attribute this to enrichment from more Pop III stars, and most likely pollution from super-AGB stars (discussed below in Section~\ref{sec:chem_aqu_kna}).

\begin{figure*}
\hspace{-0.3cm}
    \includegraphics[width=1\textwidth]{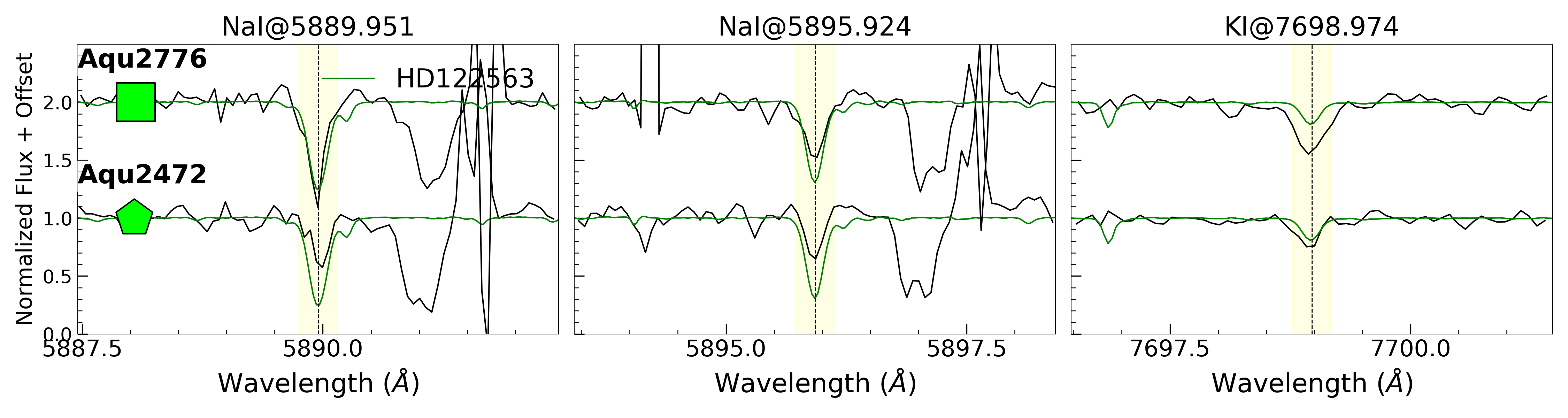}
    \caption{Comparison of the strong KI and weak NaD spectral lines for two stars in Aqu2 and the metal-poor standard star HD122563. }
    \label{fig:NaK_Aqu2}
\end{figure*}

\begin{figure*}
    \centering
    \includegraphics[width=1\linewidth]{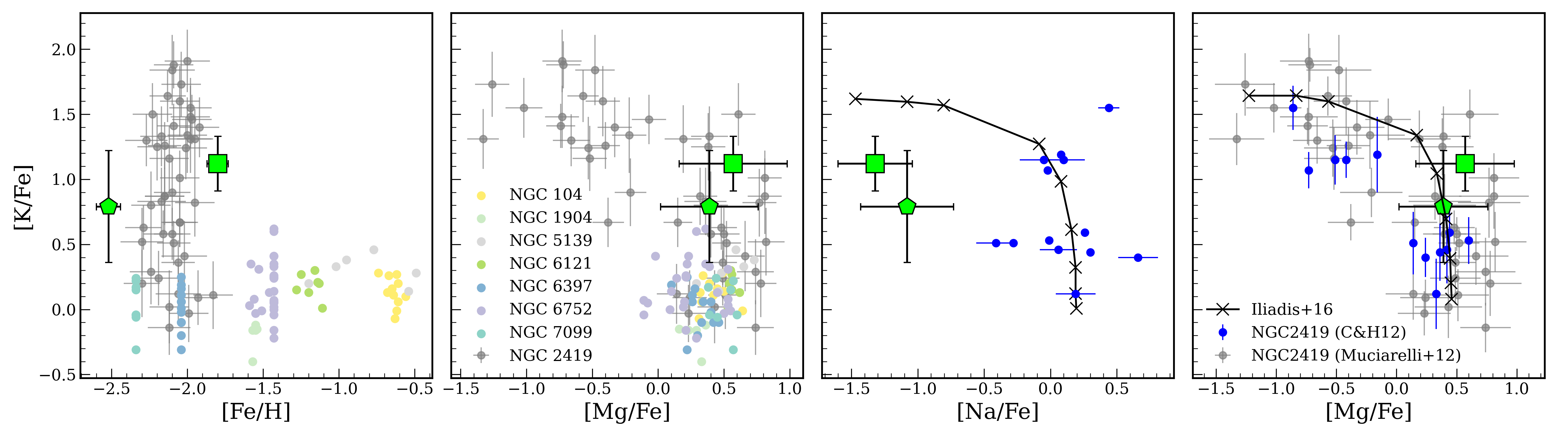}
    \caption{Comparison of [Na/Fe] and [Mg/Fe] vs [K/Fe] for the star cluster NGC 2419. 1DLTE abundances from \cite{Cohen12} are shown in blue. Abundances from \cite{Mucciarelli12}, in grey, include a unique NLTE correction of $-0.3$ applied to K abundances. They did not apply a NLTE correction to Mg, which is predicted to be negligible for Mg-poor stars; however, 1DLTE Mg abundances may be overestimated by 0.2-0.3 dex for the Mg-rich stars. Other MW GC data is from \cite{Carretta2013ApJ...769...40C}. The black line illustrates the predictions     obtained by mixing one part of processed matter with $f$ parts of pristine matter
    in a one-zone nuclear reaction network by  \cite{Iliadis2016ApJ...818...98I} at constant temperature $T = 160$ MK, density $\rho = 900$ g cm$^{-3}$, and hydrogen mass fraction $X_{\mathrm{H}} = 0.7$; 
    the crosses denote, from left to right, the abundances obtained with dilution factors of $f$ = 0.02 (i.e., purely processed matter), 0.05, 0.1, 1.0, 3, 10, 30, 100, and 1000 (i.e., almost purely pristine matter).}
    \label{fig:aqu2_GCs_plus_model}
\end{figure*}

\begin{figure*}
    \centering
    \includegraphics[width=0.9\textwidth]{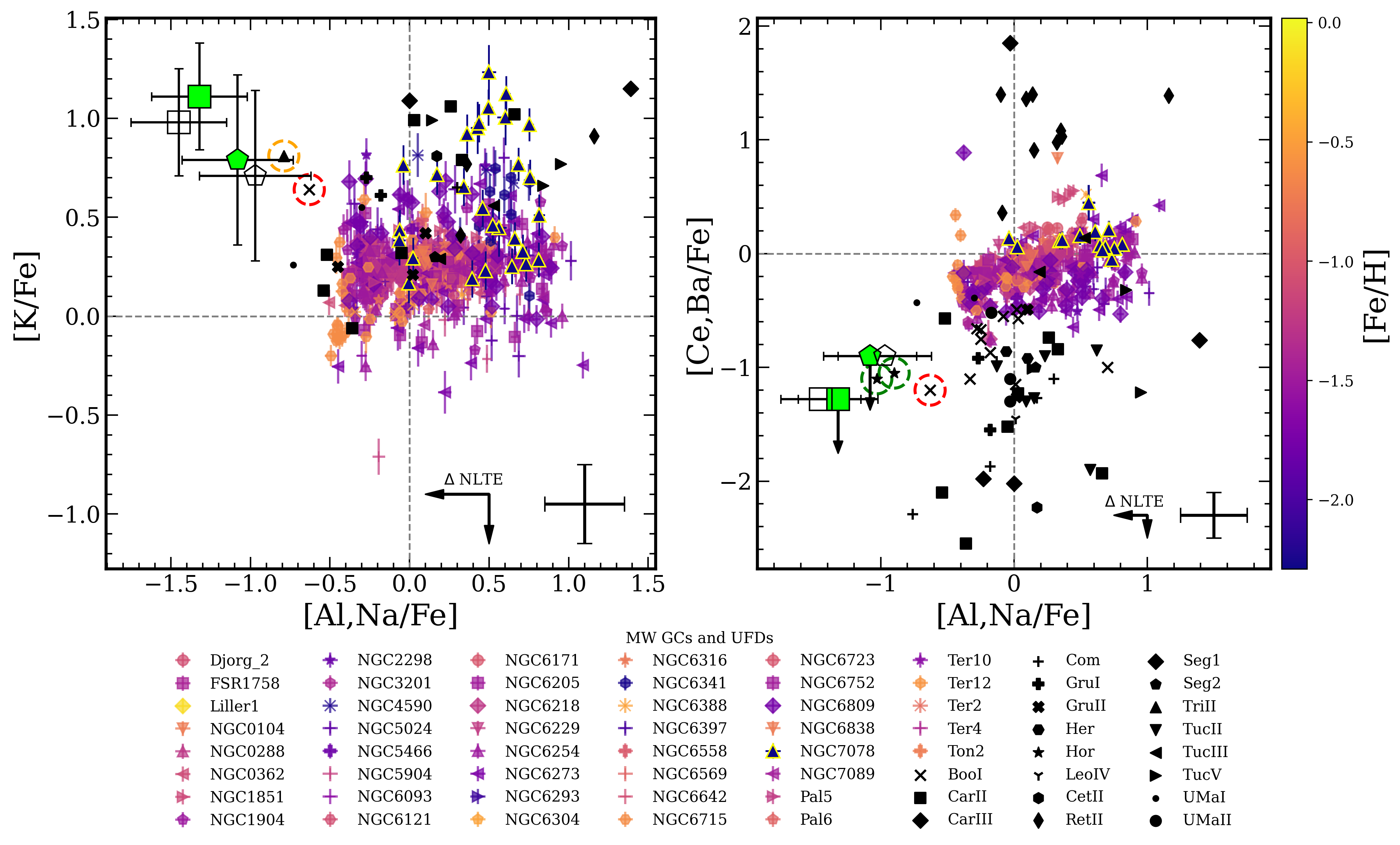}
    \caption{Comparison of Na (Al), Ba (Ce), and [K/Fe]  in Aqu2 targets with those of stars in the Milky Way GCs and UFDs. GCs are from the APOGEE Value-Added Catalogue of Galactic globular cluster stars \citep{schiavon2023apogeevalueaddedcatalogue} and are color-coded by their metallicity.  
    Here, Al (IR) is used as a proxy for Na (opt), and Ce (IR) as a proxy for Ba (opt), based on the reliability of their abundance measurements and shared nucleosynthetic production sites. Only member stars with both velocity and proper motion probabilities greater than 0.5 are included. Outliers (i\textsc{Flag} == 0) and non-giants  (\textsc{Logg} $>$ 1.5) were excluded. To ensure reliable measurements, only APOGEE stars with SNR $>$ 100 are shown, and abundance measurements were filtered using the following criteria: \textsc{X\_FE\_ERR} $<$ 0.3 and \textsc{X\_FE\_FLAG} == 0. References for UFD abundances are as in Fig.~\ref{fig:Abund_plot}. A representative uncertainty for UFDs is displayed in the bottom right corner, along with average NLTE corrections (K estimated from \citealt{Reggiani2019A&A...627A.177R}, Na from \citealt{Lind2011} and Ba from \citealt{Mashonkina19}). NLTE abundances for our targets are shown with open markers. Highlighted with dashed, colored circles are other UFD members that exhibit similar abundance patterns (see text for details). 
    }
    \label{fig:Aqu2KNaAPOGEE}
\end{figure*}

\subsubsection{Element (anti-) correlations (K-Na)}
\label{sec:chem_aqu_kna}

Our 1DLTE abundances for the two targets in Aqu2 exhibit elevated K abundances alongside significantly depleted Na, which is a very unusual abundance pattern.  To check these results, we directly compare the spectra of the NaD lines and the KI
line at 7698 \AA\ in the Aqu2 targets to the standard star HD122563, which has similar stellar parameters;
see Fig.~\ref{fig:NaK_Aqu2}.
Clearly the NaD lines are weaker, particularly in the higher metallicity Aqu2472 star, and the KI line is stronger, particularly in Aqu2776 which has a similar metallicity to HD122563.
There are no signs of contamination (telluric, interstellar, or dust shell). 
 
The [K/Fe] in our two Aqu2 members is more closely aligned with stars in some UFDs than with stars in the MW halo, as seen in the K panel of Fig.~\ref{fig:Abund_plot}.
\cite{Webber2023ApJ...959..141W} recently suggested that enhanced K could serve as a distinguishing feature of UFD stars.
This was based on a suggestion by \cite{Prantzos2018MNRAS.476.3432P} that massive rotating stars may enrich K in such environments; however, they also cautioned that all of their models produce K at levels that are far lower than the solar abundance or MW halo stars. Therefore, the source of the high K in UFDs has not yet been clearly identified.

Alternatively, the unusual  globular cluste NGC 2419 is notable for its substantial scatter in K abundances, reaching unprecedented values of [K/Fe]$\sim+2$ dex \citep{Cohen12, Mucciarelli12}; see Fig.~\ref{fig:aqu2_GCs_plus_model}.  These high-K stars are found to be correlated with depleted Mg; i.e., a Mg-K anti-correlation.
\citet{Ventura2012ApJ...761L..30V} proposed that hot-bottom burning in stars of masses around 6 \msun\ 
-- at the edge between AGB and super-AGB (SAGB) regime -- could reproduce the extreme potassium abundances in NGC 2419  if the standard cross section of Argon nuclei in $^{38} \text{Ar}(p, \gamma)  ^{39} \text{K}$ reaction
were increased by a factor of 100. In this model\footnote{The SAGB model by \cite{Ventura2012ApJ...761L..30V} also predicts a modest increase in Na and a strong depletion in O, i.e., a weak Na-O anti-correlation similar to that observed in GCs. 
The SAGB scenario has been further supported by recent high-precision measurements of the  potassium-destroying reaction $^{39}\mathrm{K}(p, \gamma)^{40}\mathrm{Ca}$ \citep{fox2024highresolutionstudy40ca}, which is critically sensitive to temperature-density conditions \citep{Dermigny_2017}.}, the reduction of Mg and the production of K are maximized by also decreasing the mass loss rate by a factor of 4. 
By exploring a nuclear reaction network sensitive to variations in temperature, density, hydrogen abundance, reaction rates, and initial composition, 
\citet{Iliadis2016ApJ...818...98I}
identified that these abundance variations are limited to a narrow temperature-density range, achievable only by SAGB stars and classical novae.

In the right two panels of Fig.~\ref{fig:aqu2_GCs_plus_model}, we show [Na/Fe] and [Mg/Fe] vs [K/Fe] for stars in NGC 2419
from \cite{Mucciarelli12} 
and \cite{Cohen12}, our 2 stars in Aqu2, and
model results from \citet{Iliadis2016ApJ...818...98I}.
While the data and models clearly show the K-Mg anti-correlation, an 
anti-correlation with Na is not clear.  The star in NGC 2419 with the highest K also has the highest Na (third panel in Fig.~\ref{fig:aqu2_GCs_plus_model}). New observations of Na in the K-rich stars in NGC 2419 would help to constrain these models.  Whether these models are applicable to Aqr2 is not currently clear, particularly as these models were developed for globular clusters and not ultra faint dwarf galaxies, e.g., the SAGB scenario for Aqu2 may be inconsistent with other observed abundance patterns in this system, which align more closely with a one-shot enrichment by Pop~III star.  Investigations into nucleosynthesis in other convective-reactive environments associated with massive stars -- e.g., C and O shell mergers, which may also be induced by rotation \citep{Ritter18} 
-- could also explain high K abundances in UFDs.

\begin{figure*}
    \includegraphics[width=0.9\textwidth]
   {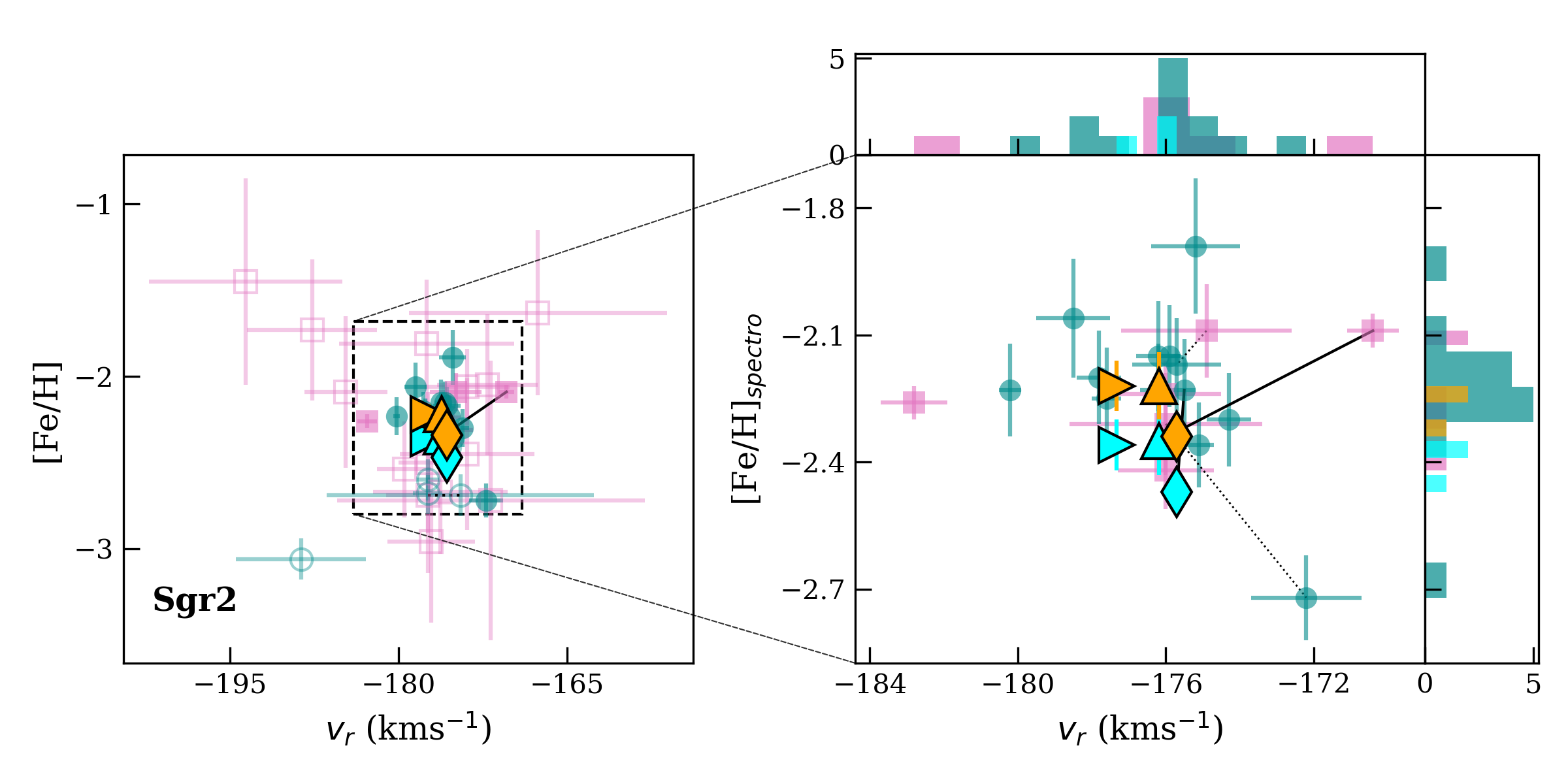}
   \caption{Radial velocity ($v_r$) and metallicity ([Fe/H]) distribution of Sgr2 members. Data from L20 and L21 are shown in pink and dark cyan, respectively. Open markers represent photometric members, and filled markers represent spectroscopic members, with one potential binary from the L20 sample excluded from the analysis. The right panel provides a zoomed-in view of the spectroscopic sample. Measurements from this work are plotted in light cyan (LTE) and orange (NLTE). Dashed lines connect the same objects between the L20 and L21 samples, while solid lines connect overlapping stars between this study and the L20, L21 datasets.} 
   \label{fig:Sgr2_rv_feh}
\end{figure*}

Finally, K in Aqu2 is compared to stars in the globular clusters available in  the APOGEE Value-Added Catalogue
of Galactic globular cluster stars \citep{schiavon2023apogeevalueaddedcatalogue}.
As Na is not reliably measured from IR spectra of metal-poor RGB stars in APOGEE, then we examine Al (IR) as a proxy for Na; this is not ideal, however they are both odd-Z elements that participate in various stages of H-burning and show abundance variations in GC stars \citep[e.g.,][]{meszaros2015exploring}.
Fig.~\ref{fig:Aqu2KNaAPOGEE} illustrates the comparison of Na (Al) and K abundances in Aqu2 targets with those from optical studies of UFDs (Na, K) and those from IR studies of MW GCs (Al, K) from the APOGEE VAC. 
The comparison reveals that very high [K/Fe] abundances are present in some globular clusters.  Notably, NGC 7078 (M15) which has a similar metallicity to Aqu2 and of course NGC 2419 (as shown in Fig.~\ref{fig:aqu2_GCs_plus_model}).
As M15 is known to show variations in neutron-capture elements \cite{Worley2013, Garcia24}, we also examine [K/Fe] vs Ce (IR) or Ba (opt) in Fig.~\ref{fig:Aqu2KNaAPOGEE}.
Unfortunately, no systems have very similar abundance patterns to our two stars in Aqu2, where the low Na is poorly matched to most stars in both the UFDs (Na) and GCs (Al).
The closest matching stars are: (i) one member in the Triangulum II UFD with high [K/Fe]$=+0.8$ and low [Na/Fe]$= -0.8$ \citep{Venn2017},
(ii) one star in Bootes I with high [K/Fe]$=+0.6$, low [Na/Fe]$= -0.6$ and low [Ba/Fe]$=-1.2$ \citep{Waller23}, and
(iii) two stars in the Horologium I UFD with low [Na/Fe]$= 0.1$  and low [Ba/Fe]$= -1$, but lacking K measurements \citep{Nagasawa_2018}. These stars are identified (circled) in Fig.~\ref{fig:Aqu2KNaAPOGEE}.

We conclude that SAGB stars are the most likely source for the enhancements in K, anti-correlated with Mg, and potentially Na.
This conclusion requires that the low n-capture abundances observed in Aqu2 from the ``one shot`` model include the Pop III yields, but limited yields from other (Pop II) massive stars, presumably due to SN feedback losses.

\section{Results for Sgr2}

Our results for three stars in Sgr2 are compared to measurements from lower resolution spectra in the literature - 
specifically, \citealt{Longeard_2020} (L20) and \citealt{Longeard_2021} (L21).

\subsection{[Fe/H] and $v_r$ dispersions}
\label{sec:sgr2_feh_rv}

The three targets in this paper have previous spectroscopic measurements by L20 (from DEIMOS) and L21 (from FLAMES), as shown in Fig.~\ref{fig:Sgr2_rv_feh}. Only two of our three targets were considered members of Sgr2 (Sg2936 and Sgr2584); however, we confirm the membership of all three stars (including Sgr2656).

\begin{figure*}
    \centering
    \includegraphics[width=0.9\textwidth]{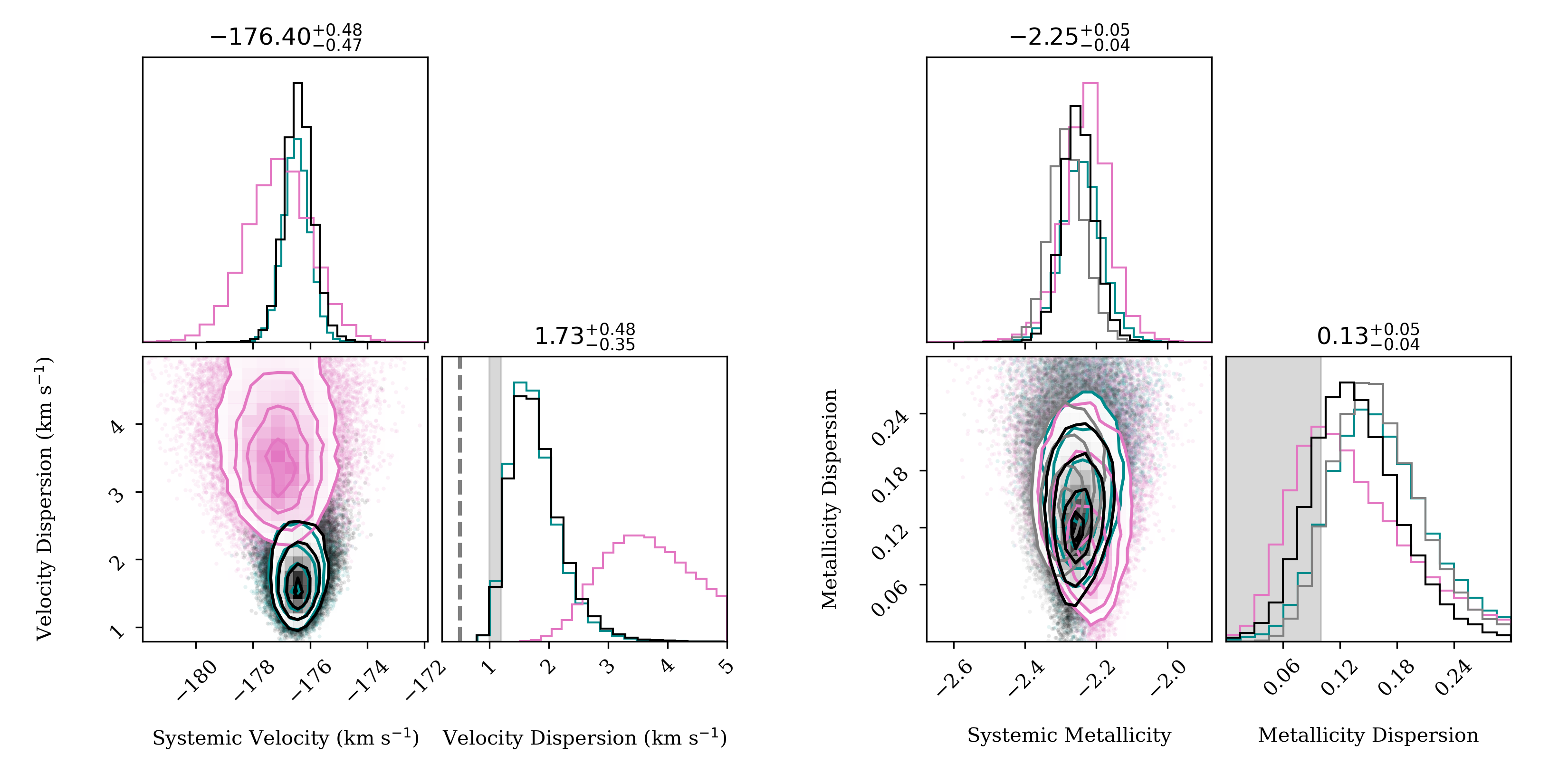}
    \caption{\textit{Left}: Two-dimensional joint PDFs of systemic velocity and its dispersion for Sgr2. Pink and dark cyan show L20 and L21 samples, respectively. Black represents the combined L21 spectroscopic data with targets from this work; overlapping targets use measurements from this study. The grey band indicates the velocity dispersion expected under a purely baryonic scenario, based on the formalism of \citet{Wolf2010} (1.1 $\pm$ 0.1 km s$^{-1}$; \citealt{Longeard_2021}). The grey dashed line represents the velocity dispersion derived from the N-body modeling of globular cluster velocity dispersion profile, calculated as described in \citet{Baumgardt2017} (0.5 \kms). \textit{Right}: Joint PDFs of systemic metallicity and its dispersion for Sgr2. Only spectroscopic members from L20 (pink) and L21 (dark cyan) were used. Grey shows combined L21 data with LTE metallicities of targets analyzed here, and black shows the same with NLTE. The grey band indicates the upper limit for metallicity dispersion in GCs ($<0.1$ dex). Contours represent the 39\%, 88\%, and 95\% volume intervals.}
    \label{fig:JPDFs_Sgr}
\end{figure*}

For two targets, our radial velocity measurements (in Table~\ref{tab:params}) are in good agreement with those from L21; $v_r$ within 0.5 $\sigma$($v_r)$ for both Sgr2936 and Sgr2656. 
For Sgr2584, we find $v_r = -176.2 \pm 0.1$ \kms, which is in excellent agreement with our other two targets, but
$\sim6$ \kms\ lower than L20 ($v_r = -170.4 \pm 0.7$ \kms). Potentially Sgr2584 is a binary star.
Similarly, our metallicity measurements (in Table~\ref{tab:params}) are in good agreement with the Ca II triplet metallicity estimates from L21; [Fe/H]$_{\rm NLTE}$ are within 1 $\sigma$([Fe/H]) for both Sgr2936 and Sgr2656.
For Sgr2584, we find [Fe/H]$_{\text{NLTE}} = -2.24 \pm 0.06$ ([Fe/H]$_{\text{LTE}} = -2.36$), which is in excellent agreement with the other two targets, but $\sim0.15$ dex lower than L20 ([Fe/H] = $-2.09 \pm 0.04$).
Generally, our NLTE metallicities are in good agreement with those derived by L21 from CaT lines (only our LTE metallicities are $\sim0.2$ dex lower).
This good agreement enables us to combine the two samples (L21 + this work) without accounting for a zero-point offset between the instruments. Using our measurements for the overlapping stars, we revisit the systemic metallicity and velocity dispersions of Sgr2, shown in Fig.~\ref{fig:JPDFs_Sgr} and as described in Section~\ref{sec:aqu2_feh_rv}.

The velocity dispersion derived from the entire L21 spectroscopic sample and this study finds
$\sigma_{v_r} = 1.73^{+0.48}_{-0.35}$ \kms.
L21 estimated the velocity dispersion for a typical Milky Way Sgr2-like globular cluster is $1.1 \pm 0.1$ \kms, thus, our PDF suggests Sgr2's velocity dispersion is resolved and just barely higher than expected for a GC. This does not clearly rule out the GC origin for Sgr2 though, as some stars may be binaries (e.g., Sgr2584).
Our examination of the metallicity dispersion shows $ \sigma_{[Fe/H]} = 0.13^{+0.05}_{-0.04}$ dex (NLTE). Thus, our PDF suggests Sgr2's metallicity dispersion is resolved, but just barely above that expected for a GC.

On further analysis, 
the L20 and L21 datasets reveal two stars in common, hereafter referred to as star34 and star83 (these numbers correspond to the target's order in Table~2 of L20; see Fig.\ref{fig:Sgr2_rv_feh}).
The measurements for star34 are consistent within the reported uncertainties; however, star83 shows a large discrepancy in radial velocity and metallicity, where RV$_{\text{DEIMOS}}^{\text{L20}} = -176.0 \pm 2.6$ km/s and 
[Fe/H]$_{\text{DEIMOS}}^{\text{L20}} = -2.31 \pm 0.12$ (SNR = 17), compared to
RV$_{\text{FLAMES}}^{\text{L21}} = -172.2 \pm 1.5$ km/s and [Fe/H]$_{\text{FLAMES}}^{\text{L21}} = -2.72 \pm 0.10$ (SNR = 11). 
To assess the influence of star83 on the systemic metallicity and velocity dispersions ($\sigma_{[Fe/H]}$, $\sigma_{v_r}$), we performed a jackknife test, presented in Table~\ref{tab:jacknife_test}.
The L21 FLAMES data for star83 has no significant influence on the velocity dispersion, however it alone can double the metallicity dispersion in Sgr2 and suggest a metallicity spread larger than seen in globular clusters, i.e., consistent with a UFD.
The discrepancy between L20 and L21, combined with the outsized influence of this star, suggests that star83 (PS1 $g_0=20.21$) should be reobserved.

\begin{table}
\caption{Metallicity and velocity dispersions for Sgr2, 
including and excluding star83.}
\label{tab:jacknife_test}
\resizebox{0.48\textwidth}{!}{
\hspace{-2.0cm}
\begin{tabular}{lcc|ccc}
\toprule
& sys [Fe/H] & $\sigma_{[Fe/H]}$ & & sys $v_r$ & $\sigma_{v_r}$ \\
& (dex) & (dex) & & (\kms) &  (\kms) \\
\hline
\multicolumn{6}{c}{\textit{L21 + this work}} \\
\hline
NLTE & $-2.25^{+0.04}_{-0.04}$ & $0.13^{+0.05}_{-0.04}$ & 
ph+sp: & $-176.52^{+0.39}_{-0.39}$ & $1.55^{+0.38}_{-0.28}$ \\
\textit{LTE} & $-2.29^{+0.05}_{-0.05}$ & $0.16^{+0.05}_{-0.04}$ & \textit{spec:} & $-176.89^{+0.63}_{-0.66}$  & $ 1.72^{+0.48}_{-0.35} $\\ 
\hline
\multicolumn{6}{c}{\textit{L21 (star83 excluded) + this work}} \\
\hline
NLTE & $-2.23^{+0.03}_{-0.03}$ & $0.06^{+0.03}_{-0.03}$ & 
ph+sp: & $-176.41^{+0.49}_{-0.47}$ &$1.45^{+0.36}_{-0.26}$ \\
\textit{LTE} & $-2.27^{+0.05}_{-0.05}$ & $0.13^{+0.05}_{-0.03}$ &  
\textit{spec:} & $-177.15^{+0.60}_{-0.65}$ & $1.59^{+0.43}_{-0.31}$ \\
\hline
\end{tabular}}
\end{table}

\begin{figure*}
\hspace{-0.3cm}
    \centering
    \includegraphics[width=1.0\textwidth]{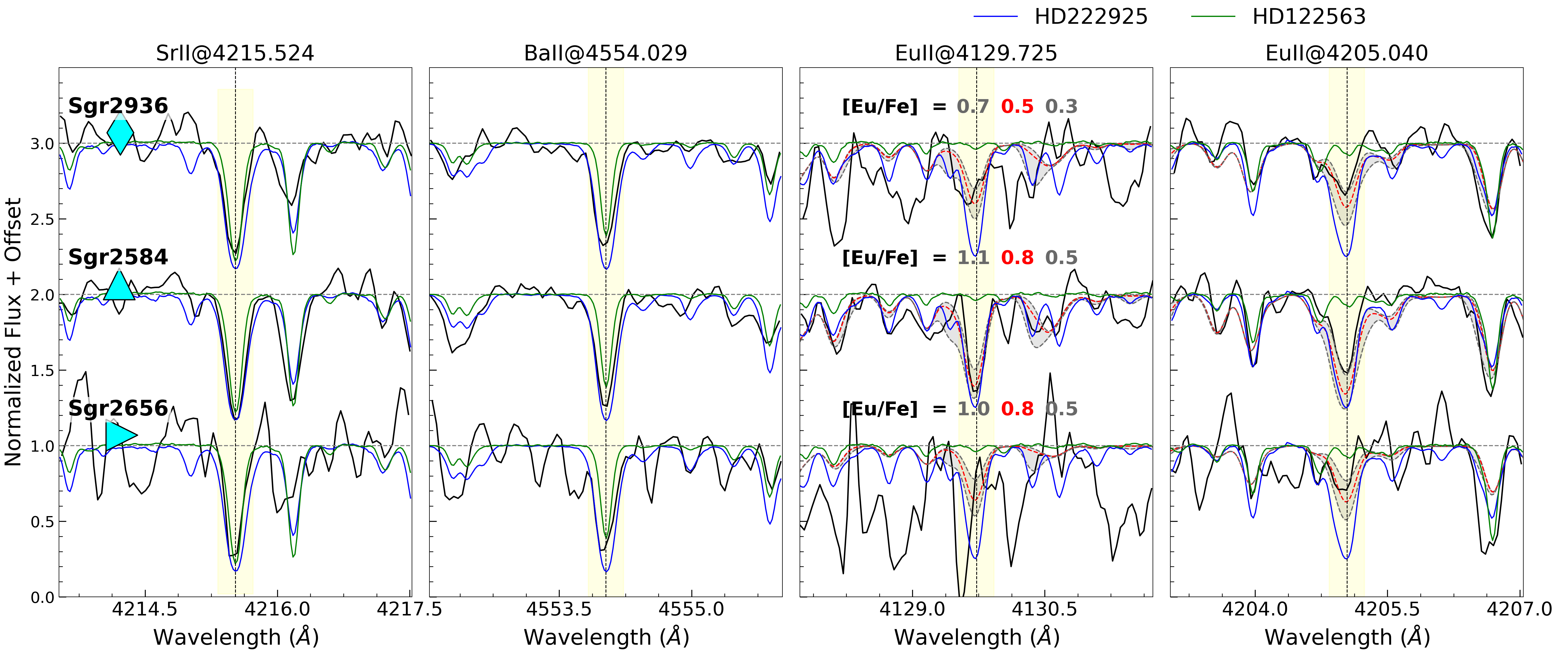}
    \caption{Comparison of n-capture element lines of SrII, BaII, and EuII between Sgr2 stars (black), the r-II standard star HD 222925 (blue), and the non-r-process-enhanced star HD 122563 (green). For the Eu lines, we also include synthetic spectra with varying [Eu/Fe] abundances, as indicated in the plots. The enhanced n-capture elemental abundances in Sgr2 stars are apparent, particularly the pronounced Eu II lines in Sgr 2584. For Sgr 2936 and Sgr 2656, only upper limits could be determined, both derived from the Eu line @4205 \AA.
}
    \label{fig:ncapt_Sgr2}
\end{figure*}

\subsection{Chemistry in Sgr2}
\label{sect:sgr2-rI}


Due to the small sample size, asserting any abundance spreads among the elements in Fig.~\ref{fig:Abund_plot} in Sgr2 is challenging.  
The only chemical abundances that catch our attention are those for the neutron-capture elements, where Sr II and Ba II are usually somewhat lower for stars in dwarf galaxies than the Galactic comparison stars.  

To emphasize this result, we show the
Sr II, Ba II, and Eu II line syntheses for our Sgr2 stars in Fig.~\ref{fig:ncapt_Sgr2}, relative to GHOST spectra of two standard stars, HD222925 and HD122563.  
While Eu II is clearly identified and synthesized in Sgr2584, we are more cautious in the analysis of the other Sgr2 stars, i.e., suggesting only upper limits for Eu II. Nevertheless, the Eu II upper limit for Sgr2936 is a valuable constraint as it is \textit{lower} than the measurement for Sgr2584.  

We also examine the absolute abundances of Sgr2584 and Sgr2936\footnote{Sgr2656 is not plotted due to having fewer abundance measurements, all of which are consistent with the other two targets.} compared to solar values and our two standard stars, HD222925 and HD122563.
In Fig.~\ref{fig:HD_abund}, the chemistry for the two standard stars in determined in two ways: (1) from the literature  for HD222925 \citep{roederer2018}
and HD122563 \citep{Honda_2006, Collet2018},  and (2) derived using the same spectral lines as for the Sgr2 stars (details discussed further in \citealt{Venn2025}). 
The solar system's s- and r-process abundance
patterns from \citet{Simmerer2004ApJ...617.1091S} are scaled to match the
Ba and Eu abundances in Sgr2584. 
The chemical abundances for both Sgr2 targets are similar, and generally fall between those of the two standard stars. 

\subsubsection{Discovery of an r-I star in Sgr2}

\begin{figure}
    \hspace{-0.3cm}
    \includegraphics[width=1.08\linewidth]
    {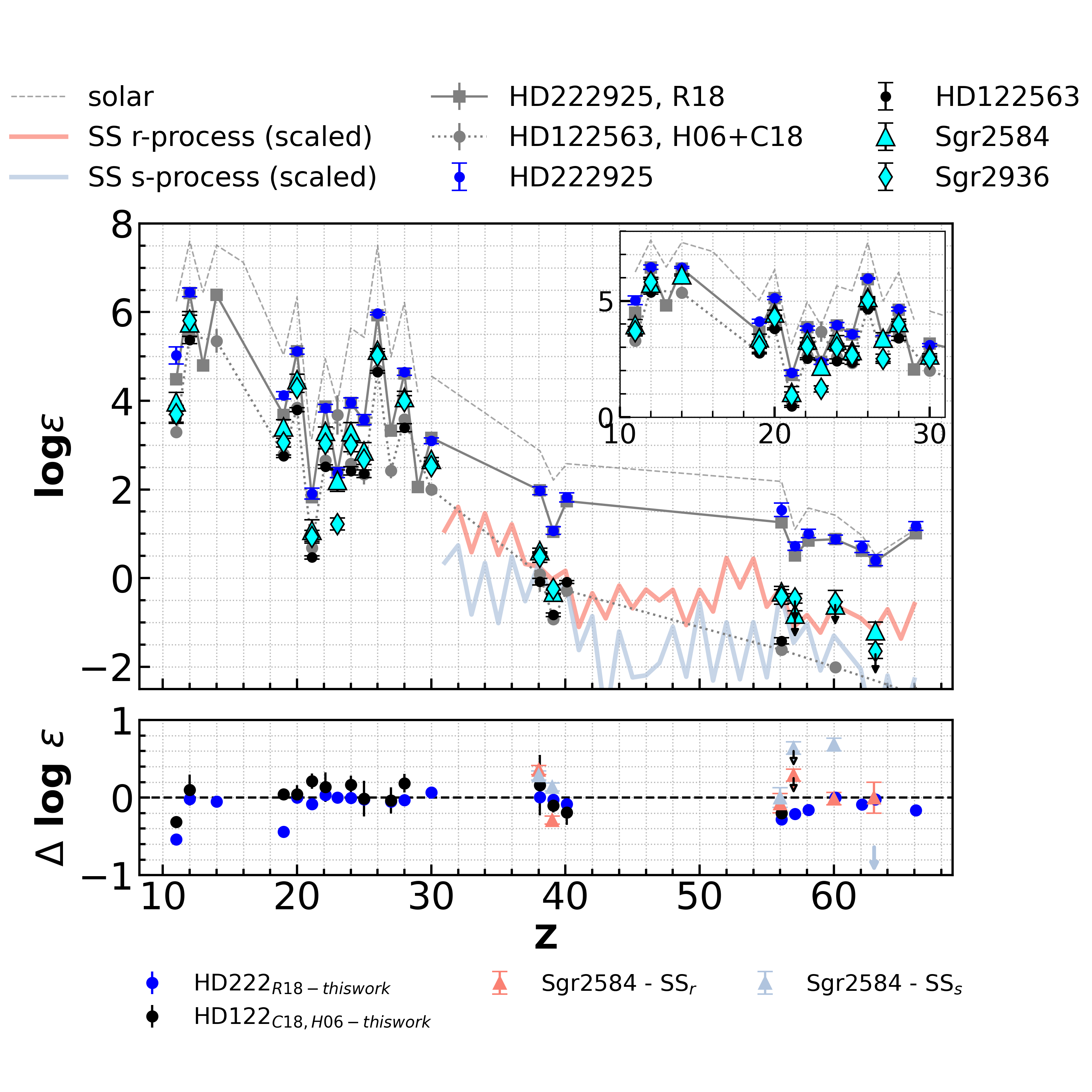}
    \caption{Comparison of abundances for two Sgr2 targets with HD222925, a standard r-process star, and HD122563, which has similar stellar parameters to the targets. The solar abundances are indicated by the grey dashed line. Abundances for the standard stars, derived using the same spectral lines as for the targets, are shown as black circles for HD122563 and blue circles for HD222925. Grey markers represent literature values for HD222925 from \citet{roederer2018} and for HD122563 from  \citet{Collet2018} and \citet{Honda_2006}. The solar system's s- and r-process abundance patterns from \citet{Simmerer2004ApJ...617.1091S}, scaled to match the Ba and Eu abundances in Sgr2584, are shown in blue and red, respectively. The lower panel displays the residuals. 
    }
    \label{fig:HD_abund}
\end{figure}

The abundance pattern of Sr, Ba, and Eu suggests that Sgr2584 is an r-I star (0.3$<$[Eu/Fe]$<$1 and [Eu/Ba]$>$0.4; c.f., \citealt{Hansen2017}).
In Fig.~\ref{fig:HD_abund} we also present the solar system s- and r-process curves from \cite{Sneden2008}, scaled to the Ba and Eu abundances of Sgr2584, respectively. Thus, a scaled solar system r-process pattern shows good agreement with the abundances in Sgr2584, based on our measurements of six n-capture elements.

\begin{figure}
    \centering
    \includegraphics[width=1\linewidth]{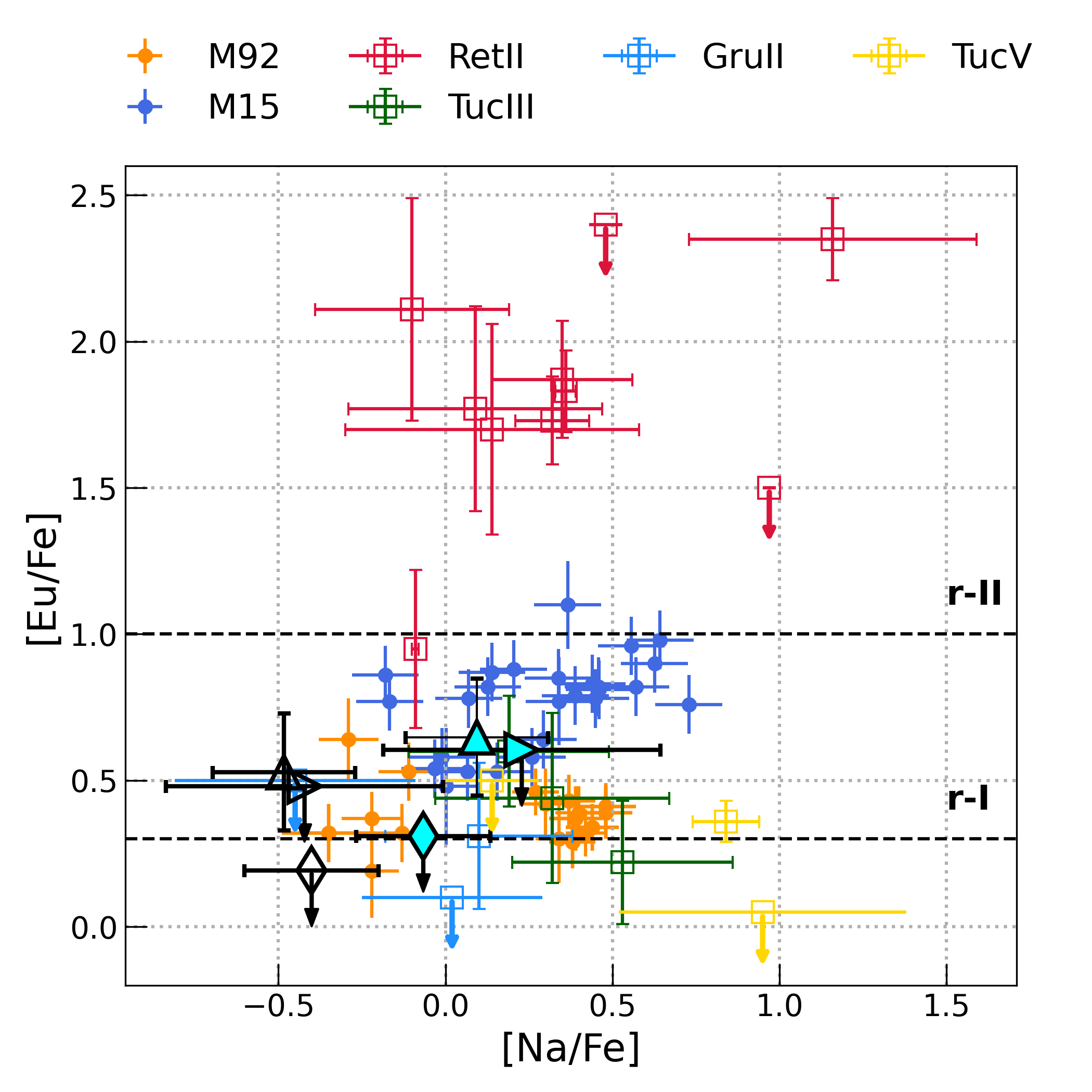}
    \caption{[Eu/Fe] as a function of [Na/Fe] for Sgr2 targets (cyan markers, downward arrows for upper limits), compared with two r-process-enriched globular clusters M15 (blue circles) and M92 (orange circles), both showing Eu abundance spreads; as well as four r-process-enriched UFDs Reticulum II (red open squares), Tucana III (green open squares), Gru~II (light-blue open squares) and Tuc~V (yellow open squares). NLTE abundances for our targets are shown as open markers. Na abundances for M15 stars are taken from \cite{sneden1997star}, \cite{sneden2000barium}, \cite{carretta2009anticorrelation}, and \cite{Sobeck11}; Eu abundances are from \cite{Garcia24}; data for M92 are sourced from \cite{Kirby_2023}; for Ret~II, from \cite{Ji16}, \cite{Hayes23}, for Tuc~III, from \cite{Hansen2017}, \cite{Marshall_2019} though only three of five stars with Na measurements are shown, for Gru~II, from \cite{Hansen2020}, and for Tuc~V, from \cite{Hansen2024}}. 
    \label{fig:EuvsNa}
\end{figure}

\begin{figure}
    \centering
    \includegraphics[width=1.0\linewidth]{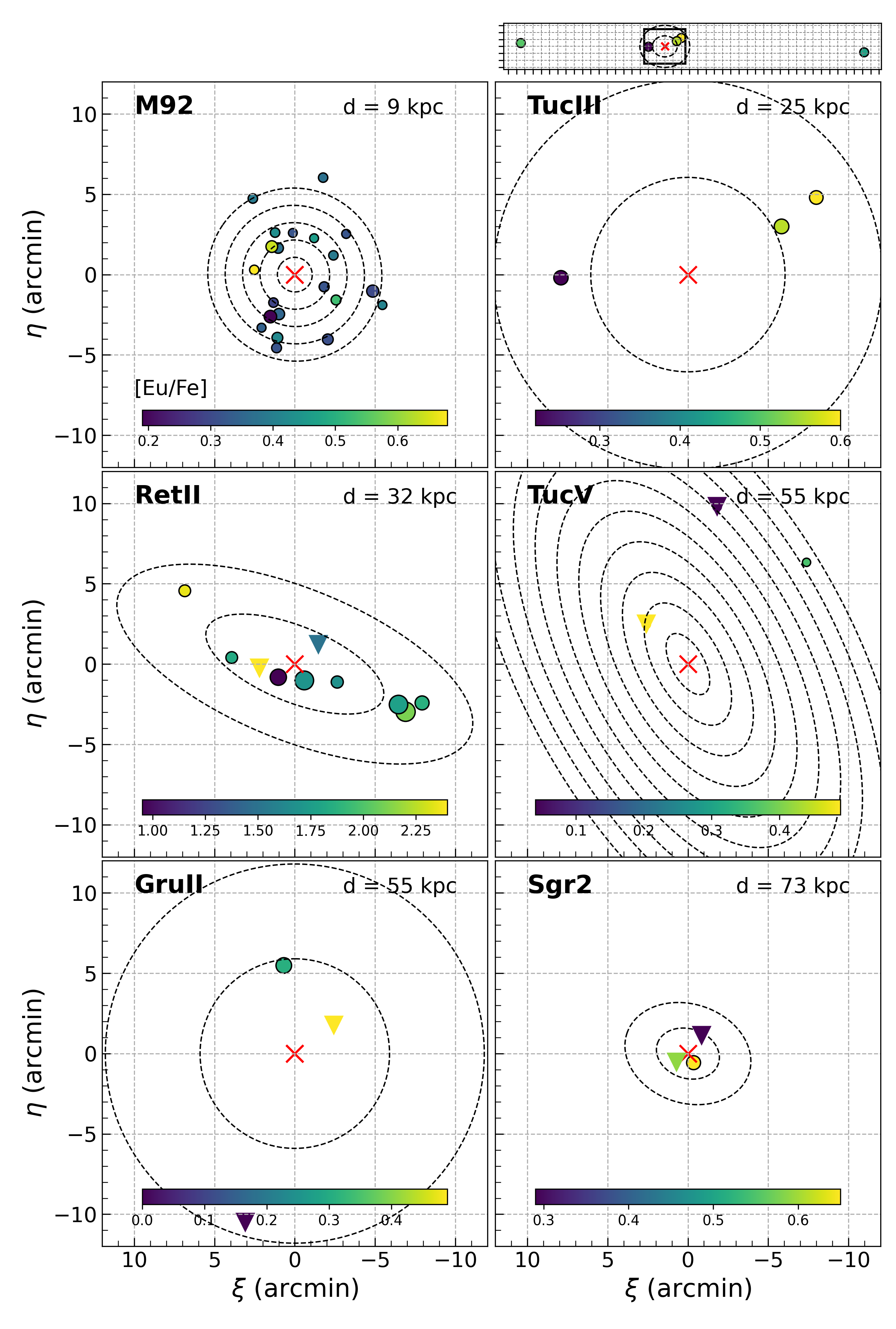}
    \caption{Positions of stars with  [Eu/Fe] abundance measurement in M92, Tucana III, Reticulum II, Tuc~V, Gru~II and Sgr2 (shown in order of increasing distance).  Marker sizes are proportional to the uncertainties, where smaller markers represent lower total errors and colours are correlated with Eu. Upper limits are indicated by downward triangles. Contours start at 1 R$_h$, with steps of 1 R$_h$. Due to the difficulty in constraining the system's ellipticities, for Tuc~III and Gru~II, circles of radii $n \times R_h$ are shown. Two very distant members of Tuc~III are noted, near $\sim$ 12 and 15 R$_{h}$. 
    Both axes are equally scaled to allow for size comparisons between systems. 
    }
    \label{fig:Eu_Ret2_Sgr2}
\end{figure}

\subsection{R-process Enrichments in Stellar Populations}

\textit{R-process enrichment in GCs: }
The globular clusters M15 and M92 show r-process enrichments. 
In M92, \citet{Kirby_2023} observed a spread in Eu between Na-enhanced and low-Na stars, with the significant scatter in [Eu/Fe] confined to the low-Na first generation (1G) stars. They proposed a scenario where a source of the main r-process polluted M92 during the early stages of star formation. The 1G stars formed quickly, in less time than it took for gas to mix fully (within a crossing time), resulting in inhomogeneous r-process enrichment. The 2G formed later ($\sim0.8$ Gyr) after the gas had homogenized, resulting in little to no variation in r-process elements among the second generation (2G) stars. 
In M15, \citet{Garcia24} report similar findings to M92.
Sgr2 exhibits abundance patterns that generally align with those of M92, as shown in Fig.~\ref{fig:EuvsNa}.

\textit{R-process enrichments in UFDs: }
R-process enrichments in UFDs are exceptionally rare.
Several UFD galaxies have been found to host r-process rich stars, including
Tuc~III, Gru~II, and Tuc~V which host r-I stars \citep{Hansen2017, Hansen2020, Hansen2024, Marshall_2019}, and Ret~II which shows that 70\% of its observed members are highly Eu-enhanced r-II stars \citep{Ji16,Hayes23}.
Fig~\ref{fig:EuvsNa} shows that the Eu and Na abundances of the UFDs with only r-I stars do not stand out compared to the r-process-enriched globular clusters. 

To test the r-process enrichment scenario further, 
we show the projected positions of stellar members in M92, Tuc~III, Gru~II, Tuc~V, Ret~II, and Sgr2, within their respective systems, in Fig.~\ref{fig:Eu_Ret2_Sgr2}, along with their corresponding [Eu/Fe] abundances.
\cite{Jeon21} has suggested that the most critical factor in the formation of Eu-enhanced stars is how quickly new stars form around the event/NSM site, predicting that the highest Eu enhancements are achieved within $\sim$300 pc of the event/NSM site.
Notably, the most Eu-rich star in Sgr2 \textit{is} located near the system's center.
However, this is not seen for any of the other systems.  In the case of Tuc~III, only core members are shown (excluding two stars from the tidal tails) as it is a disrupted galaxy. Its disruption may have altered the original positions of the three members shown, complicating efforts to trace the site of the initial Eu enrichment. 
In Gru~II, its Eu-enriched star lies within the system's half-light radius, whereas in Tuc~V, a similarly r-process rich member is located at the system's outermost edge ($\sim$ 9 R$_h$) and the closer members only have upper limits.
In Ret~II, the most Eu-rich member is found beyond one half-light radius from the system's center, while the star with the lowest [Eu/Fe] value is more central. Thus, overall, no clear trend with increasing distance is observed in these comparison systems. 
\cite{Jeon21} also suggested that if Ret~II were a satellite of the LMC, as indicated by its derived orbital history, 
this may have impacted the effects of reionization on its star formation history, 
and thereby the likelihood of forming metal-poor r-II stars throughout. 
Full cosmological simulations with detailed chemical evolution will be necessary to explore connections between r-process enhancements and locations in dwarf galaxies \citep[e.g.,][]{Manwadkar2022, Manwadkar2024}.

As a reference globular cluster for comparison, M92 does not exhibit a clear gradient in Eu abundance with respect to radial distance or along the east-west and north-south directions (see Fig.~\ref{fig:Eu_Ret2_Sgr2}).   No gradient was seen in M15 either by \cite[][see their Fig.~13]{Garcia24}.

In general, our analysis of these systems shows that r-process enrichments alone, at least via the presence of r-I stars, is insufficient to constrain a system's origin, i.e., as an UFD or globular cluster, 
whether examining the scatter in the [Eu/Fe] abundances within a system or the spatial distribution of r-process rich stars within the system.

In summary, we find that Sgr2 is a very intriguing case. It is comparable to Tuc~III, as an ambiguous system that remains difficult to classify, even with detailed chemical analyses of its brightest members. 

\subsection{Comparison of Sgr2 with Tuc~III}

Tuc~III is a particularly interesting system as it has been classified as a UFD based on its low average metallicity ([Fe/H]~$\sim -2.49$; \citealt{Li_2018a}), a velocity and metallicity dispersion\footnote{The observed metallicity gradient from CaT lines among core and tidal tail members has likewise been presented as evidence for the system's origin as a UFD \citep{Li_2018a}. We reanalyze the evidence for a metallicity dispersion and metallicity gradients in the Appendix and do not find clear evidence for Tuc~III to be classified as an UFD.}, and the absence of light-element anti-correlations amongst five of its brightest members \citep{Hansen2017, Marshall_2019}.
However, \citet{Marshall_2019} note that adopting the photometric temperature for the metallicity outlier star from \citet{Hansen2017} would eliminate the statistically significant metallicity spread among its members.
And, unfortunately, the velocity distribution for Tuc III is not representative of the system due to its disrupted state (only an upper limit of 1.5 \kms\ at the 95.5 \% confidence level for core members was derived by \citealt{Simon2017}).
Taking into account that low metallicity is also insufficient evidence for the presence of dark matter (i.e., an UFD classification), as globular clusters are typically found with [Fe/H] $= -2.5$, and even lower ([Fe/H] $= -2.9$ have been found for M31 clusters \citealt{Simpson2018, Larsen_2020}, and the stellar stream C-19 appears to be a disrupted star cluster with [Fe/H] $= -3.4$, see \citealt{MartinVenn2022, Yuan22, Venn2025}), then Tuc~III remains a challenging system to classify.

Both Sgr2 and Tuc~III have at least one star that exhibits a europium enrichment, but at levels comparable to stars in the GCs, M15 and M92, and well below the r-II stars found in the UFD, Ret~II. 
Other n-capture elements (Sr and Ba) are at levels consistent with the MW halo, similar to those in GCs, and therefore somewhat higher than typically found in stars in UFDs. 
This is noteworthy as \citet{Ji2019} has suggested that low neutron-capture element abundances are a distinguishing feature of the faintest dwarfs.
Earlier, in our Fig.~\ref{fig:n-capt_Aqu2}, we showed that the mean neutron-capture element abundances for Sgr2 and Tuc~III are higher than those of other UFDs. 
Taken together, we find that the evidence is not yet clear on the classification of either Sgr2 nor Tuc~III.

\section{Additional diagnostics for  ambiguous systems}

The present work demonstrates two possible scenarios that may arise during the review or reassessment of UFDs and UFD candidates with high-resolution spectroscopy. On the one hand, there is Aqu2, where a clear metallicity spread and distinct chemical signatures, even among just the two brightest member stars, strongly support its classification as a UFD. Combined with its extended size and the radial velocity dispersion estimated in previous studies, this provides robust evidence for the presence of a dark matter halo in this system.
On the other hand, high-resolution spectroscopic follow-up of Sgr2 has not resolved its classification, instead highlighting its "on the border" position in terms of  kinematics and metallicity, as well as its detailed chemical abundances. 

The most common approach to identify a low surface brightness system within a dark matter halo is to examine the velocity dispersion profile 
and compare it to estimates for a purely baryonic scenario. However, as noted in Fig.~\ref{fig:JPDFs_Sgr}, the method used to calculate the expected velocity dispersion for the self-gravitating case significantly affects the results. The expected velocity dispersion for Sgr2 using the \citet{Wolf2010} formalism is shown in
Fig.~\ref{fig:JPDFs_Sgr} as the grey-shaded band at $\sim1$ \kms, which is too large to clearly resolve the velocity dispersion. In contrast, adopting the star-cluster-specific approach of \citet{Baumgardt2017}, which accounts for energy equipartition and mass segregation effects commonly observed in globular clusters (e.g., \citealt{Baumgardt2003}), reduces the expected velocity dispersion to $\sim0.5$ \kms (gray dashed line in Fig.~\ref{fig:JPDFs_Sgr}).This lower value means that we do resolve our velocity dispersion, and hints at a classification for Sgr2 as an UFD.

In addition to theoretical uncertainties, the observed velocity values can also be influenced by data quality and sample selection (e.g., compare the pink profile from L20 and the green profile from L21 in Fig.~\ref{fig:JPDFs_Sgr}), as well as binary contamination. GC observations show that the binary fraction increases toward a cluster center \citep[e.g.,][]{Sollima07, Ji15}, and simulations reveal that as binaries segregate, "hard"\footnote{"hard" means that their binding energy is much higher than the average kinetic energy of stars in a cluster} binaries become even harder. Their semimajor axes shrink according to the  \citealt{Heggie75} - \citealt{Hills1975} law, making them more difficult to identify and potentially causing up to a 70\% overestimation of the velocity dispersion in the cluster's core \citep{Aros_2021}. 
As a quantitative example, \cite{Wang2024} 
estimated the observed binary fraction
for a Palomar 5-like globular cluster (which we note is close to Sgr2 on the $M_V$ vs.\ $r_h$ diagram; see Fig.~\ref{fig:Mv_rh}), with a heliocentric distance of $\sim$ 20 kpc.
They showed that most bright binaries with periods below $10^4$ days can be detected within $\sim$6 months by measuring line-of-sight velocities of $|\Delta v_r| > 0.3$ \kms.
The undetected binaries ($|\Delta v_r| < 0.3$ \kms) could still inflate the computed velocity dispersion by a factor of 1.5-2 compared to estimates based solely on single stars. 
For Sgr2, which is $>3$ times farther away, the stars have measured radial velocity uncertainties $\sim$1 \kms\ (L21), constrained by observations of only $\sim$1 month (38 days). This implies its $\sigma$($v_r$) is not resolved. 
This agrees with \citet{Baumgardt22}, who
estimated a lower velocity dispersion for Sgr2\footnote{The velocity dispersion profile for Sgr2 derived with N-body models of \citet{Baumgardt2017} can be found at \url{https://people.smp.uq.edu.au/HolgerBaumgardt/globular/fits/kin/sgrii_vel.pdf}} of $\sim0.6$ \kms, i.e.,
3x lower than L21 (see Fig.~\ref{fig:Sgr2_rv_feh}), and attributed the higher observed $\sigma$($v_r$) entirely to binary contamination.

\begin{figure}
    \includegraphics[width=0.4\textwidth]{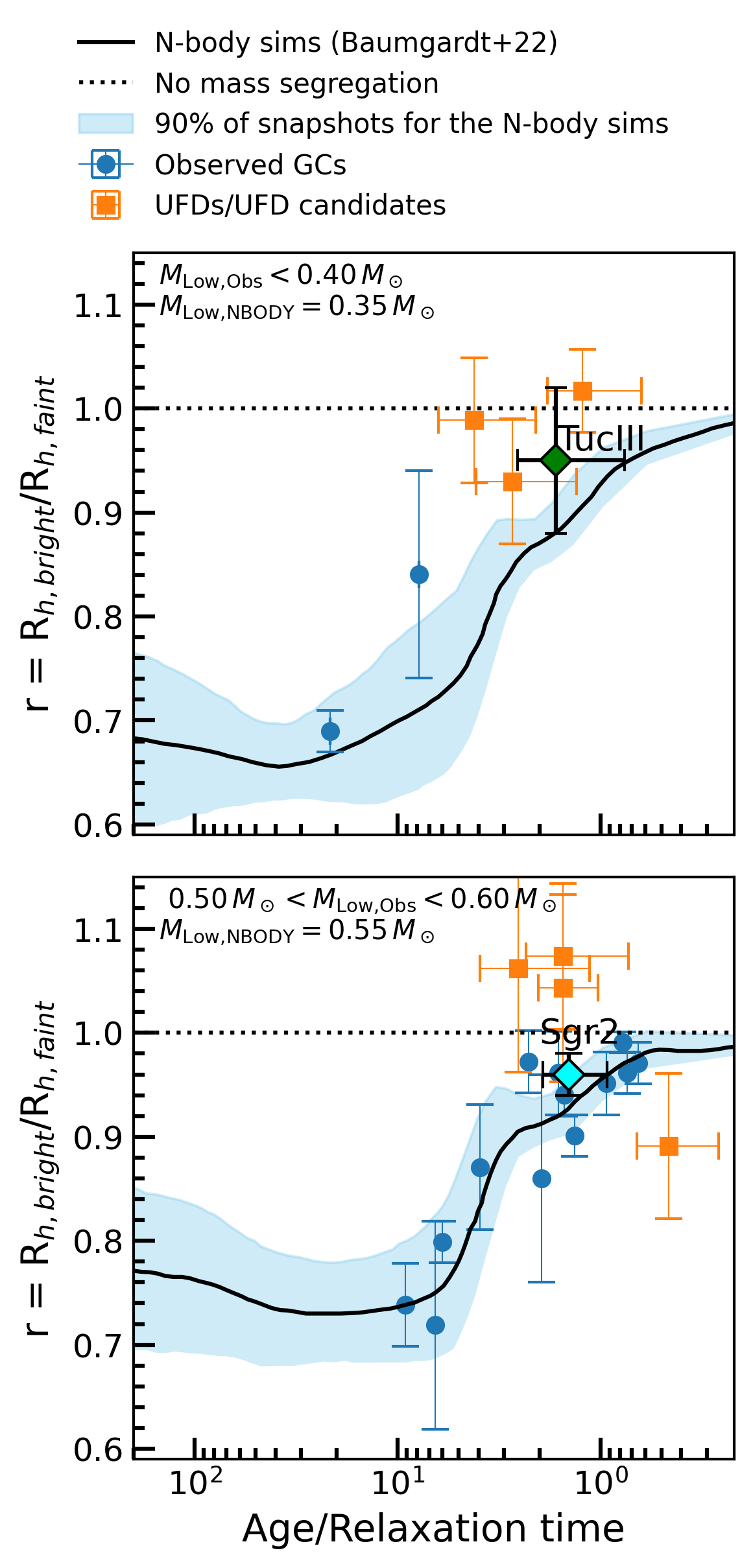}
\caption{Mass segregation ratios, defined as the ratio of the radius containing half the bright stars ($R_{h,bright}$) to the radius containing half the faint stars ($R_{h,faint}$), plotted as a function of the dynamical age ($T_{\text{Age}}/T_{\text{RH}}$) for MW globular clusters. Results are based on the analysis by \citet{Baumgardt22}. Samples are categorized by the lowest stellar mass analyzed in each cluster, with corresponding N-body model results adjusted for varying minimum masses, as indicated in the panels. The black solid line and pale blue shaded region represent results from N-body simulations. Observed GCs are shown in blue, while UFDs analyzed in the study, either confirming or establishing their classification, are shown in orange. Tuc~III and Sgr2 are highlighted as black-edged diamonds.
}

    \label{fig:Mass_seg}
\end{figure}

To take these studies further, 
\citet{Baumgardt22} have conducted a homogeneous and extensive study of mass segregation in GCs and UFD candidates 
using N-body simulations. As a measure of mass segregation, they used the ratio of the radii where 
the cumulative fraction of bright and faint stars reaches 0.5. A lower \(\frac{R_{h, \text{bright}}}{R_{h, \text{faint}}}\) ratio (deviating further from unity) indicates stronger segregation and is expected in clusters with relaxation times much shorter than their age. To get the two subsets, members below the main sequence turn-off and above a certain \textit{M$_{Low}$} mass threshold were divided into two equally sized groups based on brightness. For Sgr2, their mass segregation parameter \( r = \frac{R_{h, \text{bright}}}{R_{h, \text{faint}}} = 0.96 \pm 0.02 \), which is in  agreement with the expected value for a star cluster of Sgr2's relaxation time of 8.37 Gyr, calculated following \citet{Spitzer1987degc.book.....S}\footnote{We note that \citet{Wang2024} find that the presence of black holes can significantly shorten the two-component relaxation time.}.
 
We present the mass segregation parameters versus the ratio of age to relaxation time (dynamical age, $T_{\text{Age}}/T_{\text{RH}}$) for globular clusters in Fig.~\ref{fig:Mass_seg}, alongside the N-body predictions from \citet{Baumgardt22}. 
Two panels are shown to differentiate between the lowest mass analyzed in each cluster, and to include corresponding N-body model results for other systems adjusted for the minimum mass. 

The computed mass segregation parameter for Sgr2 from \citet{Baumgardt22} aligns well with the trend observed in GCs. While the value is close to 1 (indicating no mass segregation), the small bootstrapped uncertainties place it firmly within the region typically occupied by star clusters at this $T_{\text{Age}}/T_{\text{RH}}$. As an additional test for mass segregation, the authors performed a KS test on the cumulative distributions of bright and faint stars, finding a high significance for segregation in Sgr2 ($P_{\text{Mseg}} = 97.5\%$). This contrasts with the results for Tuc~III, where both the mass segregation ratio ($0.95 \pm 0.07$) and the KS test result ($42.8\%$) remain inconclusive. 

We support that the mass segregation parameter can provide an additional and valuable test for low surface brightness systems when trying to determine their origins (GC vs UFD), total mass, and thereby dark matter contents. The chemical patterns -- particularly in heavy elements -- and mass segregation of Sgr2 align with those of similar Milky Way GCs, supporting its classification as a star cluster.
Such diagnostics are especially important when alternative approaches, like studying system stability within the Milky Way's tidal field, are inapplicable or computationally expensive (e.g., Sgr2's distant pericenter, as noted in Section~\ref{sec:intro}, or the impact of the LMC on Tuc~III).

\section{Conclusions}

The analysis of five stars observed in two low surface brightness Milky Way satellites, Sagittarius II (Sgr2) and Aquarius II (Aqu2) is presented based on spectra taken during the commissioning of the Gemini/GHOST spectrograph. 
The spectra were taken in both the single and double IFU standard modes, binning 2x8 for the highest possible signal on these faint objects (G$<18.8$).  The spectra are exquisite in their resolution and high throughput over a wide wavelength range. 
 
From GHOST spectra of two stars in Aqu2 and data in the literature, we find: 
(i) radial velocity and metallicity dispersions consistent with membership in a dark matter-dominated UFD galaxy; 
(ii) chemical abundances that indicate inefficient star formation, i.e., low abundances of [Na/Fe], [Sr/Fe], and [Ba/Fe]; 
(iii) [K/Fe] enrichment, most likely due to the impact of super AGB stars in a low-mass and unevolved UFD galaxy.

From GHOST spectra of three stars in Sgr2 and data in the literature, we find: 
(i) radial velocity and metallicity dispersions that are just barely resolved, and inconclusive on the nature of the system, especially if there are binary stars;
(ii) chemical abundances that are exceptional in only one element (Eu), and in only one star, Sgr2584, where [Eu/Fe]$=+0.7\pm0.2$, typical of the r-I stars which are found in both globular clusters (e.g., M15, M92) and UFDs (e.g., Tuc~III). 
From these results, Sgr2 remains a challenge to classify, highlighting the difficulty in classifying some of the lowest mass and faintest MW satellites, even with detailed chemical abundances. We suggest this is also true for Tuc~III.
We discuss the value of additional diagnostics in classifying the most ambiguous systems, such as mass segregation
in exploring the origins and total mass (dark matter content) of star clusters and potentially UFDs \citep{Baumgardt22}. 
For Sgr2, we support their conclusion that Sgr2 is most likely a globular cluster with a radial velocity dispersion that is slightly inflated by binary stars. 

The spectra used in this analysis demonstrate the high quality available with the new \texttt{Gemini/GHOST} spectrograph.  It also served as a pilot program for the launch of the Gemini High-resolution Optical-UV Legacy Survey (GHOULS), where all bright (G$<18.5$) stars within R$<3$ R$_h$ without published high resolution spectra in all ultra faint MW satellites are currently queued for ongoing studies on the nature of these systems.

\section*{Acknowledgements}

DZ and KAV thank the National Sciences and Engineering Research Council of Canada and the Mitacs for funding through the Discovery Grants and Globalink programs.
We also thank the anonymous referee for their helpful comments that improved the manuscript.
We acknowledge and respect the l\textschwa\textvbaraccent {k}$^{\rm w}$\textschwa\ng{}\textschwa n peoples on whose traditional territory the University of Victoria stands and the Songhees, Esquimalt and $\ubar{\rm W}$S\'ANE\'C  peoples whose historical relationships with the land continue to this day.

This work is based on observations obtained with Gemini-South/GHOST, during the commissioning run in June 2022. The international Gemini Observatory is a program of NSF's NOIRLab, which is managed by the Association of Universities for Research in Astronomy (AURA) under a cooperative agreement with the National Science Foundation on behalf of the Gemini Observatory partnership: the National Science Foundation (United States), National Research Council (Canada), Agencia Nacional de Investigaci\'{o}n y Desarrollo (Chile), Ministerio de Ciencia, Tecnolog\'{i}a e Innovaci\'{o}n (Argentina), Minist\'{e}rio da Ci\^{e}ncia, Tecnologia, Inova\c{c}\~{o}es e Comunica\c{c}\~{o}es (Brazil), and Korea Astronomy and Space Science Institute (Republic of Korea).

This work has made use of data from the European Space Agency (ESA) mission {\it Gaia} (\url{https://www.cosmos.esa.int/gaia}), processed by the {\it Gaia} Data Processing and Analysis Consortium (DPAC, \url{https://www.cosmos.esa.int/web/gaia/dpac/consortium}). Funding for the DPAC has been provided by national institutions, in particular the institutions participating in the {\it Gaia} Multilateral Agreement.

This work made extensive use of \textsc{TOPCAT} \citep{Taylor05} and Python \citep{python}.

\section*{Data Availability}
GHOST commissioning spectra are not available in the Gemini Observatory Archive.  Please contact us for access to the original raw data.  All reduced data are incorporated into this paper.


\bibliography{ghost_planar,bibliography, refs}{}
\bibliographystyle{aasjournal}

\clearpage

\appendix \label{appendix}

\section{GHOST observations}

The GHOST exposures used from commissioning for the spectra analysed in this paper are provided in Table~\ref{tab:exp}.  This includes target information per IFU using both the single and dual observing modes, as well as exposure times, number of exposures coadded, and the SNR of the coadded exposures at various wavelengths.  We also include the calibration files used, which is important to note as the commissioning data includes many of these files as we began testing the spectrograph.  The slitview file was particularly important for proper use of the commissioning version of the GHOST data reduction pipeline.

 \begin{table*}
\caption{GHOST exposures for Sgr2 and Aqu2 targets, including the calibration files used for the data reduction pipeline. }
\label{tab:exp}
\centering
\hspace{-0.6cm}
\begin{tabular}{llllccc|llr}
\toprule
Target & Science  & IFU & Arm & t$_{\rm exp}$ & N$_{\rm exp}$ & SNR @$\lambda$ & 
Calibration Files & Type & t$_{\rm exp}$ \\
      & &     && (s) & & (\AA) & & & (s)\\
\hline
Aqu2776\ & aqu2\_sr\_2x4\_br & 1 & Blue &  1800 & x3 &  4 @4130 
& arcs\_sr\_1x1\_brs300\_20220629 & arc & 300 \\
   & 1800s300\_20220629 &&  Red  & 1800 & x3 &  26 @6050 
& flat\_sr\_1x1\_br6\_s02\_20220629 & flat & 6 \\
Aqu2472  &        & 2 &      &      &    & 4 @4130
& bias\_2x4\_20220629 & 2x4 bias \\
  &                 &   &      &      &    & 28 @6050
& bias\_1x1\_20220629  & 1x1 bias \\
  & &   &      &      &    & 
& HD122196\_HIP068460\_sr1x1\_br300s3 & slitview & 0.1 \\
\hline
Sgr2584 & Sag2\_sr\_2x4\_b3600\_ & 1 & Blue     &  3600 & x1 &   8 @4130 
& arc\_hr\_1x1\_thxe2\_20220628 & arc & 300 \\
   & r1200\_s300\_20220628 & & Red  & 1200 & x3 & 59 @6050 
& flats\_hr\_1x1\_20220628 & flat & 6 \\
   &   &        & &       &      &    & 
 bias\_2x4\_20220628 & 2x4 bias \\
   &   &        & &       &      &    & 
 bias\_1x1\_20220628  & 1x1 bias \\
   &   &        & &       &      &    & 
HD122196\_HIP068460\_sr1x1\_br300s3 & slitview & 1 \\

\hline
Sgr2656\ & Sag2\_set2\_sr\_br\ & 1 & Blue &  3600 & x3 &  5 @4130 
& arcs\_sr\_1x1\_brs300\_20220629 & arc & 300 \\
   & 3600s150\_20220630 &&  Red  & 3600 & x3 &  24 @6050 
& flat\_1x1\_sr\_br6s02\_set2\_20220630 & flat & 6 \\
 Sgr2936  &        & 2 &      &      &    & 7 @4130
& bias\_2x4\_20220630 & 2x4 bias \\
  &                 &   &      &      &    & 69 @6050
& bias\_1x1\_20220630  & 1x1 bias \\
  & &   &      &      &    & 
& HD122196\_HIP068460\_sr1x1\_br300s3 & slitview & 0.1 \\

\hline
\end{tabular}
\end{table*}

\section{\texttt{Py\_Looper} validation}

\begin{figure*}[!htbp]
    \includegraphics[width=1\textwidth]{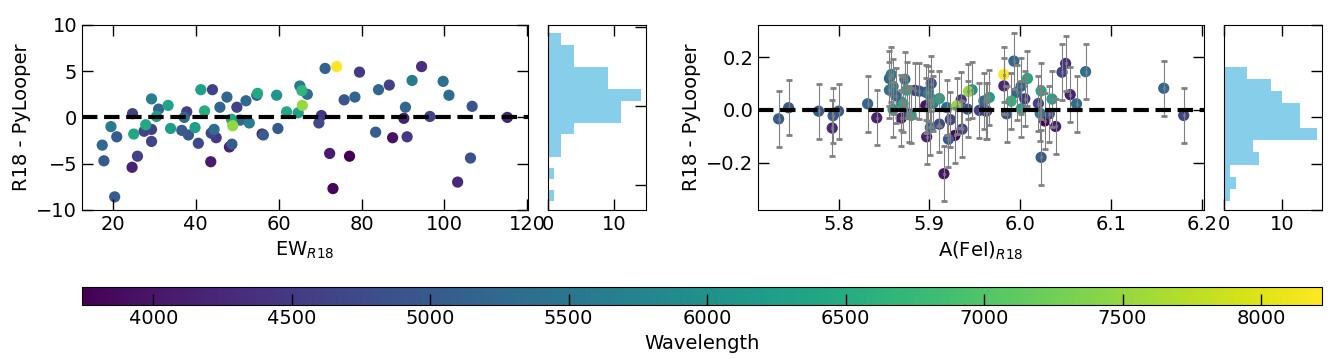}
    \caption{Line-by-line analysis of equivalent width (EW) measurements and derived abundances for FeI in HD222925, performed to validate the \texttt{Py\_Looper} method. \textit{Left}: Comparison of EWs  measured using the \texttt{Py\_Looper} notebooks with those reported by \citealt[][R18]{roederer2018}. \textit{Right}:  Absolute FeI abundances derived in this work vs abundances reported by R18. In this analysis, only lines with  $20 < $EW $< 140$ m\AA\ and EP $> 1.4$ eV were retained.}
    \label{fig:HD222_analysis}
\end{figure*}

To ensure that the new \texttt{Py\_Looper} Jupyter notebooks described in Section~\ref{sec:method} provide reliable abundance measurements, a comparative analysis was conducted using the standard star HD 222925, focusing on Fe lines. The analysis included: (i) comparing EWs measured manually with IRAF to those measured with \texttt{Py\_Looper}; (ii) comparing EWs reported by \cite[][R18]{roederer2018} to those measured with \texttt{Py\_Looper}; and (iii) comparing abundances derived with \texttt{Py\_Looper}, including error analysis, to those reported by R18. All three tests demonstrated good agreement within $\sim 2 \sigma$, with higher discrepancies observed at the bluest end due to the lower SNR of the GHOST spectrum in this region. \texttt{Py\_Looper} proved to be much more efficient and consistent compared to manual measurements. A comparison of the FeI lines EWs and A(FeI) values with data from R18 are shown in Fig.~\ref{fig:HD222_analysis}.

\section{Tuc~III metallicity dispersion}

\begin{figure*}[!htbp]
\centering
\includegraphics[width=0.5\textwidth]{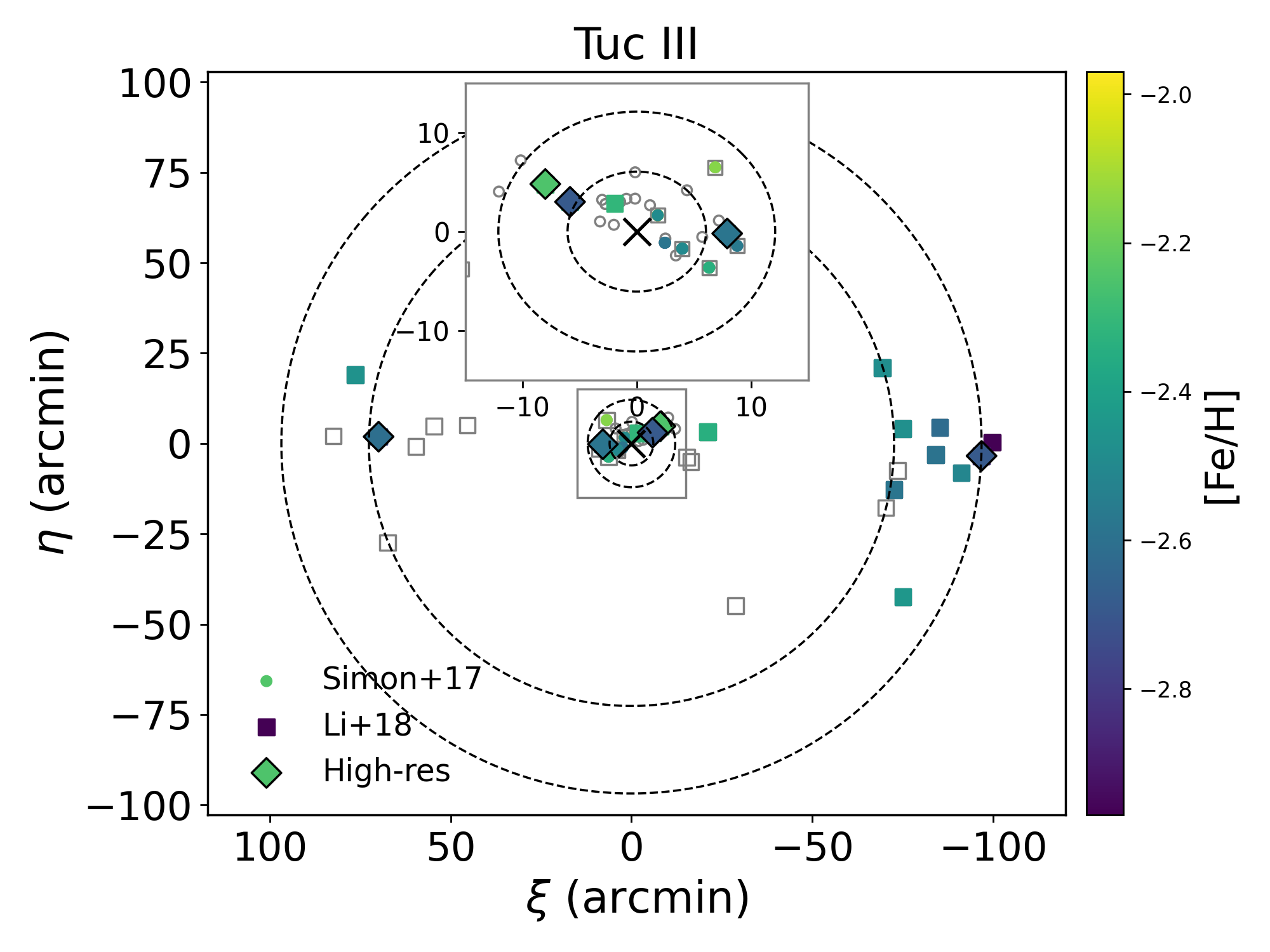}
\caption{Spatial distribution of Tuc~III members observed by \citet{Simon2017} (circles, concentrated in the core), \citet{Li2018} (squares), and high-resolution studies by \citet{Hansen2017, Marshall_2019} (diamonds) is shown. Members with available metallicities (determined from either CaT or iron lines) are color-coded by metallicity, while open gray markers represent observations without measured metallicities. Dashed circles indicate radii of 1, 2, 12, and 16 $R_h$. }
\label{fig:Tuc3_proj}

\end{figure*}

\begin{figure*}[!htbp]
\centering
\includegraphics[width=1.0\textwidth]{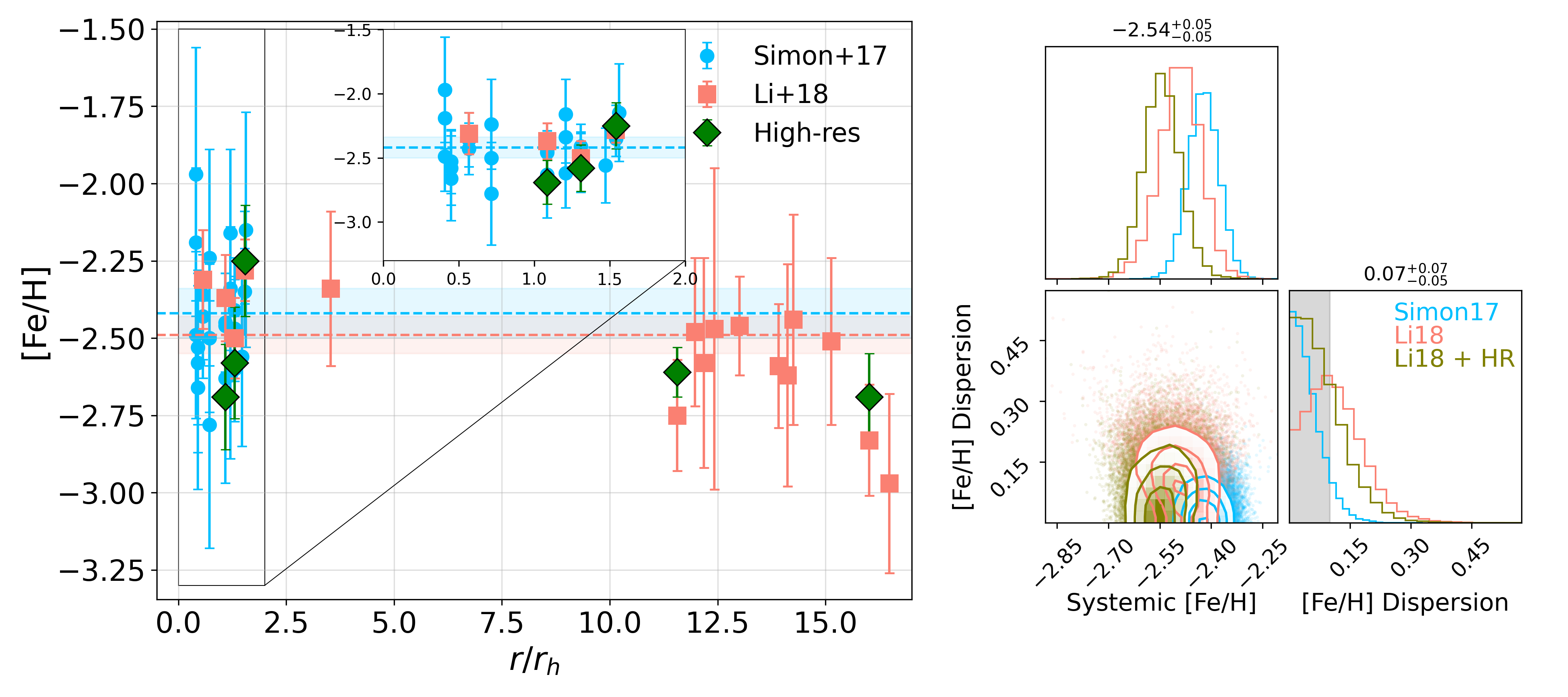}
\caption{\textit{Left:} Radial distribution of metallicities for TucIII members. Marker shapes are consistent with those in Fig.~\ref{fig:Tuc3_proj}. Notably, the measurements by \citet{Li2018} suggest a slight metallicity gradient between the core and tidal tail members. However, this trend is primarily driven by the three most metal-poor members, two of which were followed up with high-resolution spectroscopy by \citet{Marshall_2019}, weakening the strength of the reported gradient. \textit{Right:} Probability density function (PDF) of systemic metallicity and metallicity dispersion for Tuc~III, based on samples from \citet[][core members, light blue]{Simon2017}, \citet[][primarily tidal tail members, salmon]{Li2018}, and a combined sample from \citet{Li2018} and high-resolution studies by \citet{Hansen2017} and \citet{Marshall_2019}, with high-resolution metallicities used for stars present in both datasets (olive).
The analysis shows that when high-resolution metallicities are applied to the most metal-poor members in the tails, the dispersion decreases to a level comparable to that of star clusters (gray band), aligning more closely with the dispersion observed in the core. Note also he most metal-rich member from high-resolution studies shows a decrease in metallicity by $\sim$ 0.2 dex when T$_{phot}$ is applied (see discussion in \citealt{Marshall_2019})}
\label{fig:TucIIImetdisp}
\end{figure*}

In this study, we report that one of the analyzed systems, Sgr2, closely resembles the ultra-faint Milky Way satellite Tuc~III, which has been classified as a tidally disrupted UFD.
To further investigate the uncertain origin of this ambiguous system, we reanalyzed the metallicity dispersion among both its core and tidal tail members (shown in Fig.~\ref{fig:Tuc3_proj}).  The metallicities vs radial distances are shown in Fig.~\ref{fig:TucIIImetdisp}, as well as the probability distribution function for the metallicities, showing the unresolved metallicity dispersion for Tuc~III of $\sigma$([Fe/H])$=0.07^{+0.07}_{-0.05}$. 

\section{Line-by-line abundances}

The spectral lines used in this analysis are listed in Tables~\ref{tab:lbl_iron_abund} and \ref{tab:lbl_ab}.  This includes wavelength, excitation potential, and oscillator strengths as summarized in \texttt{linemake} \citep{Placco21}.  Both the EW and A(X)$_{1DLTE}$ values are provided, and the NLTE corrections calculated from online tables and the literature (see text).  These Tables will be provided as machine readable tables only.

\newpage

\begin{longrotatetable}
\small
\movetabledown=15mm
\begin{deluxetable}{lccccccccccccccccccccccc}
\tablecaption{Line-by-line equivalent widths (EW), abundances (A(X)), and NLTE corrections ($\Delta_{NLTE}$) for iron in five targets and two standard stars.}
\tablehead{
$\lambda$ & $\chi$ &  \multicolumn{3}{c}{HD222925} & \multicolumn{3}{c}{HD122563} & \multicolumn{3}{c}{Aqu2472} & \multicolumn{3}{c}{Aqu2776} &  \multicolumn{3}{c}{Sgr2936} &  \multicolumn{3}{c}{Sgr2584} &  \multicolumn{3}{c}{Sgr2656} \\
\small & \small & \small EW & \small A(X) & \small $\Delta_{nlte}$ & \small EW & \small A(X) & \small $\Delta_{nlte}$ & \small EW & \small A(X) & \small $\Delta_{nlte}$ & \small EW & \small A(X) & \small $\Delta_{nlte}$ & \small EW & \small A(X) & \small $\Delta_{nlte}$ & \small EW & \small A(X) & \small $\Delta_{nlte}$ & \small EW & \small A(X) & \small $\Delta_{nlte}$  \\
}
\startdata
\multicolumn{23}{c}{\textbf{FeI}} \\
4213.647 & 2.8 & 48 & 5.8 & 0.3 & 22 & 4.5 & 0.1 & ... & ... & ... & ... & ... & ... & ... & ... & ... & ... & ... & ... & ... & ... & ... \\
4216.184 & 0.0 & 106 & 6.1 & 0.3 & 111 & 4.9 & 0.3 & ... & ... & ... & ... & ... & ... & ... & ... & ... & ... & ... & ... & ... & ... & ... \\
4217.545 & 3.4 & 71 & 6.0 & 0.3 & 29 & 4.6 & 0.2 & ... & ... & ... & ... & ... & ... & ... & ... & ... & ... & ... & ... & ... & ... & ... \\
4222.213 & 2.5 & 93 & 5.9 & 0.2 & 69 & 4.6 & 0.1 & ... & ... & ... & ... & ... & ... & 109 & 5.3 & 0.1 & ... & ... & ... & ... & ... & ... \\
4227.426 & 3.3 & ... & ... & ... & 93 & 5.1 & 0.2 & ... & ... & ... & ... & ... & ... & ... & ... & ... & ... & ... & ... & ... & ... & ... \\
4233.603 & 2.5 & 110 & 5.9 & 0.2 & 83 & 4.6 & 0.1 & ... & ... & ... & ... & ... & ... & 110 & 5.0 & 0.2 & ... & ... & ... & ... & ... & ... \\
4238.810 & 3.4 & 83 & 5.9 & 0.2 & 43 & 4.6 & 0.1 & ... & ... & ... & ... & ... & ... & ... & ... & ... & ... & ... & ... & ... & ... & ... \\
4250.119 & 2.5 & 124 & 6.0 & 0.1 & 94 & 4.6 & 0.1 & ... & ... & ... & ... & ... & ... & ... & ... & ... & ... & ... & ... & ... & ... & ... \\
4250.787 & 1.6 & ... & ... & ... & 127 & 4.6 & 0.1 & ... & ... & ... & ... & ... & ... & ... & ... & ... & ... & ... & ... & ... & ... & ... \\
4260.474 & 2.4 & ... & ... & ... & 117 & 4.5 & 0.1 & ... & ... & ... & ... & ... & ... & 126 & 4.4 & 0.1 & ... & ... & ... & 122 & 4.8 & 0.1 \\
4271.153 & 2.5 & 127 & 6.0 & 0.1 & 103 & 4.7 & 0.1 & ... & ... & ... & ... & ... & ... & ... & ... & ... & ... & ... & ... & ... & ... & ... \\
4282.403 & 2.2 & 126 & 6.2 & 0.2 & 90 & 4.6 & 0.2 & ... & ... & ... & ... & ... & ... & ... & ... & ... & ... & ... & ... & ... & ... & ... \\
4337.045 & 1.6 & 100 & 5.9 & 0.3 & 95 & 4.8 & 0.2 & ... & ... & ... & ... & ... & ... & 135 & 5.3 & 0.2 & ... & ... & ... & ... & ... & ... \\
4352.735 & 2.2 & 89 & 5.9 & 0.3 & 73 & 4.7 & 0.1 & ... & ... & ... & ... & ... & ... & ... & ... & ... & ... & ... & ... & ... & ... & ... \\
4375.930 & 0.0 & 119 & 6.0 & 0.3 & 126 & 4.8 & 0.3 & ... & ... & ... & ... & ... & ... & ... & ... & ... & ... & ... & ... & ... & ... & ... \\
4388.407 & 3.6 & 45 & 5.9 & 0.3 & ... & ... & ... & ... & ... & ... & ... & ... & ... & ... & ... & ... & ... & ... & ... & ... & ... & ... \\
4415.122 & 1.6 & ... & ... & ... & 133 & 4.5 & 0.2 & ... & ... & ... & ... & ... & ... & ... & ... & ... & ... & ... & ... & ... & ... & ... \\
4427.310 & 0.1 & 122 & 6.0 & 0.4 & 126 & 4.7 & 0.3 & ... & ... & ... & ... & ... & ... & ... & ... & ... & ... & ... & ... & ... & ... & ... \\
4430.614 & 2.2 & 70 & 5.9 & 0.3 & 52 & 4.7 & 0.1 & ... & ... & ... & ... & ... & ... & 79 & 5.0 & 0.1 & ... & ... & ... & ... & ... & ... \\
4442.339 & 2.2 & 99 & 6.0 & 0.2 & 77 & 4.7 & 0.1 & ... & ... & ... & ... & ... & ... & 126 & 5.4 & 0.1 & ... & ... & ... & ... & ... & ... \\
4443.194 & 2.9 & 97 & 6.5 & 0.3 & 36 & 4.5 & 0.1 & ... & ... & ... & ... & ... & ... & ... & ... & ... & ... & ... & ... & ... & ... & ... \\
4447.717 & 2.2 & 93 & 6.0 & 0.3 & 70 & 4.7 & 0.1 & ... & ... & ... & ... & ... & ... & 83 & 4.8 & 0.1 & ... & ... & ... & ... & ... & ... \\
4459.117 & 2.2 & 122 & 6.5 & 0.2 & 90 & 5.0 & 0.2 & ... & ... & ... & ... & ... & ... & ... & ... & ... & ... & ... & ... & ... & ... & ... \\
4461.653 & 0.1 & 108 & 6.0 & 0.3 & 117 & 4.8 & 0.2 & ... & ... & ... & ... & ... & ... & ... & ... & ... & ... & ... & ... & ... & ... & ... \\
4466.551 & 2.8 & 96 & 6.0 & 0.3 & 83 & 5.0 & 0.2 & ... & ... & ... & ... & ... & ... & 99 & 5.1 & 0.2 & ... & ... & ... & ... & ... & ... \\
4484.220 & 3.6 & 47 & 5.9 & 0.3 & ... & ... & ... & ... & ... & ... & ... & ... & ... & ... & ... & ... & ... & ... & ... & ... & ... & ... \\
4489.739 & 0.1 & 63 & 6.0 & 0.3 & 82 & 4.9 & 0.2 & ... & ... & ... & ... & ... & ... & ... & ... & ... & ... & ... & ... & ... & ... & ... \\
4494.563 & 2.2 & 104 & 6.0 & 0.3 & 81 & 4.6 & 0.2 & ... & ... & ... & ... & ... & ... & 82 & 4.5 & 0.2 & ... & ... & ... & ... & ... & ... \\
4528.614 & 2.2 & ... & ... & ... & 99 & 4.7 & 0.2 & ... & ... & ... & ... & ... & ... & ... & ... & ... & ... & ... & ... & ... & ... & ... \\
4531.148 & 1.5 & 92 & 6.0 & 0.3 & 81 & 4.8 & 0.2 & ... & ... & ... & ... & ... & ... & ... & ... & ... & ... & ... & ... & ... & ... & ... \\
4547.847 & 3.5 & 37 & 5.8 & 0.3 & ... & ... & ... & ... & ... & ... & ... & ... & ... & ... & ... & ... & ... & ... & ... & ... & ... & ... \\
4592.651 & 1.6 & 81 & 6.2 & 0.3 & 66 & 4.8 & 0.2 & ... & ... & ... & ... & ... & ... & ... & ... & ... & ... & ... & ... & ... & ... & ... \\
4595.358 & 3.3 & 35 & 6.5 & 0.3 & ... & ... & ... & ... & ... & ... & ... & ... & ... & ... & ... & ... & ... & ... & ... & ... & ... & ... \\
4602.000 & 1.6 & 24 & 6.0 & 0.1 & 23 & 4.8 & 0.1 & ... & ... & ... & ... & ... & ... & 46 & 5.2 & 0.1 & ... & ... & ... & ... & ... & ... \\
4602.941 & 1.5 & 83 & 5.9 & 0.1 & 81 & 4.8 & 0.1 & ... & ... & ... & ... & ... & ... & 120 & 5.3 & 0.1 & ... & ... & ... & ... & ... & ... \\
4607.647 & 3.3 & 31 & 6.0 & 0.3 & ... & ... & ... & ... & ... & ... & ... & ... & ... & ... & ... & ... & ... & ... & ... & ... & ... & ... \\
4619.288 & 3.6 & 30 & 6.0 & 0.3 & ... & ... & ... & ... & ... & ... & ... & ... & ... & ... & ... & ... & ... & ... & ... & ... & ... & ... \\
4630.120 & 2.3 & 29 & 6.2 & 0.0 & ... & ... & ... & ... & ... & ... & ... & ... & ... & ... & ... & ... & ... & ... & ... & ... & ... & ... \\
4637.503 & 3.3 & 32 & 6.0 & 0.3 & ... & ... & ... & ... & ... & ... & ... & ... & ... & ... & ... & ... & ... & ... & ... & ... & ... & ... \\
4643.463 & 3.7 & 20 & 5.9 & 0.3 & ... & ... & ... & ... & ... & ... & ... & ... & ... & ... & ... & ... & ... & ... & ... & ... & ... & ... \\
4647.434 & 2.9 & 50 & 6.0 & 0.3 & 26 & 4.7 & 0.1 & ... & ... & ... & ... & ... & ... & 56 & 5.2 & 0.1 & ... & ... & ... & ... & ... & ... \\
4668.134 & 3.3 & 45 & 6.0 & 0.3 & 22 & 4.7 & 0.1 & ... & ... & ... & ... & ... & ... & ... & ... & ... & ... & ... & ... & ... & ... & ... \\
4733.591 & 1.5 & 43 & 6.1 & 0.3 & 39 & 4.8 & 0.1 & ... & ... & ... & 114 & 6.1 & 0.2 & 57 & 5.0 & 0.1 & ... & ... & ... & ... & ... & ... \\
4736.773 & 3.2 & 70 & 6.0 & 0.3 & 41 & 4.7 & 0.1 & ... & ... & ... & 97 & 5.7 & 0.1 & 74 & 5.2 & 0.1 & ... & ... & ... & ... & ... & ... \\
4786.807 & 3.0 & 38 & 6.1 & 0.2 & ... & ... & ... & ... & ... & ... & ... & ... & ... & ... & ... & ... & ... & ... & ... & ... & ... & ... \\
4789.651 & 3.5 & 36 & 6.0 & 0.1 & ... & ... & ... & ... & ... & ... & ... & ... & ... & ... & ... & ... & ... & ... & ... & 38 & 5.5 & 0.1 \\
4871.318 & 2.9 & 106 & 5.8 & 0.2 & 78 & 4.5 & 0.1 & ... & ... & ... & ... & ... & ... & 91 & 4.6 & 0.1 & 117 & 5.0 & 0.2 & 90 & 5.0 & 0.1 \\
4872.138 & 2.9 & ... & ... & ... & 66 & 4.6 & 0.1 & ... & ... & ... & ... & ... & ... & 105 & 5.1 & 0.1 & 102 & 5.0 & 0.1 & 77 & 5.0 & 0.1 \\
4890.755 & 2.9 & 111 & 6.0 & 0.1 & 75 & 4.5 & 0.1 & ... & ... & ... & ... & ... & ... & ... & ... & ... & 137 & 5.4 & 0.2 & ... & ... & ... \\
4891.492 & 2.8 & 119 & 5.9 & 0.1 & 87 & 4.5 & 0.1 & 126 & 5.2 & 0.1 & ... & ... & ... & ... & ... & ... & ... & ... & ... & ... & ... & ... \\
4903.310 & 2.9 & 74 & 5.8 & 0.2 & 48 & 4.5 & 0.1 & ... & ... & ... & ... & ... & ... & ... & ... & ... & 103 & 5.3 & 0.1 & ... & ... & ... \\
4918.994 & 2.9 & ... & ... & ... & 77 & 4.5 & 0.1 & 70 & 4.5 & 0.1 & ... & ... & ... & 106 & 4.9 & 0.1 & 126 & 5.1 & 0.2 & ... & ... & ... \\
4920.502 & 2.8 & ... & ... & ... & 98 & 4.4 & 0.1 & ... & ... & ... & ... & ... & ... & ... & ... & ... & ... & ... & ... & ... & ... & ... \\
4939.687 & 0.9 & ... & ... & ... & 68 & 4.7 & 0.1 & ... & ... & ... & ... & ... & ... & 97 & 5.0 & 0.1 & 86 & 4.6 & 0.1 & ... & ... & ... \\
4946.390 & 3.4 & 40 & 6.0 & 0.2 & ... & ... & ... & ... & ... & ... & ... & ... & ... & 66 & 5.6 & 0.1 & 70 & 5.6 & 0.1 & ... & ... & ... \\
4950.110 & 3.4 & ... & ... & ... & ... & ... & ... & ... & ... & ... & ... & ... & ... & ... & ... & ... & 26 & 5.3 & 0.1 & ... & ... & ... \\
4966.090 & 3.3 & 52 & 5.8 & 0.3 & 30 & 4.7 & 0.1 & ... & ... & ... & ... & ... & ... & 70 & 5.3 & 0.1 & 91 & 5.6 & 0.1 & ... & ... & ... \\
4994.130 & 0.9 & ... & ... & ... & 79 & 4.7 & 0.2 & ... & ... & ... & 126 & 5.4 & 0.2 & ... & ... & ... & 118 & 4.9 & 0.3 & ... & ... & ... \\
5001.860 & 3.9 & 61 & 5.8 & 0.3 & 30 & 4.5 & 0.2 & ... & ... & ... & 106 & 6.0 & 0.1 & 90 & 5.5 & 0.1 & 56 & 4.9 & 0.1 & ... & ... & ... \\
5002.790 & 3.4 & 22 & 6.0 & 0.3 & ... & ... & ... & ... & ... & ... & ... & ... & ... & ... & ... & ... & ... & ... & ... & ... & ... & ... \\
5005.710 & 3.9 & 58 & 5.8 & 0.2 & 24 & 4.5 & 0.1 & 40 & 5.0 & 0.2 & ... & ... & ... & 44 & 4.9 & 0.1 & 76 & 5.3 & 0.1 & ... & ... & ... \\
5006.119 & 2.8 & ... & ... & ... & 68 & 4.5 & 0.1 & ... & ... & ... & 121 & 5.4 & 0.1 & 94 & 4.8 & 0.1 & 124 & 5.3 & 0.1 & 70 & 4.8 & 0.2 \\
5012.068 & 0.9 & ... & ... & ... & 106 & 4.8 & 0.2 & 80 & 4.5 & 0.2 & ... & ... & ... & ... & ... & ... & ... & ... & ... & 116 & 5.4 & 0.2 \\
5014.940 & 3.9 & 52 & 5.8 & 0.3 & 20 & 4.6 & 0.2 & ... & ... & ... & ... & ... & ... & ... & ... & ... & ... & ... & ... & ... & ... & ... \\
5022.240 & 4.0 & 38 & 5.8 & 0.3 & ... & ... & ... & ... & ... & ... & ... & ... & ... & ... & ... & ... & 40 & 5.1 & 0.1 & ... & ... & ... \\
5028.130 & 3.6 & 23 & 5.8 & 0.2 & ... & ... & ... & ... & ... & ... & ... & ... & ... & ... & ... & ... & ... & ... & ... & ... & ... & ... \\
5039.250 & 3.4 & 29 & 6.2 & 0.0 & ... & ... & ... & ... & ... & ... & ... & ... & ... & ... & ... & ... & ... & ... & ... & ... & ... & ... \\
5041.756 & 1.5 & ... & ... & ... & 87 & 4.8 & 0.1 & ... & ... & ... & ... & ... & ... & ... & ... & ... & 127 & 5.1 & 0.1 & ... & ... & ... \\
5049.820 & 2.3 & 85 & 5.9 & 0.2 & 69 & 4.6 & 0.2 & ... & ... & ... & 131 & 5.7 & 0.2 & 95 & 4.9 & 0.2 & 117 & 5.1 & 0.2 & 98 & 5.4 & 0.2 \\
5051.635 & 0.9 & ... & ... & ... & 94 & 4.8 & 0.2 & 82 & 4.8 & 0.2 & 128 & 5.2 & 0.3 & ... & ... & ... & ... & ... & ... & ... & ... & ... \\
5079.224 & 2.2 & ... & ... & ... & 41 & 4.8 & 0.2 & ... & ... & ... & ... & ... & ... & ... & ... & ... & 100 & 5.5 & 0.2 & ... & ... & ... \\
5083.339 & 1.0 & ... & ... & ... & 83 & 4.7 & 0.2 & ... & ... & ... & ... & ... & ... & 105 & 4.8 & 0.2 & 130 & 5.0 & 0.3 & ... & ... & ... \\
5110.413 & 0.0 & ... & ... & ... & 111 & 4.9 & 0.2 & ... & ... & ... & ... & ... & ... & ... & ... & ... & ... & ... & ... & 125 & 5.7 & 0.3 \\
5123.720 & 1.0 & ... & ... & ... & 72 & 4.7 & 0.2 & ... & ... & ... & ... & ... & ... & 126 & 5.5 & 0.2 & 137 & 5.4 & 0.3 & ... & ... & ... \\
5127.360 & 0.9 & ... & ... & ... & 66 & 4.7 & 0.2 & 86 & 5.3 & 0.2 & ... & ... & ... & 81 & 4.8 & 0.2 & 119 & 5.1 & 0.3 & 65 & 5.2 & 0.2 \\
5133.690 & 4.2 & 66 & 5.8 & 0.3 & 30 & 4.5 & 0.2 & ... & ... & ... & 80 & 5.4 & 0.1 & ... & ... & ... & 75 & 5.2 & 0.1 & ... & ... & ... \\
5150.840 & 1.0 & ... & ... & ... & 70 & 4.7 & ... & ... & ... & ... & ... & ... & ... & 104 & 5.0 & ... & 117 & 5.0 & ... & ... & ... & ... \\
5166.282 & 0.0 & ... & ... & ... & 90 & 4.8 & 0.2 & ... & ... & ... & ... & ... & ... & 121 & 5.1 & 0.3 & ... & ... & ... & 90 & 5.4 & 0.2 \\
5171.596 & 1.5 & 111 & 5.9 & 0.3 & 105 & 4.6 & 0.2 & ... & ... & ... & ... & ... & ... & ... & ... & ... & ... & ... & ... & ... & ... & ... \\
5191.454 & 3.0 & 108 & 6.2 & 0.1 & 56 & 4.5 & 0.1 & ... & ... & ... & 106 & 5.3 & 0.1 & 103 & 5.1 & 0.1 & 92 & 4.9 & 0.1 & ... & ... & ... \\
5192.344 & 3.0 & 96 & 5.8 & 0.1 & 65 & 4.5 & 0.1 & ... & ... & ... & ... & ... & ... & 80 & 4.6 & 0.1 & 110 & 5.0 & 0.1 & ... & ... & ... \\
5194.940 & 1.6 & 90 & 5.9 & 0.3 & 85 & 4.7 & 0.2 & 92 & 5.0 & 0.2 & ... & ... & ... & 126 & 5.2 & 0.2 & 139 & 5.2 & 0.2 & ... & ... & ... \\
5198.710 & 2.2 & 45 & 5.9 & 0.3 & 34 & 4.7 & 0.1 & ... & ... & ... & ... & ... & ... & 75 & 5.3 & 0.1 & ... & ... & ... & ... & ... & ... \\
5202.340 & 2.2 & 76 & 6.1 & 0.3 & 56 & 4.7 & 0.1 & ... & ... & ... & ... & ... & ... & 81 & 5.0 & 0.1 & 114 & 5.4 & 0.1 & ... & ... & ... \\
5215.180 & 3.3 & 51 & 5.8 & 0.3 & 25 & 4.5 & 0.1 & ... & ... & ... & ... & ... & ... & 50 & 4.9 & 0.1 & 76 & 5.2 & 0.1 & ... & ... & ... \\
5216.274 & 1.6 & 81 & 5.8 & 0.3 & 79 & 4.6 & 0.2 & ... & ... & ... & ... & ... & ... & 96 & 4.7 & 0.2 & 128 & 5.1 & 0.2 & 76 & 5.0 & 0.2 \\
5217.390 & 3.2 & 45 & 5.9 & 0.3 & 22 & 4.6 & 0.1 & ... & ... & ... & ... & ... & ... & ... & ... & ... & 60 & 5.1 & 0.1 & ... & ... & ... \\
5232.940 & 2.9 & 114 & 5.9 & 0.0 & 88 & 4.6 & 0.1 & 104 & 4.9 & 0.1 & 129 & 5.2 & 0.1 & 116 & 4.9 & 0.1 & 123 & 4.9 & 0.2 & 102 & 5.1 & 0.1 \\
5242.490 & 3.6 & 30 & 5.9 & 0.3 & ... & ... & ... & ... & ... & ... & ... & ... & ... & ... & ... & ... & ... & ... & ... & ... & ... & ... \\
5253.460 & 3.3 & 20 & 6.0 & 0.2 & ... & ... & ... & ... & ... & ... & ... & ... & ... & ... & ... & ... & ... & ... & ... & ... & ... & ... \\
5263.310 & 3.3 & 50 & 5.8 & 0.3 & 26 & 4.5 & 0.1 & ... & ... & ... & ... & ... & ... & ... & ... & ... & 40 & 4.7 & 0.1 & ... & ... & ... \\
5266.555 & 3.0 & 95 & 5.9 & 0.1 & 68 & 4.6 & 0.1 & ... & ... & ... & ... & ... & ... & ... & ... & ... & 126 & 5.3 & 0.1 & ... & ... & ... \\
5283.621 & 3.2 & ... & ... & ... & 48 & 4.5 & 0.1 & ... & ... & ... & 105 & 5.4 & 0.1 & ... & ... & ... & 121 & 5.5 & 0.1 & ... & ... & ... \\
5307.360 & 1.6 & 37 & 6.0 & ... & 32 & 4.7 & ... & ... & ... & ... & 110 & 5.9 & ... & ... & ... & ... & 74 & 5.1 & ... & ... & ... & ... \\
5324.180 & 3.2 & 99 & 5.8 & 0.2 & 68 & 4.5 & 0.1 & 65 & 4.5 & 0.2 & ... & ... & ... & 100 & 4.9 & 0.1 & 129 & 5.2 & 0.1 & ... & ... & ... \\
5332.900 & 1.6 & 46 & 5.9 & ... & 43 & 4.7 & ... & ... & ... & ... & 123 & 5.9 & ... & 81 & 5.1 & ... & 98 & 5.2 & ... & ... & ... & ... \\
5339.930 & 3.3 & 64 & 5.8 & 0.3 & 38 & 4.5 & 0.1 & ... & ... & ... & ... & ... & ... & ... & ... & ... & 84 & 5.1 & 0.1 & 66 & 5.3 & 0.2 \\
5341.020 & 1.6 & 96 & 6.0 & 0.0 & 92 & 4.7 & ... & ... & ... & ... & ... & ... & ... & ... & ... & ... & ... & ... & ... & ... & ... & ... \\
5371.489 & 1.0 & ... & ... & ... & 147 & 4.6 & 0.2 & ... & ... & ... & ... & ... & ... & ... & ... & ... & ... & ... & ... & ... & ... & ... \\
5383.369 & 4.3 & 73 & 5.9 & 0.4 & 35 & 4.6 & 0.2 & 50 & 5.0 & 0.3 & ... & ... & ... & 58 & 5.0 & 0.2 & ... & ... & ... & ... & ... & ... \\
5397.128 & 0.9 & 130 & 5.9 & 0.1 & 134 & 4.6 & 0.2 & 111 & 4.4 & 0.1 & ... & ... & ... & ... & ... & ... & ... & ... & ... & 122 & 4.9 & 0.1 \\
5405.775 & 1.0 & 134 & 6.0 & 0.0 & 134 & 4.6 & 0.2 & 121 & 4.5 & 0.1 & ... & ... & ... & ... & ... & ... & ... & ... & ... & ... & ... & ... \\
5424.068 & 4.3 & 77 & 5.9 & 0.4 & 36 & 4.6 & 0.2 & ... & ... & ... & 85 & 5.4 & 0.1 & 80 & 5.3 & 0.2 & 83 & 5.3 & 0.1 & ... & ... & ... \\
5429.696 & 1.0 & 139 & 6.1 & 0.1 & 138 & 4.7 & 0.2 & 140 & 4.9 & 0.1 & ... & ... & ... & ... & ... & ... & ... & ... & ... & 127 & 4.9 & 0.1 \\
5434.524 & 1.0 & 118 & 6.0 & 0.1 & 122 & 4.7 & 0.2 & 133 & 5.0 & 0.2 & ... & ... & ... & ... & ... & ... & ... & ... & ... & ... & ... & ... \\
5497.516 & 1.0 & ... & ... & ... & 87 & 4.7 & 0.0 & 115 & 5.4 & 0.0 & ... & ... & ... & 120 & 5.0 & 0.0 & ... & ... & ... & 93 & 5.3 & 0.0 \\
5501.465 & 1.0 & ... & ... & ... & 80 & 4.7 & ... & ... & ... & ... & 117 & 5.2 & ... & 98 & 4.9 & ... & 135 & 5.1 & ... & 84 & 5.3 & ... \\
5506.779 & 1.0 & ... & ... & ... & 90 & 4.7 & 0.0 & 94 & 5.0 & 0.0 & ... & ... & ... & 125 & 5.0 & 0.0 & ... & ... & ... & ... & ... & ... \\
5563.600 & 4.2 & 20 & 6.0 & 0.1 & ... & ... & ... & ... & ... & ... & ... & ... & ... & ... & ... & ... & ... & ... & ... & ... & ... & ... \\
5569.620 & 3.4 & 62 & 5.8 & 0.3 & 36 & 4.5 & 0.1 & ... & ... & ... & ... & ... & ... & 52 & 4.8 & 0.1 & 76 & 5.0 & 0.1 & ... & ... & ... \\
5586.760 & 3.4 & 88 & 5.8 & 0.2 & 58 & 4.5 & 0.1 & ... & ... & ... & 128 & 5.6 & 0.1 & 86 & 4.8 & 0.1 & 110 & 5.1 & 0.1 & ... & ... & ... \\
5624.542 & 3.4 & 54 & 5.9 & 0.3 & 26 & 4.6 & 0.1 & ... & ... & ... & 74 & 5.3 & 0.1 & 34 & 4.7 & 0.1 & 64 & 5.1 & 0.1 & ... & ... & ... \\
5662.516 & 4.2 & 27 & 5.8 & 0.3 & ... & ... & ... & ... & ... & ... & ... & ... & ... & ... & ... & ... & ... & ... & ... & ... & ... & ... \\
5762.992 & 4.2 & 36 & 6.0 & 0.3 & ... & ... & ... & ... & ... & ... & ... & ... & ... & ... & ... & ... & ... & ... & ... & ... & ... & ... \\
6008.560 & 3.9 & 42 & 6.4 & 0.3 & ... & ... & ... & ... & ... & ... & 53 & 5.7 & 0.1 & ... & ... & ... & ... & ... & ... & ... & ... & ... \\
6065.482 & 2.6 & 57 & 5.9 & 0.3 & 44 & 4.6 & 0.1 & ... & ... & ... & 110 & 5.6 & 0.1 & 81 & 5.1 & 0.1 & ... & ... & ... & 65 & 5.3 & 0.2 \\
6191.558 & 2.4 & 73 & 6.0 & 0.2 & 60 & 4.7 & 0.1 & ... & ... & ... & 120 & 5.6 & 0.2 & 84 & 5.0 & 0.1 & ... & ... & ... & 47 & 4.9 & 0.2 \\
6219.280 & 2.2 & 35 & 6.0 & ... & 25 & 4.7 & ... & ... & ... & ... & 101 & 5.7 & ... & 48 & 5.1 & ... & ... & ... & ... & ... & ... & ... \\
6230.722 & 2.6 & ... & ... & ... & 62 & 4.6 & 0.1 & 71 & 5.0 & 0.2 & ... & ... & ... & 100 & 5.1 & 0.1 & ... & ... & ... & 61 & 5.0 & 0.2 \\
6246.320 & 3.6 & 41 & 5.9 & ... & ... & ... & ... & ... & ... & ... & ... & ... & ... & ... & ... & ... & ... & ... & ... & 41 & 5.3 & ... \\
6252.560 & 2.4 & 64 & 5.9 & 0.3 & 51 & 4.6 & 0.1 & 51 & 4.9 & 0.2 & ... & ... & ... & 71 & 4.9 & 0.1 & ... & ... & ... & ... & ... & ... \\
6265.134 & 2.2 & 30 & 6.0 & 0.3 & 20 & 4.6 & 0.1 & ... & ... & ... & 73 & 5.3 & 0.2 & 56 & 5.2 & 0.1 & ... & ... & ... & ... & ... & ... \\
6301.500 & 3.6 & 38 & 5.8 & 0.3 & ... & ... & ... & ... & ... & ... & ... & ... & ... & ... & ... & ... & ... & ... & ... & ... & ... & ... \\
6336.820 & 3.7 & 32 & 5.9 & 0.3 & ... & ... & ... & ... & ... & ... & ... & ... & ... & ... & ... & ... & ... & ... & ... & ... & ... & ... \\
6408.020 & 3.7 & 27 & 5.9 & 0.3 & ... & ... & ... & ... & ... & ... & 68 & 5.7 & 0.1 & ... & ... & ... & ... & ... & ... & ... & ... & ... \\
6411.650 & 3.6 & 49 & 5.9 & 0.3 & ... & ... & ... & ... & ... & ... & 88 & 5.6 & 0.1 & ... & ... & ... & ... & ... & ... & ... & ... & ... \\
6421.350 & 2.3 & 52 & 6.0 & ... & 44 & 4.7 & ... & ... & ... & ... & 131 & 6.0 & ... & 71 & 5.0 & ... & ... & ... & ... & ... & ... & ... \\
6430.846 & 2.2 & 61 & 5.9 & ... & 50 & 4.6 & ... & ... & ... & ... & ... & ... & ... & 78 & 4.9 & ... & ... & ... & ... & ... & ... & ... \\
6481.869 & 2.3 & ... & ... & ... & ... & ... & ... & ... & ... & ... & 72 & 5.9 & ... & ... & ... & ... & ... & ... & ... & ... & ... & ... \\
6494.980 & 2.4 & ... & ... & ... & 74 & 4.6 & 0.2 & ... & ... & ... & 136 & 5.4 & 0.2 & 122 & 5.1 & 0.2 & ... & ... & ... & 83 & 5.1 & 0.2 \\
6677.985 & 2.7 & 63 & 6.0 & 0.2 & 47 & 4.7 & 0.1 & ... & ... & ... & 130 & 5.9 & 0.1 & 77 & 5.1 & 0.1 & ... & ... & ... & ... & ... & ... \\
6750.152 & 2.4 & ... & ... & ... & ... & ... & ... & ... & ... & ... & 91 & 6.0 & ... & ... & ... & ... & ... & ... & ... & ... & ... & ... \\
7495.067 & 4.2 & 50 & 5.9 & 0.2 & ... & ... & ... & ... & ... & ... & 104 & 6.0 & -0.0 & ... & ... & ... & ... & ... & ... & ... & ... & ... \\
7511.021 & 4.2 & 64 & 5.9 & 0.2 & 24 & 4.5 & 0.1 & ... & ... & ... & 91 & 5.5 & -0.0 & ... & ... & ... & ... & ... & ... & ... & ... & ... \\
8220.379 & 4.3 & 68 & 5.9 & ... & 28 & 4.5 & ... & ... & ... & ... & ... & ... & ... & ... & ... & ... & ... & ... & ... & ... & ... & ... \\
8688.624 & 2.2 & 127 & 6.0 & ... & 118 & 4.7 & ... & ... & ... & ... & ... & ... & ... & ... & ... & ... & ... & ... & ... & 106 & 4.9 & ... \\
\multicolumn{23}{c}{\textbf{FeII}} \\
4178.854 & 2.6 & 102 & 6.1 & ... & 55 & 4.8 & ... & ... & ... & ... & ... & ... & ... & ... & ... & ... & ... & ... & ... & ... & ... & ... \\
4233.163 & 2.6 & ... & ... & ... & 85 & 4.8 & -0.0 & ... & ... & ... & ... & ... & ... & ... & ... & ... & ... & ... & ... & ... & ... & ... \\
4491.400 & 2.9 & 75 & 6.0 & -0.0 & 29 & 4.7 & -0.0 & ... & ... & ... & ... & ... & ... & ... & ... & ... & ... & ... & ... & ... & ... & ... \\
4515.334 & 2.8 & 91 & 6.0 & -0.0 & 40 & 4.6 & -0.0 & ... & ... & ... & ... & ... & ... & 80 & 5.2 & -0.0 & ... & ... & ... & ... & ... & ... \\
4522.628 & 2.8 & ... & ... & ... & 60 & 4.7 & ... & ... & ... & ... & ... & ... & ... & ... & ... & ... & ... & ... & ... & 80 & 5.1 & ... \\
4555.888 & 2.8 & 102 & 6.1 & -0.0 & 48 & 4.7 & -0.0 & ... & ... & ... & ... & ... & ... & ... & ... & ... & ... & ... & ... & ... & ... & ... \\
4576.328 & 2.8 & 63 & 6.0 & -0.0 & 22 & 4.8 & -0.0 & ... & ... & ... & ... & ... & ... & ... & ... & ... & ... & ... & ... & ... & ... & ... \\
4582.835 & 2.8 & 49 & 6.0 & -0.0 & ... & ... & ... & ... & ... & ... & ... & ... & ... & ... & ... & ... & ... & ... & ... & ... & ... & ... \\
4583.829 & 2.8 & 131 & 6.1 & -0.0 & 78 & 4.7 & -0.0 & 70 & 4.4 & -0.0 & 116 & 5.8 & -0.0 & 105 & 5.0 & -0.0 & ... & ... & ... & ... & ... & ... \\
4620.513 & 2.8 & 41 & 5.9 & ... & ... & ... & ... & ... & ... & ... & ... & ... & ... & ... & ... & ... & ... & ... & ... & ... & ... & ... \\
4629.332 & 2.8 & 96 & 5.9 & -0.0 & 52 & 4.7 & -0.0 & ... & ... & ... & ... & ... & ... & ... & ... & ... & ... & ... & ... & ... & ... & ... \\
4731.439 & 2.9 & 56 & 6.2 & ... & ... & ... & ... & ... & ... & ... & ... & ... & ... & ... & ... & ... & ... & ... & ... & ... & ... & ... \\
4923.922 & 2.9 & ... & ... & ... & 99 & 4.6 & 0.1 & 120 & 4.8 & 0.1 & ... & ... & ... & 125 & 4.8 & 0.0 & 120 & 4.8 & 0.0 & ... & ... & ... \\
5018.435 & 2.9 & ... & ... & ... & 110 & 4.9 & ... & ... & ... & ... & ... & ... & ... & ... & ... & ... & ... & ... & ... & ... & ... & ... \\
5197.568 & 3.2 & 87 & 5.9 & -0.0 & 36 & 4.6 & -0.0 & ... & ... & ... & ... & ... & ... & ... & ... & ... & 104 & 5.6 & -0.0 & ... & ... & ... \\
5234.624 & 3.2 & 88 & 6.1 & ... & 39 & 4.8 & ... & ... & ... & ... & 70 & 5.7 & ... & 56 & 5.0 & ... & 75 & 5.3 & ... & 71 & 5.5 & ... \\
5275.997 & 3.2 & 102 & 6.0 & ... & 45 & 4.6 & ... & ... & ... & ... & 86 & 5.6 & ... & ... & ... & ... & 77 & 5.0 & ... & ... & ... & ... \\
5325.543 & 3.2 & 33 & 6.2 & ... & ... & ... & ... & ... & ... & ... & ... & ... & ... & ... & ... & ... & ... & ... & ... & ... & ... & ... \\
5534.839 & 3.2 & 55 & 6.3 & -0.0 & ... & ... & ... & ... & ... & ... & ... & ... & ... & ... & ... & ... & ... & ... & ... & 40 & 5.7 & -0.0 \\
6247.559 & 3.9 & 39 & 6.1 & -0.0 & ... & ... & ... & ... & ... & ... & ... & ... & ... & ... & ... & ... & ... & ... & ... & ... & ... & ... \\
6432.677 & 2.9 & 29 & 6.0 & ... & ... & ... & ... & ... & ... & ... & ... & ... & ... & ... & ... & ... & ... & ... & ... & ... & ... & ... \\
6456.381 & 3.9 & 53 & 6.2 & ... & ... & ... & ... & ... & ... & ... & 50 & 6.0 & ... & ... & ... & ... & 44 & 5.5 & ... & ... & ... & ... \\
6516.077 & 2.9 & 41 & 6.1 & ... & ... & ... & ... & ... & ... & ... & 66 & 6.2 & ... & ... & ... & ... & ... & ... & ... & ... & ... & ... \\
\enddata
\label{tab:lbl_iron_abund}
\end{deluxetable}
\end{longrotatetable}
\newpage

\newpage

\begin{longrotatetable}
\small
\movetabledown=15mm
\begin{deluxetable}{lccccccccccccccccccccccc}
\tablecaption{Same as Table~\ref{tab:lbl_iron_abund}, but for other elements}
\label{tab:lbl_ab}

\tablehead{
$\lambda$ & $\chi$ &  \multicolumn{3}{c}{HD222925} & \multicolumn{3}{c}{HD122563} & \multicolumn{3}{c}{Aqu2472} & \multicolumn{3}{c}{Aqu2776} &  \multicolumn{3}{c}{Sgr2936} &  \multicolumn{3}{c}{Sgr2584} &  \multicolumn{3}{c}{Sgr2656} \\
\small & \small & \small EW & \small A(X) & \small $\Delta_{nlte}$ & \small EW & \small A(X) & \small $\Delta_{nlte}$ & \small EW & \small A(X) & \small $\Delta_{nlte}$ & \small EW & \small A(X) & \small $\Delta_{nlte}$ & \small EW & \small A(X) & \small $\Delta_{nlte}$ & \small EW & \small A(X) & \small $\Delta_{nlte}$ & \small EW & \small A(X) & \small $\Delta_{nlte}$  \\
}
\startdata
\multicolumn{23}{c}{\textbf{NaI}} \\
5889.950 & 0.0 & 243 & 5.2 & -0.6 & 184 & 3.7 & -0.4 & 90 & 2.4 & -0.03 & 169 & 3.2 & -0.2 & 195 & 3.6 & -0.3 & 253 & 4.0 & -0.2 & 166 & 3.7 & -0.5 \\
5895.920 & 0.0 & 212 & 5.2 & -0.6 & 154 & 3.5 & -0.3 & 80 & 2.6 & -0.03 & 131 & 2.9 & -0.2 & 209 & 4.0 & -0.3 & synth & 3.5 & -0.2 & 202 & 4.5 & -0.6 \\
\multicolumn{23}{c}{\textbf{MgI}} \\
4167.271 & 4.3 & 103 & 6.7 & 0.3 & 51 & 5.5 & 0.2 & ... & ... & ... & ... & ... & ... & 139 & 6.9 & 0.2 & ... & ... & ... & ... & ... & ... \\
4571.096 & 0.0 & 62 & 6.4 & 0.4 & 82 & 5.3 & 0.3 & ... & ... & ... & ... & ... & ... & 130 & 5.9 & 0.2 & 124 & 5.5 & 0.4 & ... & ... & ... \\
4702.991 & 4.3 & 120 & 6.5 & 0.2 & 70 & 5.4 & 0.1 & ... & ... & ... & 93 & 5.7 & -0.0 & 114 & 6.0 & 0.1 & 108 & 5.9 & 0.0 & 71 & 5.5 & 0.2 \\
5172.680 & 2.7 & 257 & 6.2 & 0.3 & 204 & 5.0 & 0.1 & 237 & 5.4 & 0.2 & ... & ... & ... & 281 & 5.5 & 0.2 & 304 & 5.5 & 0.2 & 312 & 5.9 & 0.2 \\
5183.600 & 2.7 & 310 & 6.2 & 0.3 & 226 & 4.9 & 0.1 & 228 & 5.2 & 0.2 & ... & ... & ... & 287 & 5.3 & 0.2 & 372 & 5.6 & 0.2 & ... & ... & ... \\
5528.400 & 4.3 & 121 & 6.4 & 0.1 & 74 & 5.2 & 0.1 & 78 & 5.3 & 0.1 & 185 & 6.5 & -0.1 & 100 & 5.5 & 0.0 & 115 & 5.7 & -0.0 & 79 & 5.4 & 0.1 \\
5711.090 & 4.3 & 33 & 6.3 & 0.2 & ... & ... & ... & ... & ... & ... & 109 & 6.9 & -0.1 & ... & ... & ... & 45 & 5.9 & 0.0 & ... & ... & ... \\
\multicolumn{23}{c}{\textbf{KI}} \\
7664.911 & 0.0 & 124 & 4.5 & ... & 253 & 5.0 & ... & 57 & 2.8 & ... & ... & ... & ... & ... & ... & ... & ... & ... & ... & ... & ... & ... \\
7698.974 & 0.0 & 81 & 4.1 & ... & 44 & 2.7 & ... & 83 & 3.5 & -0.2 & 161 & 4.3 & -0.2 & 70 & 3.1 & -0.2 & 110 & 3.4 & -0.16 & 84 & 3.6 & -0.25 \\
\multicolumn{23}{c}{\textbf{CaI}} \\
5265.560 & 2.5 & 67 & 5.2 & ... & 31 & 4.0 & ... & ... & ... & ... & 99 & 5.1 & ... & ... & ... & ... & 129 & 5.6 & ... & 80 & 5.0 & ... \\
5349.470 & 2.7 & 43 & 5.1 & 0.2 & ... & ... & ... & ... & ... & ... & ... & ... & ... & ... & ... & ... & ... & ... & ... & 48 & 4.8 & 0.3 \\
5512.990 & 2.9 & 25 & 5.1 & 0.1 & ... & ... & ... & ... & ... & ... & ... & ... & ... & ... & ... & ... & 34 & 4.7 & 0.1 & ... & ... & ... \\
5581.970 & 2.5 & 42 & 5.1 & ... & ... & ... & ... & ... & ... & ... & ... & ... & ... & ... & ... & ... & ... & ... & ... & ... & ... & ... \\
5588.750 & 2.5 & 94 & 5.2 & 0.1 & 47 & 3.8 & 0.3 & ... & ... & ... & ... & ... & ... & 90 & 4.4 & 0.2 & 90 & 4.4 & 0.3 & 140 & 5.6 & 0.3 \\
5590.110 & 2.5 & 38 & 5.0 & 0.2 & ... & ... & ... & ... & ... & ... & ... & ... & ... & 31 & 4.3 & 0.3 & 42 & 4.4 & 0.3 & 61 & 5.0 & 0.3 \\
5594.460 & 2.5 & 82 & 5.1 & ... & 35 & 3.7 & ... & ... & ... & ... & 90 & 4.5 & ... & 62 & 4.1 & ... & 110 & 4.8 & ... & ... & ... & ... \\
5598.480 & 2.5 & 74 & 5.1 & ... & 28 & 3.8 & ... & 45 & 4.2 & ... & ... & ... & ... & 60 & 4.3 & ... & 86 & 4.6 & ... & 76 & 4.7 & ... \\
5601.280 & 2.5 & 42 & 5.1 & ... & ... & ... & ... & 34 & 4.4 & ... & 63 & 4.7 & ... & 29 & 4.2 & ... & 34 & 4.2 & ... & ... & ... & ... \\
5857.450 & 2.9 & 67 & 5.1 & 0.1 & 23 & 3.8 & 0.2 & ... & ... & ... & 78 & 4.7 & 0.1 & 60 & 4.4 & 0.1 & 51 & 4.2 & 0.1 & ... & ... & ... \\
6102.720 & 1.9 & 67 & 5.1 & 0.2 & 37 & 3.8 & 0.2 & ... & ... & ... & 101 & 4.7 & 0.1 & 69 & 4.2 & 0.2 & 100 & 4.6 & 0.2 & ... & ... & ... \\
6122.220 & 1.9 & 104 & 5.2 & 0.1 & 66 & 3.8 & 0.2 & 103 & 4.5 & 0.2 & ... & ... & ... & 118 & 4.5 & 0.2 & 139 & 4.7 & 0.2 & 89 & 4.4 & 0.2 \\
6162.170 & 1.9 & 117 & 5.2 & ... & 78 & 3.8 & 0.2 & 54 & 3.6 & 0.2 & 118 & 4.3 & -0.0 & 126 & 4.4 & 0.1 & 141 & 4.5 & 0.2 & 102 & 4.4 & 0.1 \\
6166.439 & 2.5 & 20 & 5.2 & ... & ... & ... & ... & ... & ... & ... & 58 & 5.1 & ... & ... & ... & ... & ... & ... & ... & ... & ... & ... \\
6169.040 & 2.5 & 31 & 5.1 & 0.1 & ... & ... & ... & ... & ... & ... & ... & ... & ... & ... & ... & ... & 39 & 4.5 & 0.1 & ... & ... & ... \\
6169.560 & 2.5 & 46 & 5.1 & 0.1 & ... & ... & ... & ... & ... & ... & 100 & 5.2 & 0.1 & ... & ... & ... & 52 & 4.4 & 0.1 & ... & ... & ... \\
6439.070 & 2.5 & 104 & 5.1 & -0.1 & 58 & 3.7 & 0.2 & ... & ... & ... & 93 & 4.1 & -0.2 & 96 & 4.1 & 0.1 & 113 & 4.3 & 0.2 & 50 & 3.8 & 0.1 \\
6449.810 & 2.5 & 47 & 5.1 & 0.0 & ... & ... & ... & ... & ... & ... & ... & ... & ... & ... & ... & ... & ... & ... & ... & ... & ... & ... \\
6499.650 & 2.5 & 30 & 5.1 & ... & ... & ... & ... & ... & ... & ... & 46 & 4.6 & ... & ... & ... & ... & ... & ... & ... & ... & ... & ... \\
6717.680 & 2.7 & 39 & 5.2 & ... & ... & ... & ... & 31 & 4.5 & ... & ... & ... & ... & ... & ... & ... & 46 & 4.5 & ... & ... & ... & ... \\
\multicolumn{23}{c}{\textbf{ScII}} \\
4314.083 & 0.6 & 146 & 2.3 & ... & ... & ... & ... & ... & ... & ... & ... & ... & ... & ... & ... & ... & ... & ... & ... & ... & ... & ... \\
4324.996 & 0.6 & 128 & 2.2 & ... & 85 & 0.5 & ... & ... & ... & ... & ... & ... & ... & 131 & 1.2 & ... & 223 & 2.8 & ... & 125 & 1.5 & ... \\
4374.457 & 0.6 & 116 & 2.0 & ... & 82 & 0.4 & ... & ... & ... & ... & ... & ... & ... & ... & ... & ... & 145 & 1.5 & ... & ... & ... & ... \\
4400.389 & 0.6 & 110 & 1.9 & ... & 78 & 0.4 & ... & ... & ... & ... & ... & ... & ... & 160 & 1.8 & ... & ... & ... & ... & ... & ... & ... \\
4415.557 & 0.6 & 101 & 1.8 & ... & 74 & 0.5 & ... & ... & ... & ... & 128 & 2.0 & ... & 104 & 0.8 & ... & 160 & 2.0 & ... & 93 & 1.0 & ... \\
5031.010 & 1.4 & 64 & 1.7 & ... & 41 & 0.5 & ... & ... & ... & ... & ... & ... & ... & 71 & 0.8 & ... & 75 & 0.9 & ... & ... & ... & ... \\
5526.770 & 1.8 & 71 & 1.8 & ... & 40 & 0.5 & ... & $<$38 & $<$0.6 & ... & 69 & 1.2 & ... & 74 & 0.9 & ... & 72 & 0.8 & ... & 56 & 1.0 & ... \\
\multicolumn{23}{c}{\textbf{TiI}} \\
4840.870 & 0.9 & 21 & 3.6 & 0.7 & ... & ... & ... & ... & ... & ... & ... & ... & ... & ... & ... & ... & 44 & 2.8 & 0.6 & ... & ... & ... \\
4981.730 & 0.8 & 80 & 3.6 & 0.2 & 60 & 2.2 & 0.4 & 93 & 2.9 & 0.3 & 95 & 2.5 & 0.4 & 91 & 2.6 & 0.3 & 114 & 2.7 & 0.4 & ... & ... & ... \\
4999.500 & 0.8 & 78 & 3.8 & 0.3 & 48 & 2.2 & 0.4 & ... & ... & ... & 108 & 3.0 & 0.6 & 98 & 2.9 & 0.4 & 104 & 2.8 & 0.5 & 85 & 3.2 & 0.4 \\
5007.210 & 0.8 & 83 & 4.0 & 0.3 & 51 & 2.4 & 0.4 & 113 & 3.6 & 0.4 & ... & ... & ... & 95 & 3.0 & 0.4 & 116 & 3.1 & 0.4 & ... & ... & ... \\
5022.870 & 0.8 & 30 & 3.6 & 0.3 & ... & ... & ... & ... & ... & ... & ... & ... & ... & 54 & 2.9 & 0.4 & 68 & 2.9 & 0.4 & ... & ... & ... \\
5036.460 & 1.4 & 25 & 3.7 & 0.5 & ... & ... & ... & 37 & 3.2 & 0.5 & 53 & 2.9 & 0.5 & ... & ... & ... & 58 & 3.1 & 0.4 & ... & ... & ... \\
5039.960 & 0.0 & 32 & 3.6 & 0.7 & 28 & 2.2 & 0.8 & ... & ... & ... & ... & ... & ... & 76 & 2.9 & 0.7 & 92 & 2.9 & 0.8 & ... & ... & ... \\
5064.650 & 0.1 & 36 & 3.6 & 0.6 & 35 & 2.2 & 0.8 & 93 & 3.4 & 0.8 & ... & ... & ... & 56 & 2.5 & 0.7 & 86 & 2.7 & 0.8 & 43 & 2.9 & 0.8 \\
5173.740 & 0.0 & 39 & 3.7 & 0.7 & 32 & 2.2 & 0.8 & 115 & 3.8 & 0.8 & 86 & 2.8 & 0.7 & 65 & 2.7 & 0.7 & 110 & 3.1 & 0.8 & 52 & 3.0 & 0.8 \\
5192.970 & 0.0 & 61 & 3.9 & 0.7 & 37 & 2.2 & 0.8 & 65 & 3.0 & 0.8 & 90 & 2.8 & 0.8 & 102 & 3.1 & 0.7 & 92 & 2.7 & 0.9 & 108 & 3.9 & 0.8 \\
5210.380 & 0.1 & 45 & 3.6 & 0.6 & 42 & 2.2 & 0.8 & ... & ... & ... & ... & ... & ... & 78 & 2.7 & 0.7 & 109 & 2.9 & 0.9 & ... & ... & ... \\
\multicolumn{23}{c}{\textbf{TiII}} \\
4762.780 & 1.1 & 35 & 3.9 & 0.0 & 21 & 2.6 & 0.1 & ... & ... & ... & ... & ... & ... & ... & ... & ... & 207 & 5.8 & 0.0 & ... & ... & ... \\
4763.880 & 1.2 & 62 & 3.9 & 0.0 & 34 & 2.6 & 0.2 & 62 & 3.2 & 0.1 & ... & ... & ... & 66 & 3.0 & 0.1 & 90 & 3.4 & 0.1 & ... & ... & ... \\
4764.520 & 1.2 & 33 & 3.8 & 0.0 & ... & ... & ... & ... & ... & ... & 60 & 3.5 & 0.0 & 47 & 3.1 & 0.1 & 61 & 3.2 & 0.1 & ... & ... & ... \\
4798.530 & 1.1 & 46 & 3.8 & -0.0 & 30 & 2.6 & 0.1 & 33 & 2.8 & 0.1 & ... & ... & ... & 55 & 3.0 & 0.0 & 102 & 3.7 & 0.0 & ... & ... & ... \\
4849.170 & 1.1 & 41 & 4.1 & 0.0 & ... & ... & ... & 23 & 3.0 & 0.1 & ... & ... & ... & ... & ... & ... & 88 & 3.8 & 0.1 & ... & ... & ... \\
4874.010 & 3.1 & 32 & 3.8 & 0.1 & ... & ... & ... & ... & ... & ... & ... & ... & ... & ... & ... & ... & 41 & 3.4 & 0.1 & ... & ... & ... \\
4911.190 & 3.1 & 40 & 3.7 & ... & ... & ... & ... & ... & ... & ... & ... & ... & ... & ... & ... & ... & ... & ... & ... & ... & ... & ... \\
5013.690 & 1.6 & 45 & 3.8 & -0.0 & 21 & 2.5 & 0.1 & 32 & 2.8 & 0.0 & ... & ... & ... & 52 & 3.0 & 0.1 & 76 & 3.3 & 0.1 & ... & ... & ... \\
5129.160 & 1.9 & 78 & 3.8 & -0.2 & 41 & 2.5 & 0.1 & ... & ... & ... & ... & ... & ... & 74 & 2.9 & 0.0 & 72 & 2.8 & 0.0 & ... & ... & ... \\
5154.068 & 1.6 & 74 & 3.8 & -0.1 & 43 & 2.5 & 0.1 & ... & ... & ... & 108 & 4.0 & -0.1 & 104 & 3.3 & 0.0 & 106 & 3.4 & 0.1 & ... & ... & ... \\
5185.900 & 1.9 & 73 & 3.8 & -0.2 & 35 & 2.4 & 0.1 & ... & ... & ... & ... & ... & ... & 88 & 3.1 & 0.0 & 80 & 3.0 & 0.0 & 36 & 2.7 & -0.0 \\
5188.687 & 1.6 & 135 & 4.2 & -0.3 & 85 & 2.5 & 0.0 & 140 & 3.4 & -0.0 & ... & ... & ... & 131 & 3.1 & -0.0 & 158 & 3.6 & 0.0 & 237 & 5.0 & -0.1 \\
5211.530 & 2.6 & 26 & 3.7 & 0.0 & ... & ... & ... & ... & ... & ... & ... & ... & ... & ... & ... & ... & ... & ... & ... & ... & ... & ... \\
5226.539 & 1.6 & 109 & 3.9 & -0.3 & 70 & 2.4 & 0.1 & ... & ... & ... & 110 & 3.5 & -0.2 & ... & ... & ... & 115 & 3.0 & 0.0 & ... & ... & ... \\
5336.790 & 1.6 & 81 & 3.7 & -0.1 & 51 & 2.5 & 0.1 & ... & ... & ... & ... & ... & ... & 87 & 2.9 & 0.1 & 127 & 3.5 & 0.1 & ... & ... & ... \\
5381.020 & 1.6 & 62 & 3.8 & -0.1 & 31 & 2.5 & 0.1 & ... & ... & ... & 129 & 4.6 & -0.1 & 61 & 2.9 & 0.1 & 97 & 3.4 & 0.1 & 25 & 2.6 & 0.0 \\
5418.770 & 1.6 & 50 & 3.8 & -0.1 & 24 & 2.5 & 0.1 & ... & ... & ... & ... & ... & ... & 68 & 3.1 & 0.1 & 64 & 3.0 & 0.1 & 57 & 3.3 & 0.0 \\
\multicolumn{23}{c}{\textbf{VI}} \\
4379.230 & 0.3 & 38 & 2.4 & ... & 28 & 1.0 & ... & ... & ... & ... & ... & ... & ... & 42 & 1.2 & ... & 111 & 2.2 & ... & ... & ... & ... \\
\multicolumn{23}{c}{\textbf{CrI}} \\
4254.352 & 0.0 & 145 & 4.1 & 0.1 & 108 & 2.2 & 0.6 & 150 & 3.1 & 0.4 & ... & ... & ... & 154 & 2.8 & 0.6 & 200 & 3.5 & 0.9 & ... & ... & ... \\
4274.812 & 0.0 & 140 & 4.1 & 0.1 & 94 & 2.0 & 0.6 & ... & ... & ... & 149 & 3.1 & 0.4 & 204 & 3.7 & 0.6 & 189 & 3.4 & 0.9 & ... & ... & ... \\
4289.731 & 0.0 & 147 & 4.4 & 0.1 & 97 & 2.2 & 0.6 & 100 & 2.3 & 0.5 & ... & ... & ... & 168 & 3.3 & 0.6 & ... & ... & ... & ... & ... & ... \\
4545.953 & 0.9 & 28 & 4.0 & 0.4 & ... & ... & ... & ... & ... & ... & 42 & 2.9 & 0.3 & 32 & 3.0 & 0.5 & ... & ... & ... & ... & ... & ... \\
4600.749 & 1.0 & 28 & 3.9 & 0.4 & ... & ... & ... & ... & ... & ... & ... & ... & ... & 48 & 3.2 & 0.4 & 75 & 3.5 & 0.5 & ... & ... & ... \\
4616.124 & 1.0 & 33 & 3.9 & 0.4 & ... & ... & ... & ... & ... & ... & 83 & 3.6 & 0.2 & ... & ... & ... & 46 & 2.9 & 0.5 & ... & ... & ... \\
4626.173 & 1.0 & 26 & 3.9 & 0.4 & ... & ... & ... & ... & ... & ... & 44 & 2.9 & 0.2 & ... & ... & ... & 55 & 3.2 & 0.5 & ... & ... & ... \\
4646.162 & 1.0 & 53 & 3.9 & 0.3 & 35 & 2.5 & 0.5 & ... & ... & ... & 119 & 4.0 & 0.3 & 75 & 3.1 & 0.5 & 93 & 3.3 & 0.5 & ... & ... & ... \\
4652.157 & 1.0 & 39 & 3.9 & 0.4 & 23 & 2.5 & 0.5 & ... & ... & ... & 80 & 3.4 & 0.2 & ... & ... & ... & 56 & 2.9 & 0.5 & ... & ... & ... \\
5206.020 & 0.9 & 113 & 4.0 & 0.0 & 83 & 2.4 & 0.5 & 118 & 3.1 & 0.4 & ... & ... & ... & 109 & 2.6 & 0.4 & 165 & 3.4 & 0.6 & ... & ... & ... \\
5208.409 & 0.9 & 137 & 4.3 & 0.0 & 117 & 2.9 & 0.4 & 104 & 2.7 & 0.3 & ... & ... & ... & 180 & 3.7 & 0.4 & 209 & 3.9 & 0.6 & 111 & 3.1 & 0.2 \\
5296.690 & 1.0 & 25 & 3.9 & 0.4 & ... & ... & ... & ... & ... & ... & 85 & 3.6 & 0.2 & ... & ... & ... & 57 & 3.1 & 0.4 & 72 & 4.0 & 0.4 \\
5298.270 & 1.0 & 39 & 3.9 & 0.3 & 23 & 2.5 & 0.4 & ... & ... & ... & 135 & 4.4 & 0.2 & 41 & 2.8 & 0.4 & 113 & 3.7 & 0.4 & 34 & 3.2 & 0.4 \\
5345.800 & 1.0 & 50 & 3.9 & 0.3 & 30 & 2.5 & 0.4 & ... & ... & ... & 115 & 3.8 & 0.3 & 86 & 3.3 & 0.4 & 122 & 3.7 & 0.5 & ... & ... & ... \\
5348.310 & 1.0 & 31 & 3.9 & 0.3 & ... & ... & ... & ... & ... & ... & ... & ... & ... & ... & ... & ... & 82 & 3.3 & 0.5 & ... & ... & ... \\
5409.780 & 1.0 & 61 & 3.9 & 0.2 & 40 & 2.4 & 0.4 & 68 & 3.1 & 0.4 & ... & ... & ... & 77 & 2.9 & 0.4 & 113 & 3.3 & 0.5 & 70 & 3.3 & 0.3 \\
\multicolumn{23}{c}{\textbf{MnI}} \\
4030.753 & 0.0 & 172 & 4.4 & 0.2 & 135 & 2.5 & 0.1 & ... & ... & ... & ... & ... & ... & ... & ... & ... & 350 & 4.3 & 0.2 & ... & ... & ... \\
4033.062 & 0.0 & 148 & 4.1 & 0.1 & 121 & 2.4 & 0.1 & ... & ... & ... & ... & ... & ... & ... & ... & ... & 275 & 4.1 & 0.1 & ... & ... & ... \\
4034.483 & 0.0 & 122 & 3.7 & 0.2 & 112 & 2.4 & 0.1 & ... & ... & ... & ... & ... & ... & ... & ... & ... & ... & ... & ... & ... & ... & ... \\
4762.370 & 2.9 & 23 & 3.5 & 0.4 & ... & ... & ... & ... & ... & ... & ... & ... & ... & ... & ... & ... & ... & ... & ... & ... & ... & ... \\
4783.430 & 2.3 & 40 & 3.5 & ... & 23 & 2.3 & ... & ... & ... & ... & 69 & 2.9 & ... & 46 & 2.7 & ... & 72 & 3.0 & ... & ... & ... & ... \\
4823.520 & 2.3 & 47 & 3.5 & 0.4 & 25 & 2.2 & 0.4 & ... & ... & ... & ... & ... & ... & ... & ... & ... & 64 & 2.8 & 0.4 & ... & ... & ... \\
\multicolumn{23}{c}{\textbf{NiI}} \\
4604.988 & 3.5 & 24 & 4.6 & ... & ... & ... & ... & ... & ... & ... & ... & ... & ... & 24 & 3.9 & ... & 28 & 3.9 & ... & ... & ... & ... \\
4648.652 & 3.4 & 32 & 4.6 & ... & ... & ... & ... & 56 & 4.4 & ... & 51 & 4.1 & ... & ... & ... & ... & 42 & 4.0 & ... & ... & ... & ... \\
4756.515 & 3.5 & 24 & 4.7 & ... & ... & ... & ... & ... & ... & ... & 52 & 4.4 & ... & 57 & 4.5 & ... & ... & ... & ... & ... & ... & ... \\
4786.530 & 3.4 & 42 & 4.8 & ... & ... & ... & ... & ... & ... & ... & ... & ... & ... & ... & ... & ... & ... & ... & ... & ... & ... & ... \\
4904.412 & 3.5 & 28 & 4.7 & ... & ... & ... & ... & ... & ... & ... & 49 & 4.3 & ... & ... & ... & ... & 71 & 4.6 & ... & ... & ... & ... \\
4980.170 & 3.6 & 32 & 4.6 & ... & ... & ... & ... & ... & ... & ... & 68 & 4.5 & ... & 32 & 3.9 & ... & 31 & 3.8 & ... & ... & ... & ... \\
5035.360 & 3.6 & 42 & 4.6 & ... & 20 & 3.4 & ... & ... & ... & ... & ... & ... & ... & 41 & 3.8 & ... & 66 & 4.2 & ... & ... & ... & ... \\
5080.530 & 3.6 & 43 & 4.6 & ... & 21 & 3.4 & ... & ... & ... & ... & ... & ... & ... & 53 & 4.0 & ... & 58 & 4.0 & ... & 48 & 4.2 & ... \\
5081.107 & 3.8 & 35 & 4.6 & ... & ... & ... & ... & ... & ... & ... & 46 & 4.1 & ... & ... & ... & ... & ... & ... & ... & 24 & 4.0 & ... \\
5084.096 & 3.7 & 29 & 4.6 & ... & ... & ... & ... & ... & ... & ... & 41 & 4.1 & ... & ... & ... & ... & 49 & 4.2 & ... & ... & ... & ... \\
5137.070 & 1.7 & 34 & 4.7 & ... & 33 & 3.5 & ... & 31 & 3.8 & ... & ... & ... & ... & ... & ... & ... & 85 & 4.1 & ... & ... & ... & ... \\
5476.900 & 1.8 & 96 & 4.7 & ... & 67 & 3.1 & ... & 43 & 2.9 & ... & 94 & 3.4 & ... & 124 & 3.9 & ... & 140 & 3.9 & ... & ... & ... & ... \\
6643.630 & 1.7 & 25 & 4.7 & ... & 25 & 3.5 & ... & ... & ... & ... & 70 & 4.0 & ... & 56 & 4.0 & ... & 80 & 4.1 & ... & ... & ... & ... \\
\multicolumn{23}{c}{\textbf{ZnI}} \\
4722.150 & 4.0 & 33 & 3.1 & ... & ... & ... & ... & ... & ... & ... & ... & ... & ... & ... & ... & ... & 46 & 2.7 & ... & ... & ... & ... \\
4810.530 & 4.1 & 41 & 3.1 & ... & ... & ... & ... & ... & ... & ... & 58 & 3.0 & ... & 46 & 2.5 & ... & 52 & 2.6 & ... & ... & ... & ... \\
\multicolumn{23}{c}{\textbf{SrII}} \\
4077.714 & 0.0 & 323 & 2.0 & ... & 158 & -0.1 & ... & ... & ... & ... & $<$88 & $<$-1.2 & ... & ... & ... & ... & ... & ... & ... & ... & ... & ... \\
4215.524 & 0.0 & 256 & 2.0 & ... & 149 & -0.0 & ... & $<$86 & $<$-1.4 & ... & $<$92 & $<$-0.9 & ... & 207 & 0.5 & ... & 225 & 0.6 & ... & 208 & 0.9 & ... \\
\multicolumn{23}{c}{\textbf{YII}} \\
4398.010 & 0.1 & 72 & 1.1 & ... & 27 & -0.8 & ... & ... & ... & ... & ... & ... & ... & ... & ... & ... & ... & ... & ... & ... & ... & ... \\
4682.325 & 0.4 & 26 & 1.1 & ... & ... & ... & ... & ... & ... & ... & ... & ... & ... & ... & ... & ... & ... & ... & ... & ... & ... & ... \\
4854.870 & 1.0 & 50 & 1.0 & ... & ... & ... & ... & ... & ... & ... & ... & ... & ... & 36 & -0.3 & ... & ... & ... & ... & ... & ... & ... \\
4883.680 & 1.1 & 80 & 1.1 & ... & 22 & -0.8 & ... & ... & ... & ... & ... & ... & ... & 61 & -0.2 & ... & 51 & -0.5 & ... & ... & ... & ... \\
4883.684 & 1.1 & 78 & 1.1 & ... & 22 & -0.8 & ... & ... & ... & ... & ... & ... & ... & ... & ... & ... & ... & ... & ... & ... & ... & ... \\
5087.420 & 1.1 & 60 & 1.0 & ... & ... & ... & ... & $<$26 & $<$-0.4 & ... & ... & ... & ... & ... & ... & ... & 45 & -0.4 & ... & 45 & 0.1 & ... \\
5123.220 & 1.0 & 34 & 1.1 & ... & ... & ... & ... & ... & ... & ... & ... & ... & ... & ... & ... & ... & ... & ... & ... & 34 & 0.5 & ... \\
5200.410 & 1.0 & 43 & 1.0 & ... & ... & ... & ... & ... & ... & ... & ... & ... & ... & ... & ... & ... & ... & ... & ... & ... & ... & ... \\
5205.730 & 1.0 & 60 & 1.1 & ... & ... & ... & ... & ... & ... & ... & ... & ... & ... & ... & ... & ... & ... & ... & ... & ... & ... & ... \\
5662.925 & 1.9 & 38 & 1.2 & ... & ... & ... & ... & ... & ... & ... & ... & ... & ... & ... & ... & ... & ... & ... & ... & ... & ... & ... \\
\multicolumn{23}{c}{\textbf{BaII}} \\
4554.029 & 0.0 & 238 & 1.5 & ... & 96 & -1.4 & ... & ... & ... & ... & ... & ... & ... & 220 & 0.3 & 0.0 & 190 & -0.3 & 0.0 & ... & ... & ... \\
5853.670 & 0.6 & 99 & 1.4 & ... & ... & ... & ... & ... & ... & ... & ... & ... & ... & 67 & -0.4 & 0.0 & 76 & -0.4 & 0.0 & ... & ... & ... \\
6141.710 & 0.7 & 161 & 1.5 & ... & 42 & -1.5 & ... & $<$59 & $<$-1.1 & ... & $<$40 & $<$-1.4 & ... & 118 & -0.5 & ... & 133 & -0.4 & ... & 93 & -0.3 & ... \\
6496.900 & 0.6 & 173 & 1.8 & ... & 35 & -1.5 & ... & $<$31 & $<$-1.4 & ... & ... & ... & ... & 123 & -0.3 & -0.2 & 143 & -0.2 & -0.2 & 72 & -0.5 & -0.2 \\
\multicolumn{23}{c}{\textbf{LaII}} \\
4662.498 & 0.0 & 28 & 0.7 & ... & ... & ... & ... & ... & ... & ... & ... & ... & ... & $<$31 & $<$-0.5 & ... & $<$20 & $<$-0.8 & ... & $<$20 & $<$-0.8 & ... \\
\multicolumn{23}{c}{\textbf{NdII}} \\
4061.080 & 0.5 & 78 & 1.0 & ... & ... & ... & ... & ... & ... & ... & 114 & 1.4 & ... & ... & ... & ... & ... & ... & ... & ... & ... & ... \\
4825.480 & 0.2 & 38 & 0.9 & ... & ... & ... & ... & ... & ... & ... & ... & ... & ... & 22 & -0.8 & ... & ... & ... & ... & ... & ... & ... \\
4959.120 & 0.1 & 27 & 0.9 & ... & ... & ... & ... & $<$57 &$<$0.3 & ... & ... & ... & ... & ... & ... & ... & $<$27 & $<$-0.6 & ... & ... & ... & ... \\
5130.590 & 1.3 & 22 & 0.8 & ... & ... & ... & ... & ... & ... & ... & ... & ... & ... & 21 & -0.3 & ... & ... & ... & ... & ... & ... & ... \\
5234.190 & 0.6 & 20 & 0.9 & ... & ... & ... & ... & ... & ... & ... & ... & ... & ... & ... & ... & ... & ... & ... & ... & 64 & 0.9 & ... \\
5249.580 & 1.0 & 31 & 0.9 & ... & ... & ... & ... & ... & ... & ... & 27 & -0.2 & ... & ... & ... & ... & ... & ... & ... & ... & ... & ... \\
5255.510 & 0.2 & 28 & 0.9 & ... & ... & ... & ... & ... & ... & ... & ... & ... & ... & ... & ... & ... & ... & ... & ... & ... & ... & ... \\
5293.160 & 0.8 & 28 & 0.8 & ... & ... & ... & ... & ... & ... & ... & ... & ... & ... & ... & ... & ... & ... & ... & ... & ... & ... & ... \\
5319.810 & 0.6 & 31 & 0.8 & ... & ... & ... & ... & ... & ... & ... & 31 & -0.3 & ... & ... & ... & ... & 32 & -0.6 & ... & ... & ... & ... \\
4844.210 & 0.3 & 21 & 0.8 & ... & ... & ... & ... & ... & ... & ... & ... & ... & ... & ... & ... & ... & ... & ... & ... & ... & ... & ... \\
\multicolumn{23}{c}{\textbf{EuII}} \\
4129.725 & 0.0 & synth & 0.4 & ... & ... & ... & ... & ... & ... & ... & ... & ... & ... & ... & ... & ... & synth & -1.0 & ... & ... & ... & ... \\
4205.042 & 0.0 & synth & 0.4 & ... & ... & ... & ... & ... & ... & ... & ... & ... & ... & synth & -1.65 & ... & synth & -1.3 & ... & synth & -1.23 & ... \\
\multicolumn{23}{c}{\textbf{DyII}} \\
4449.700 & 0.0 & 50 & 1.2 & ... & ... & ... & ... & ... & ... & ... & ... & ... & ... & 23 & -0.6 & ... & 74 & 0.2 & ... & ... & ... & ... \\
\enddata
\end{deluxetable}
\end{longrotatetable}
\newpage

\end{document}